\pdfoutput=1
\documentclass[11pt]{article} 

\usepackage[]{ajtex} 

\usepackage{todonotes}
\usepackage{csquotes}

\DeclareRobustCommand{\DIEP}{\ensuremath{%
    \mathchoice{\includegraphics[height=2ex]{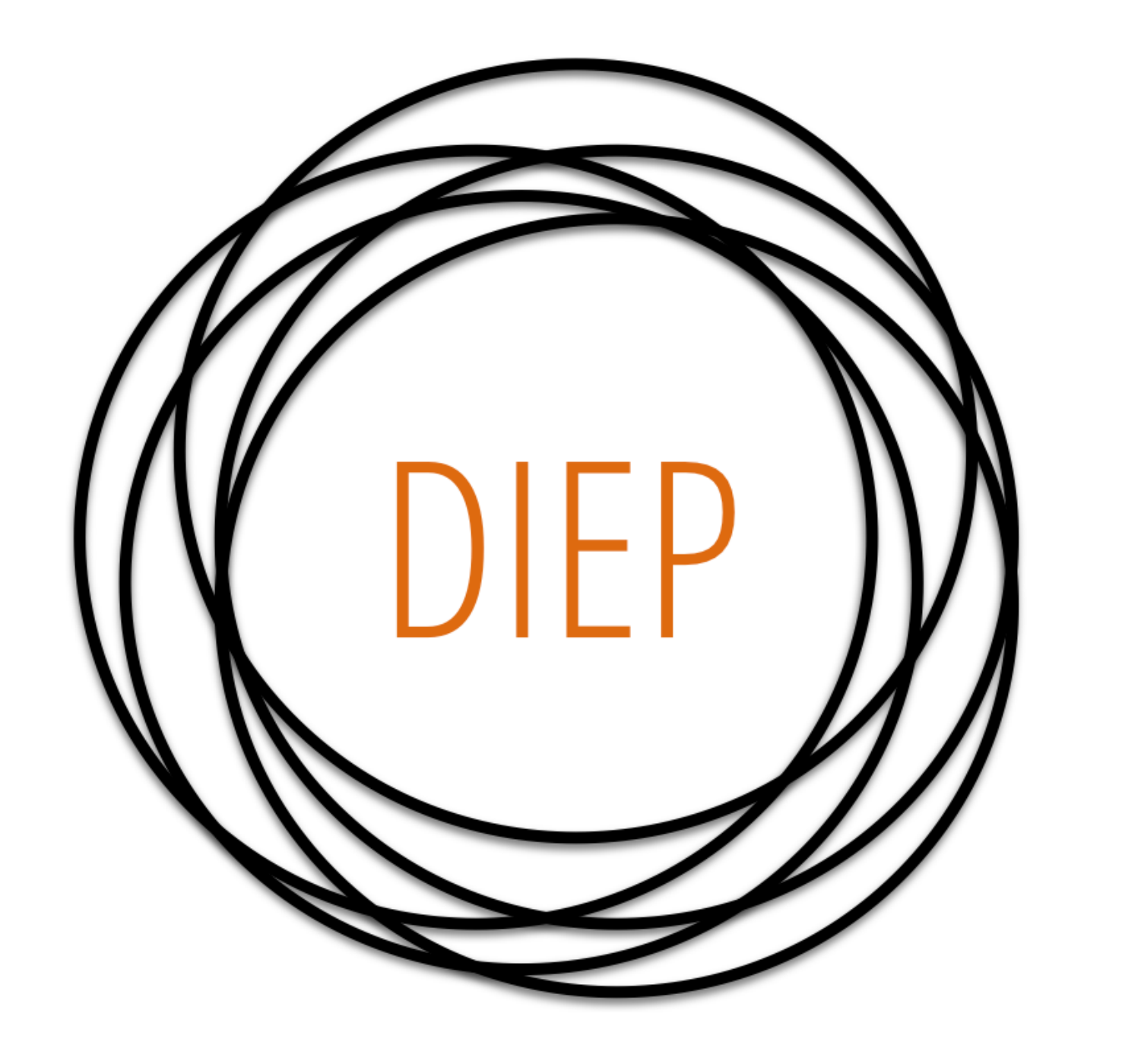}}
    {\includegraphics[height=2ex]{DIEPs.pdf}}
    {\includegraphics[height=1.5ex]{DIEPs.pdf}}
    {\includegraphics[height=1ex]{DIEPs.pdf}}
  }}

\crefname{equation}{eq.}{eqs.}
\crefname{section}{sec.}{secs.}
\crefname{appendix}{app.}{apps.}
\crefname{figure}{fig.}{figs.}
\crefname{table}{tab.}{tabs.} 

\title{One-form superfluids \& magnetohydrodynamics}

\author[a,\DIEP]{Jay Armas}\email{j.armas@uva.nl} 
\author[b]{Akash Jain}\email{ajain@uvic.ca}

\affiliation[a]{Institute for Theoretical Physics, University of Amsterdam, 1090
  GL Amsterdam, The Netherlands}
\affiliation[\DIEP]{Dutch Institute for Emergent Phenomena, The Netherlands}
\affiliation[b]{Department of Physics \& Astronomy, University of Victoria, PO
  Box 1700 STN CSC, Victoria, BC, V8W 2Y2, Canada}

\newcommand\ext{\text{ext}}

\renewcommand\fbox{\fcolorbox{black}{lightgray}}

\abstract{We use the framework of generalised global symmetries to study various
  hydrodynamic regimes of hot electromagnetism. We formulate the hydrodynamic
  theories with an unbroken or a spontaneously broken U(1) one-form
  symmetry. The latter of these describes a one-form superfluid, which is
  characterised by a vector Goldstone mode and a two-form superfluid
  velocity. Two special limits of this theory have been studied in detail: the
  string fluid limit where the U(1) one-form symmetry is partly restored, and
  the electric limit in which the symmetry is completely broken. The transport
  properties of these theories are investigated in depth by studying the
  constraints arising from the second law of thermodynamics and Onsager's
  relations at first order in derivatives. We also construct a hydrostatic
  effective action for the Goldstone modes in these theories and use it to
  characterise the space of all equilibrium configurations. To make explicit
  contact with hot electromagnetism, the traditional treatment of
  magnetohydrodynamics, where the electromagnetic photon is incorporated as
  dynamical degrees of freedom, is extended to include parity-violating
  contributions. We argue that the chemical potential and electric fields are
  not independently dynamical in magnetohydrodynamics, and illustrate how to
  eliminate these within the hydrodynamic derivative expansion using Maxwell's
  equations. Additionally, a new hydrodynamic theory of non-conducting, but
  polarised, plasmas is formulated, focusing primarily on the magnetically
  dominated sector. Finally, it is shown that the different limits of one-form
  superfluids formulated in terms of generalised global symmetries are exactly
  equivalent to magnetohydrodynamics and the hydrodynamics of non-conducting
  plasmas at the non-linear level. }

\begin{document}

\maketitle 

\section{Introduction}

Hot electromagnetism is the theory that describes the interaction between
electromagnetic and thermal degrees of freedom of matter at finite
temperature. At sufficiently long wavelengths and time scales, this theory
admits certain hydrodynamic regimes within which these interactions are well
approximated by the physics of plasmas. Magnetohydrodynamics (MHD) is one of the
most well studied of these regimes, applicable to conducting plasmas for which
the electric fields are short range/Debye screened and the plasma is
electrically neutral at hydrodynamic length
scales~\cite{davidson2001introduction}. Over the past decades, MHD has developed
into a framework capable of describing a wide range of phenomena, from the
modelling of accretion disks surrounding astrophysical black holes to the
magnetic confinement of hot plasmas at fusion
reactors~\cite{goedbloed2004principles}.

Despite its historical success as a phenomenological theory, the
traditional treatments of MHD have only
recently began to incorporate some of the modern developments in
hydrodynamics~\cite{Hernandez:2017mch}, which have proven to be extremely useful
to further our understanding of ordinary fluid and superfluid
dynamics~\cite{Bhattacharya:2011tra, Bhattacharyya:2012nq}. These developments,
among many others, include: the understanding of hydrodynamics as an effective
field theory~\cite{Kovtun:2012rj}; the relevance of hydrostatic partition
functions that describe all equilibrium states in
hydrodynamics~\cite{Jensen:2012jh, Banerjee:2012iz}; the role of symmetries and
classification schemes in constraining transport properties~\cite{Haehl:2014zda,
  Haehl:2015pja}; the usefulness of black hole physics and holography in the
evaluation of transport coefficients~\cite{Bhattacharyya:2008jc,
  Rangamani:2009xk}; a Lagrangian formulation of dissipative
hydrodynamics~\cite{Glorioso:2017lcn, Haehl:2018lcu, Jensen:2018hse}; the
incorporation of boundaries/surfaces in hydrodynamic
descriptions~\cite{Armas:2015ssd, Armas:2016xxg}; a novel understanding of
non-relativistic limits~\cite{Jensen:2014ama, Banerjee:2015hra}; and the
application of the framework of generalised global symmetries to reformulate
hydrodynamic theories~\cite{Schubring:2014iwa, Grozdanov:2016tdf, Armas:2018ibg,
  Grozdanov:2018ewh, Armas:2018atq, Grozdanov:2018fic}.

The overarching goal of this work is to further develop the effective
hydrodynamic theories of hot electromagnetism under the light of some of these
recent developments, and to investigate another of its hydrodynamic regimes
besides MHD. In particular, we provide a new formulation of 
dissipative MHD in terms of a system with higher-form conservation laws, which
is better suited for numerical studies, classify all dissipative transport coefficients
that appear at first order in a long-wavelength expansion 
and resolve standing issues related to the definition of hydrostatic equilibrium.
Besides providing a new framework for understanding
the MHD regime, this work also focuses on a novel formulation of the
hydrodynamic description of non-conducting plasmas that can nevertheless be
polarised, which we refer to as \emph{bound-charge plasmas}. Physical examples
of such systems include a polarised neutral gas of atoms interacting with
a bath of photons.

The main tool used throughout this work is the framework of generalised global
symmetries~\cite{Gaiotto:2014kfa}, which has recently been used in the context
of MHD, recasting it as a theory of hydrodynamics with a global $\rmU(1)$
one-form symmetry~\cite{Schubring:2014iwa, Grozdanov:2016tdf,
  Armas:2018atq}.\footnote{Throughout this work, we often refer to to this
  formulation as the string fluid formulation of MHD.} The traditional treatment
of MHD involves incorporating the electromagnetic photon $A_\mu$ as a dynamical
degree of freedom in the hydrodynamic description, coupled to an external
conserved current $J^\mu_{\text{ext}}$ (see e.g. \cite{Hernandez:2017mch}). On
the other hand, the corresponding string fluid formulation, originates from the
insight that electromagnetism admits a two-form current
$J^{\mu\nu} = \epsilon^{\mu\nu\lambda\rho}\dow_\lambda A_\rho$, where
$F_{\mu\nu}=2\partial_{[\mu}A_{\nu]}$ is the electromagnetic field strength,
that is conserved due to the Bianchi identity
$\nabla_{[\mu} F_{\nu\lambda]}=0$.\footnote{This process of dualisation is
  commonly applied in the context of numerical studies of MHD
  \cite{Gammie:2003rj}. The conservation of the two-form current splits into
  what is usually denoted as the induction equation and the no-monopole
  constraint. However, no formal study of the hydrodynamic properties and
  expansion in this context had been performed. This is one of the goals of this
  paper.} This two-form current gives rise to a dipole charge that counts the
number of magnetic field lines crossing any two-dimensional surface, and couples
to an external two-form gauge field $b_{\mu\nu}$. The three-form field strength
$H_{\mu\nu\lambda}=3\partial_{[\mu}b_{\nu\lambda]}$ associated with $b_{\mu\nu}$
is seen as related to the external conserved current as
$J^\mu_{\text{ext}}=\epsilon^{\mu\nu\lambda\rho}H_{\nu\lambda\rho}/6$. Both
these formulations are developed and extended in this work and, in order to
avoid any ambiguity, one of main results obtained here can be summarised as
follows: \vspace{-0.5em}
\begin{displayquote}
  \begin{itshape}
    Under the identification
    $J^{\mu\nu} = \half \epsilon^{\mu\nu\rho\sigma}F_{\rho\sigma}$ and
    $J^\mu_{\text{ext}} = \frac16
    \epsilon^{\mu\nu\lambda\rho}H_{\nu\lambda\rho}$, the formulation of MHD in
    terms of generalised global symmetries is exactly equivalent to the
    traditional treatment of MHD with a dynamical gauge field.
  \end{itshape}
\end{displayquote}
\vspace{-0.5em}
A few remarks are now in order: this equivalence is proven here at the full
non-linear level including parity-violating terms; both the formulations make no
assumptions regarding the strength of the magnetic fields; and both the
formulations are developed using the principles of effective field theory and
hydrodynamic expansions. Finally, the traditional treatment as developed here,
following \cite{Hernandez:2017mch}, is more general than its corresponding
formulation in terms of generalised global symmetries, as it is capable of
describing plasmas that are not necessarily electrically neutral at the
hydrodynamic length scales.\footnote{It may be possible to relax the assumptions
  of the string fluid formulation in order to be able to describe plasmas that
  are not electrically neutral. Further comments on this point are left to a
  more speculative discussion in \cref{sec:discussion}.}
  
Despite the formulation of MHD in terms of generalised global symmetries, as
thus far developed, being less general than the corresponding traditional
treatment, it should be noted that there are several important reasons why this
different formulation is actually more useful. Most applications of MHD, 
specially in the context of astrophysics, concern themselves
with plasmas that are electrically neutral at the hydrodynamic length
scales~\cite{davidson2001introduction}, in which case both of these formulations
are equally applicable in general, but the formulation in
terms of generalised global symmetries is easier to implement in numerical simulations \cite{Gammie:2003rj}.
 Moreover, when expressed in terms of
generalised global symmetries, the formulation rests solely on the symmetry
principles (and their breaking), without having to incorporate a microscopic
dynamical gauge field. Additionally, the chemical potential $\mu$ and electric
fields $E^\mu$ that enter the traditional formulation, but not the string fluid
formulation, are superfluous and not independently dynamical in the hydrodynamic
regime. As a matter of fact, we show in the course of this work how 
Maxwell's equations can be exactly solved within a derivative expansion, so as
to completely remove these fields from the hydrodynamic description. Finally,
within this string fluid formulation, we directly obtain the fluid constitutive
relations for the physically observable electromagnetic fields in terms of the
background current sources, which allow for a cleaner extraction of the
respective correlation functions.

Earlier formulations of MHD within the framework of generalised global
symmetries~\cite{Schubring:2014iwa, Grozdanov:2016tdf} (see
also~\cite{Hernandez:2017mch}) take the viewpoint that MHD is a theory of long
fluctuating strings (i.e. magnetic field lines). The string direction $h^\mu$
and their chemical potential $\varpi$ serve as fundamental degrees of freedom in
the theory, while assuming that the one-form symmetry is unbroken. As has
already been explained in~\cite{Armas:2018atq}, while this treatment is
phenomenologically sufficient to understand the hydrodynamic fluctuations around
a given initial equilibrium fluid configuration, it does not allow for a precise
understanding of the space of allowed equilibrium configurations by means of a
hydrostatic effective action (or partition function). This problem can be
resolved, as advocated in \cite{Armas:2018atq}, by carefully breaking the
one-form symmetry along the direction of the fluid flow, which leads to the
exact same description for string fluids out of equilibrium as presented
in~\cite{Schubring:2014iwa, Grozdanov:2016tdf}. However, it is now possible to
properly define \emph{equilibrium} configurations by constructing a hydrostatic
effective action for the a magnetic scalar potential $\varphi$, which can be
understood as the Goldstone scalar associated with the partially broken one-form
symmetry.\footnote{The conventional formulation of MHD has a massless
  propagating degree of freedom, namely the photon. However, electric fields in
  MHD are screened. This means that in the dual formulation of MHD in terms of
  generalised global symmetries, not all the components of the \emph{dual
    photon}, which can be seen as the Goldstone of the spontaneously broken
  one-form symmetry \cite{Gaiotto:2014kfa, Lake:2018dqm, Hofman:2018lfz}, are
  actually present. However, since the magnetic fields are still unscreened, at
  least some components of the dual photon must still exist. Therefore, we refer
  to this phase as a partially broken phase of a one-form symmetry.}  The theory
is thus better understood as a theory of one-form superfluidity.

This work introduces a novel framework of one-form superfluids in which the
one-form symmetry is completely broken, giving rise to a vector Goldstone mode
$\varphi_\mu$ \cite{Gaiotto:2014kfa, Lake:2018dqm, Hofman:2018lfz}. A specific
sector of this theory, where part of the one-form symmetry is restored,
describes MHD. In general, however, one-form superfluids characterise many
hydrodynamic regimes of hot electromagnetism without any assumption on the
relative strength of electric and magnetic fields. As an example, the theory
will be used to describe the hydrodynamic regime of magnetically dominated
bound-charge plasmas (BCP), whose traditional treatment has also been developed
here and shown to be equivalent. Below, the different connections between
one-form superfluids and different aspects of hot electromagnetism are described
in more detail, together with the organisation of this paper and some of its
main results. A comparison between the number of transport coefficient in
various phases of neutral, zero-form, and one-form hydrodynamics is given in
\cref{tab:transportcount}.

\begin{table}[t]
  \centering
  \begin{tabular}{lcccccc}
    \toprule
    & Neutral
    & Zero-form & Zero-form & One-form & One-form  & One-form \\
    & Fluid
    & Ord. Fluid & Superfluid & Ord. Fluid & String Fluid
                                                   & Superfluid \\
    \midrule
    Hydrostatic     & 0 & 0 & 4 & 1 & 5 & 22 \\
    Non-hs Non-diss & 0 & 0 & 11 & 0 & 8 & 66  \\
    Dissipative     & 2 & 3 & 15 & 4 & 11 & 78 \\
    \midrule
    Total           & 2 & 3 & 30 & 5 & 24 & 166  \\
    \bottomrule
  \end{tabular}
  \caption{Comparison between counting of independent transport coefficients at
    one-derivative order for various phases of neutral, zero-form, and one-form
    charged fluids. CPT and Onsager's relations have not been implemented in this
    count. Here \emph{Non-hs} refers to non-hydrostatic and \emph{Non-diss} to
    non-dissipative.}
  \label{tab:transportcount}
\end{table}

\subsubsection*{One-form hydrodynamics and hot electromagnetism}

One of main purposes of this work is to contribute to a systematic study of
one-form hydrodynamics and its applications. As such, this paper begins in
\cref{sec:setup} with a discussion on the proper identification of the degrees
of freedom in one-form hydrodynamics, motivated from considerations in
equilibrium thermal field theories. This section also introduces the general
methodology of one-form hydrodynamics (adiabaticity equation, second law of
thermodynamics, hydrostatic effective actions, etc) that will be used in later
sections to formulate novel theories of hydrodynamics with generalised global
symmetries. The identification of the correct degrees of freedom of one-form
hydrodynamics leads to a warm up exercise: the formulation of one-form
hydrodynamics for which the one-form symmetry is unbroken, in
\cref{sec:ordinary}.
\begin{figure}[h!]
  \begin{center}
    \includegraphics[width=0.8\textwidth]{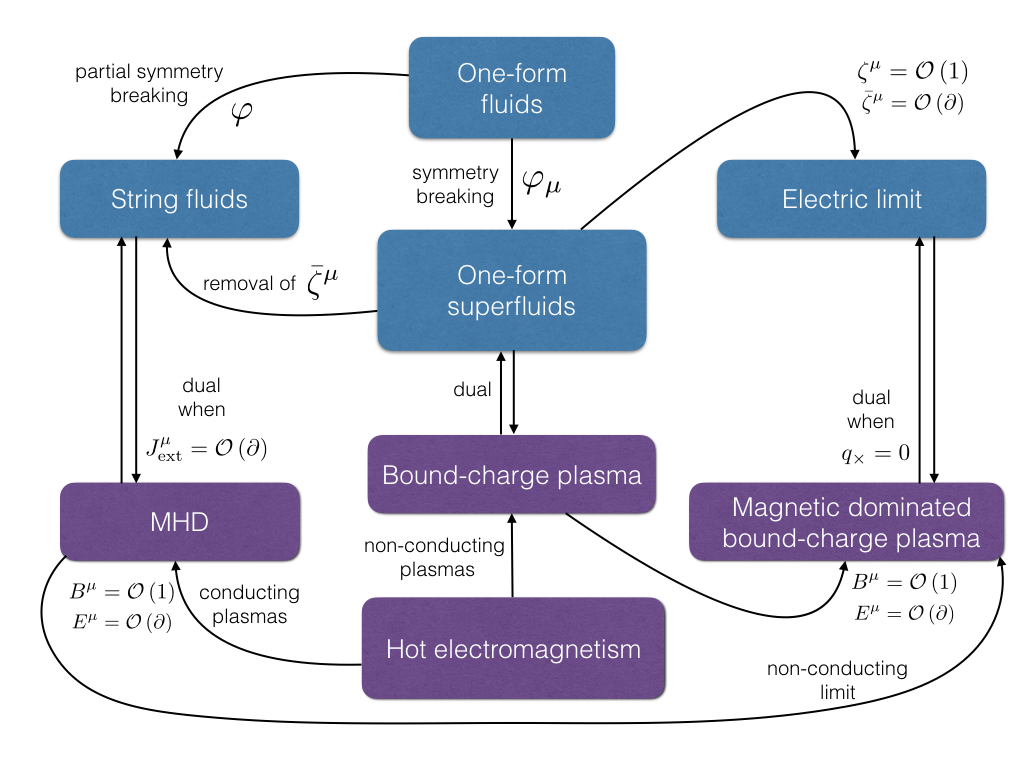} 
  \end{center}
  \caption{Schematic representation of the connections between one-form
    (super)fluids and hot electromagnetism} \label{fig:diagram}
\end{figure}
This theory turns out to be quite different from string fluids as formulated in
previous works, which are naively assumed to have the one-form symmetry
unbroken, and had not previously been considered in the literature.  Having
formulated the theory of one-form hydrodynamics, this work progresses by
incorporating the vector Goldstone mode $\varphi_\mu$ arising due to the
spontaneous breaking the one-form symmetry (see \cref{fig:diagram}).  This makes
up the core of \cref{sec:super}, where a theory of one-form superfluids is
developed and its different limits described. This theory introduces a two-form
superfluid velocity $\xi_{\mu\nu}$ (the gauge-invariant covariant derivative of
$\varphi_\mu$) which in four spacetime dimensions can be decomposed into two
vectors, $\zeta^{\mu}$ and $\bar\zeta^\mu$. These can be understood as electric
and magnetic fields associated with $\xi_{\mu\nu}$, respectively.

We study two limits of one-form superfluids in detail: the string fluid limit
and the electric limit. The string fluid limit, discussed in
sec.~\ref{sec:stringfluids}, can be obtained by partially breaking the one-form
symmetry along the fluid velocity $u^\mu$, which results in the appearance of a
scalar Goldstone mode $\varphi$. The same theory can also be obtained directly
from one-form superfluids by dropping any dependence on $\bar\zeta^\mu$ from the
constitutive relations (see \cref{fig:diagram}). The scalar Goldstone $\varphi$,
in this interpretation, is understood as the time component of the vector
Goldstone mode, that is $\varphi=u^\mu\varphi_\mu/T$, where $T$ is the fluid
temperature. On the other hand, the electric limit taken in \cref{sec:electric}
does not switch off the $\bar\zeta^\mu$ dependence. Rather, it assumes a
derivative hierarchy $\zeta^\mu=\mathcal{O}(1)$ and
$\bar\zeta^\mu=\mathcal{O}(\partial)$ between the components of $\xi_{\mu\nu}$,
rendering $\bar\zeta^\mu$ subleading in the hydrodynamic derivative
expansion. Though equivalent at ideal order, string fluids and the electric limit of
one-form superfluids deviate considerably upon including one-derivative
corrections.

In \cref{sec:hotEM}, the connections between one-form superfluids, including its
limits, and different hydrodynamic regimes of hot electromagnetism are
discussed. We have specially focused on two regimes: MHD and bound-charge
plasmas. The MHD regime, applicable to conducting plasmas for which the magnetic
fields are arbitrary $B^\mu=\mathcal{O}(1)$ and electric fields are weak
$E^\mu=\mathcal{O}(\partial)$, is shown to be exactly equivalent to string
fluids when $J^\mu_{\text{ext}}=\mathcal{O}(\partial)$, as advertised
earlier. The full map between the transport coefficients in the two formulations
at first order in derivatives is given, together with the solution to the
Maxwell's equations that eliminates non-propagating degrees of freedom from the
hydrodynamical description. Here, the traditional treatment of MHD is also
extended to include all transport coefficients at first order in derivatives,
taking into account parity-violating terms.  Also in \cref{sec:hotEM}, the
traditional treatment of the bound-charge plasma regime is formulated for the
first time, and is applicable to non-conducting plasmas (i.e. plasmas with
only bound-charges and no free charge carriers). These are argued to be exactly
equivalent to one-form superfluids, with the explicit mapping of constitutive
relations worked out at ideal order. At first order in derivatives, attention is
given to the magnetic dominated bound-charge plasma, where
$B^\mu=\mathcal{O}(1)$ and $E^\mu=\mathcal{O}(\partial)$, similarly to MHD. These
are shown to be exactly equivalent to the electric limit of one-form superfluids,
provided that a certain transport coefficient $q_\times$ is set to zero. These
connections have been summarised in \cref{fig:diagram}.

Finally, in \cref{sec:discussion} a discussion of some of these results is given
together with interesting future research directions. Some of the calculational
details relevant to this work have been assembled into
\cref{app:hydrostatic}. We have provided a comparison of our results with the
effective action approach of~\cite{Glorioso:2018kcp} in
\cref{app:action}. We also clarify the constraints imposed by discrete
symmetries, such as parity and CPT, in various phases of one-form hydrodynamics
in \cref{app:CPT}.

\subsubsection*{Comments on related work}

During the completion of this work, we became aware of an upcoming related work
that investigates different aspects of
magnetohydrodynamics~\cite{Glorioso:2018kcp}, and which has considerable overlap with \cite{Armas:2018atq}. 
We have provided a comparison between our work and that of \cite{Glorioso:2018kcp} in app.~\ref{app:action}. We also generalised parts of \cite{Glorioso:2018kcp} as to construct
an ideal order effective Lagrangian for the hydrodynamic theories of sec.~\ref{sec:ordinary} and \ref{sec:super}. Additionally,
we have also formulated an order parameter that describes the partial breaking of the one-form symmetry required to formulate MHD in the language
of generalised global symmetries.

\section{The setup of one-form hydrodynamics}
\label{sec:setup}

In this section we introduce the fundamental degrees of freedom associated with
one-form hydrodynamics and the conservation equations that constrain and govern
their dynamical evolution, including in the presence of gapless modes. These
degrees of freedom are motivated by extending the degrees of freedom characterising
thermal equilibrium partition functions into the out-of-equilibrium
context. Analogous to the case of usual zero-form charged hydrodynamics, the
symmetry properties of the background fields to which these fluids couple to are
key guiding principles in the identification of the correct degrees of
freedom. The requirement that one-form fluids satisfy the second law of
thermodynamics leads to a generalised adiabaticity equation that can be used to
constrain the transport properties of one-form fluids. The formalism described
here and associated set of tools (hydrostatic effective action, adiabaticity
equation, etc) is the point of departure for the construction of novel theories
of hydrodynamics with generalised global symmetries that we provide in later
sections of this paper.

\subsection{Symmetries, conservation, and hydrodynamic variables}

The Noether theorem ascertains that any theory that is invariant under global
Poincar\'e transformations and $\rmU(1)$ zero-form transformations must admit a
conserved energy-momentum tensor $T^{\mu\nu}$ and charge current
$J^\mu$. Coupling the theory to a spacetime background with metric $g_{\mu\nu}$
and gauge field $A_\mu$, the conservation equations associated with these
symmetries take the form\footnote{In writing \cref{0formCons} we have assumed
  that the symmetries are non-anomalous. }
\begin{equation}\label{0formCons}
  \nabla_\mu T^{\mu\nu} = F^{\nu\rho} J_\rho~~,\qquad
  \nabla_\mu J^\mu = 0~~.
\end{equation}
Here $\nabla_\mu$ is the covariant derivative associated with $g_{\mu\nu}$ and
$F_{\mu\nu} = 2\dow_{[\mu}A_{\nu]}$ is the field strength associated with
$A_\mu$. Focusing on the case of four spacetime dimensions, \cref{0formCons}
consists of a system of five conservation equations. Hydrodynamics is the
low-energy effective description at finite temperature of such systems and its
formulation requires picking an arbitrary set of five dynamical fields, as in
\cref{table:dyn0},
\begin{table}[h!]
  \begin{center}
    \begin{tabular}{|c|c|}
      \hline
      \textbf{Dynamical field} & \textbf{symbol} \\
      \hline
      Fluid velocity & $u^\mu ~ \text{with}~ u^\mu u_\mu = -1$ \\
      Temperature& $T$ \\
      Zero-form chemical potential & $\mu$ \\
      \hline
    \end{tabular}   
    \caption {Dynamical fields for zero-form charged fluids.}
    \label{table:dyn0}
  \end{center}
\end{table}
whose dynamics is governed by \cref{0formCons}. If, besides the hydrodynamic
modes, the system admits gapless modes at low energy, collectively represented
by $\Phi$, then \cref{0formCons} must be supplied with additional equations of
motion describing the evolution of $\Phi$. Once the dynamical fields have been
chosen and the gapless modes identified, the hydrodynamic theory is obtained by
writing down the most generic ``constitutive relations'' for $T^{\mu\nu}$ and
$J^\mu$ in terms of $u^\mu$, $T$, $\mu$, and $\Phi$ in a long-wavelength
derivative expansion. Empirical physical requirements, such as the second law of
thermodynamics and Onsager's relations, impose constrains on these constitutive
relations.

The motivation for the choice of hydrodynamic fields as in \cref{table:dyn0}
originates from considerations in equilibrium thermal field theories, as we now
outline. Under a generic infinitesimal symmetry transformation
parametrised by $\scX = (\chi^\mu,\Lambda^\chi)$, where $\chi^\mu$ is associated
with diffeomorphisms and $\Lambda^\chi$ with gauge transformations, the
background fields transform according to
\begin{gather}
  \delta_\scX g_{\mu\nu} = \lie_\chi g_{\mu\nu} = 2 \nabla_{(\mu} \chi_{\nu)}~~,
  \nn\\
  \delta_\scX A_\mu = \lie_\chi A_\mu + \dow_\mu \Lambda^\chi
  = \dow_\mu \lb \Lambda^\chi + \chi^\mu A_\mu \rb + \chi^\nu F_{\nu\mu}~~,
\end{gather}
while the symmetry parameters themselves transform
as\footnote{\label{foot:symmAlgebra} Symmetry transformations of the background
  are required to form a Lie algebra such that
  $[\delta_\scX,\delta_{\scX'}] g_{\mu\nu} = \delta_{[\scX,\scX']} g_{\mu\nu}$
  and $[\delta_\scX,\delta_{\scX'}] A_{\mu} = \delta_{[\scX,\scX']} A_{\mu}$,
  which fixes \cref{eq:symmAlgebra_0form}. Similarly in the case of one-form
  symmetries, requiring
  $[\delta_\scX,\delta_{\scX'}] g_{\mu\nu} = \delta_{[\scX,\scX']} g_{\mu\nu}$
  and
  $[\delta_\scX,\delta_{\scX'}] b_{\mu\nu} = \delta_{[\scX,\scX']} b_{\mu\nu}$
  fixes the transformation properties \cref{eq:symmAlgebra_1form} provided that
  we require the fields to transform appropriately under diffeomorphisms.}
\begin{equation}\label{eq:symmAlgebra_0form}
  \delta_\scX \scX' = [\scX,\scX'] = \lb \lie_\chi \chi'^\mu ~,~
  \lie_\chi \Lambda'^\chi - \lie_{\chi'} \Lambda^\chi \rb~~.
\end{equation}
We assume that the background manifold admits a timelike isometry
$\scK = (\textbf{k}^\mu, \Lambda^{\textbf{k}})$ with
$\textbf{k}^\mu \textbf{k}_\mu < 0$, i.e.
$\delta_\scK g_{\mu\nu} = \delta_\scK A_\mu = 0$. On such backgrounds, we can
define a global thermal state by the grand-canonical partition
function\footnote{For example, in the standard case of a static fluid coupled to
  a flat background $g_{\mu\nu} = \eta_{\mu\nu}$ and no external gauge fields
  $A_\mu = 0$, one works with $\scK = (\mathbf k^\mu = \delta^\mu_t/T_0,
  \Lambda^{\mathbf k} = \mu_0/T_0)$, where $T_0$ and $\mu_0$ are the temperature
and chemical potential of the global thermal state. In this case we get the
conventional expression for the grand-canonical partition function $\mathcal{Z}
= \tr\exp\lB -T_0^{-1} \int \df^3 x \lb T^{00} - \mu_0 J^0 \rb \rB$. Note that
we can always perform a gauge transformation to set $\Lambda^{\mathbf k} = 0$ at
the expense of $A_\mu = \mu_0 \delta_\mu^t$, leading to the same result.}
\begin{equation} \label{eq:part1}
  \mathcal{Z}[g_{\mu\nu},A_\mu] = \tr\exp\lB \int_\Sigma \df\sigma_\mu
  \lb T^{\mu\nu} \textbf{k}_\nu
  + (\Lambda^{\textbf{k}} + \textbf{k}^\lambda A_\lambda) J^\mu \rb
  \rB ~~,
\end{equation}
where the trace is taken over all the equilibrium configurations of $\Phi$ which
satisfy $\delta_\scK \Phi = 0$.  In \cref{eq:part1}, $\Sigma$ denotes an
arbitrary Cauchy slice with volume element $\df\sigma_\mu$. Using
\cref{0formCons}, it may be verified that $\cZ$ is independent of the choice of
$\Sigma$ and it is also manifestly invariant under the symmetries of the
theory. It is the aim of hydrodynamics to describe slight departures from the
global thermal state by replacing the background isometry $\scK$ with an
arbitrary set of slowly varying dynamical fields
$\scB = (\beta^\mu, \Lambda^\beta)$, which are related to those in
\cref{table:dyn0} via
\begin{equation} \label{eq:hydro0fields}
  \beta^\mu = \frac{u^\mu}{T} ~~,~~
  \Lambda^\beta + \beta^\mu A_\mu = \frac{\mu}{T}~~.
\end{equation}
This is the more natural way to think of the hydrodynamic degrees of freedom.
As detailed below, the identification of the correct degrees of freedom in the
case of one-form fluids follows a similar reasoning whose starting point is the
equilibrium partition function.

Analogous to systems invariant under zero-form $\rmU(1)$ transformations, physical systems that are invariant under global Poincar\'e and $\rmU(1)$ one-form
transformations admit a conserved energy-momentum tensor $T^{\mu\nu}$ and two-form charge current $J^{\mu\nu}$ such that
\begin{equation} \label{eq:1formcon}
  \nabla_\mu T^{\mu\nu} = \half H^{\nu\rho\sigma} J_{\rho\sigma} ~~,\qquad
  \nabla_\mu J^{\mu\nu} = 0~~,
\end{equation}
where $H_{\mu\nu\rho} = 3 \dow_{[\mu} b_{\nu\rho]}$ is the field strength
associated with a two-form gauge field $b_{\mu\nu}$. In order to describe the
effective low-energy hydrodynamic theory for systems with a global $\rmU(1)$
one-form symmetry, a suitable choice of dynamical fields is required. As in the
case of zero-form fluids, it is noted that under a generic infinitesimal
one-form symmetry transformation parametrised by
$\scX = (\chi^\mu,\Lambda^\chi_\mu)$, with $\Lambda^\chi_\mu$ being the
parameter associated with one-form gauge transformations, the background fields
transform according to
\begin{gather}
  \delta_\scX g_{\mu\nu} = \lie_\chi g_{\mu\nu} = 2 \nabla_{(\mu} \chi_{\nu)}~~,~~
  \nn\\
  \delta_\scX b_{\mu\nu} = \lie_\chi b_{\mu\nu} + 2\dow_{[\mu} \Lambda^\chi_{\nu]}
  = 2\dow_{[\mu} \lb \Lambda^\chi_{\nu]} + \chi^\lambda b_{\lambda\mu} \rb
  + \chi^\lambda H_{\lambda\mu\nu}~~,
\end{gather}
while the symmetry parameters transform as (see \cref{foot:symmAlgebra})
\begin{equation}\label{eq:symmAlgebra_1form}
  \delta_\scX \scX' = [\scX,\scX'] = \lb \lie_\chi \chi'^\mu ,
  \lie_\chi \Lambda'^\chi_\mu - \lie_{\chi'} \Lambda^\chi_\mu \rb~~.
\end{equation}
When coupled to spacetime backgrounds that admit a timelike isometry
$\scK = (\textbf{k}^\mu, \Lambda^\textbf{k}_\mu)$, we can define a global
thermal state by means of the grand-canonical partition function
\begin{equation}\label{PF1formTr}
  \mathcal{Z}[g_{\mu\nu},b_{\mu\nu}] = \tr\exp\lB \int_\Sigma \df\sigma_\mu
  \lb T^{\mu\nu} \textbf{k}_\nu + (\Lambda^\textbf{k}_\nu + \textbf{k}^\lambda b_{\lambda\nu}) J^{\mu\nu} \rb
  \rB~~.
\end{equation}
Following the same chain of reasoning as for zero-form symmetries, we are led to
the natural choice of hydrodynamic fields for one-form hydrodynamics as
$\scB = (\beta^\mu, \Lambda^\beta_\mu)$. By defining
\begin{equation} \label{eq:mumu}
  \beta^\mu = \frac{u^\mu}{T}~~,~~
  \Lambda^\beta_\mu + \beta^\nu b_{\nu\mu} = \frac{\mu_\mu}{T}~~,
\end{equation}
these fields can be recast in a more conventional form as in \cref{table:dyn1}.
\begin{table}
  \begin{center}
    \begin{tabular}{|c|c|}
      \hline
      \textbf{Dynamical field} & \textbf{symbol} \\
      \hline
      Fluid velocity & $u^\mu \quad \text{with}~ u^\mu u_\mu = -1$ \\
      Temperature& $T$ \\
      One-form chemical potential & $\mu_\mu$ \\
      \hline
    \end{tabular}   
    \caption {Dynamical fields for one-form charged fluids.}
    \label{table:dyn1}
  \end{center}
\end{table}
However, unlike zero-form fluids, $\mu_\mu$ defined in this way is not gauge invariant. Instead, it transforms akin to a one-form gauge field
\begin{equation}
  \delta_\scX \frac{\mu_\mu}{T}
  = \lie_\chi \frac{\mu_\mu}{T}
  - \dow_\mu \lb \beta^\nu \Lambda^\chi_\nu \rb~~.
\end{equation}
This should not come as a surprise since the time-component of the one-form
conservation equation, $\nabla_\mu J^{\mu t}$, is not a dynamical equation but
merely a constraint \cite{Armas:2018ibg}. Correspondingly, one degree of freedom
in $\mu_\mu$ is rendered unphysical due to the gauge transformation. Note that
since $\mu_\mu/T$ transforms as a one-form gauge field, all of its
gauge-invariant physical information can be captured by the antisymmetric
derivative $\dow_{[\mu}(\mu_{\nu]}/T)$. Having identified the dynamical fields
for one-form fluids, we proceed by defining the hydrostatic partition function
and deriving the adiabaticity equation.

\subsection{Hydrostatic effective action and the second law of thermodynamics}

The hydrostatic effective action is an important cornerstone of
hydrodynamics. It describes the entire set of equilibrium configurations
admissible by the fluid for a given arrangement of the background sources. These
configurations can then be used as a starting point for studying deviations away
from equilibrium order by order in the derivative expansion (e.g. dispersion
relations for Alfv\'en waves in MHD). In this subsection we introduce the
generalities of hydrostatic effective action relevant to this work and
illustrate their connection to the second law of thermodynamics.

For the purposes of this paper, it is assumed that the microscopic field
theories underlying the hydrodynamic regime are sufficiently well behaved, so
that the equilibrium partition function \bref{PF1formTr} can be computed via a
Euclidean path integral
\begin{equation} \label{eq:Eu}
  \mathcal{Z}[g_{\mu\nu},b_{\mu\nu}]
  = \int \cD\Phi \exp \lb - S^{\text{hs}}[g_{\mu\nu},b_{\mu\nu};\Phi] \rb~~.
\end{equation}
$S^{\text{hs}}[g_{\mu\nu},b_{\mu\nu};\Phi]$, known as the hydrostatic effective
action, contains all the possible diffeomorphism and gauge-invariant terms
composed of $g_{\mu\nu}$, $b_{\mu\nu}$, and $\Phi$ in the presence of a timelike
isometry $\scK$. The variation of the effective action with respect to the
background sources yields the conserved currents
\begin{equation}\label{eqbConsti}
  T^{\mu\nu}_{\text{hs}}
  = \frac{2}{\sqrt{-g}} \frac{\delta S^{\text{hs}}[g_{\mu\nu},b_{\mu\nu};\Phi]}{\delta
    g_{\mu\nu}} ~~,\qquad
  J^{\mu\nu}_{\text{hs}}
  = \frac{2}{\sqrt{-g}} \frac{\delta S^{\text{hs}}[g_{\mu\nu},b_{\mu\nu};\Phi]}{\delta
    b_{\mu\nu}}~~,
\end{equation}
while the equilibrium configurations of the gapless modes are obtained by
extremising the effective action with respect to $\Phi$ leading to
\begin{equation} \label{eq:hydroK}
  K^\Phi_{\text{hs}} = \frac{\delta S^{\text{hs}}[g_{\mu\nu},b_{\mu\nu};\Phi]}{\delta \Phi} = 0~~.
\end{equation}
As a consistency condition on the general hydrodynamic constitutive relations (including dissipative effects), we
require them to match with \cref{eqbConsti} when we revert to the global
thermal state by setting $\scB = \scK$. This requirement yields strict
constraints on their form at every derivative order~\cite{Jensen:2012jh,
  Banerjee:2012iz}.

Schematically, the hydrostatic effective action appearing in \eqref{eq:Eu} can be parametrised as
\begin{equation}
  \label{eq:hydroN}
  S^{\text{hs}}[g_{\mu\nu},b_{\mu\nu};\Phi] =
  \int_\Sigma \df\sigma_\mu N^\mu_{\text{hs}}~~,
\end{equation}
where $N^\mu_{\text{hs}}$ is the hydrostatic free energy current that satisfies
$\nabla_\mu N^\mu_{\text{hs}} = 0$. As we leave the global thermal state, the
free energy current is no longer conserved. To see this, let us slightly depart
from equilibrium by replacing $\scK$ with $\scB$ and performing a
$\scB$-variation of $S^{\text{hs}}$. We obtain the hydrostatic adiabaticity
equation
\begin{equation} \label{eq:ada}
  \nabla_\mu N^\mu_{\text{hs}}
  = \half T^{\mu\nu}_{\text{hs}} \delta_\scB g_{\mu\nu}
  + \half J^{\mu\nu}_{\text{hs}} \delta_\scB b_{\mu\nu}
  + K^\Phi_{\text{hs}} \delta_\scB \Phi~~.
\end{equation}
Physically, it is equivalent to the statement that entropy is conserved in a
hydrostatic configuration. To wit, defining the entropy current as
\begin{equation}\label{eq:EC_hs}
  S^\mu_{\text{hs}} = N_{\text{hs}}^\mu
  - \frac{1}{T} T^{\mu\nu}_{\text{hs}}u_\nu
  - \frac{1}{T} J^{\mu\nu}_{\text{hs}}\mu_\nu~~,
\end{equation}
and using the conservation equations \bref{eq:1formcon}, the adiabaticity
equation can be rewritten as $\nabla_\mu S^\mu_{\text{hs}} = 0$. However, in a
generic out-of-equilibrium hydrodynamic configuration with entropy current
$S^\mu$, we expect entropy to be produced, leading to the second law of
thermodynamics
\begin{equation}
  \nabla_\mu S^\mu = \Delta \geq 0~~.
\end{equation}
Here $\Delta$ is a non-negative quadratic form which vanishes in a hydrostatic
configuration. Correspondingly, the generic adiabaticity equation \eqref{eq:ada}
in the out-of-equilibrium context is an extrapolation of its hydrostatic
counterpart
\begin{equation}\label{eq:adiabaticity} 
  \nabla_\mu N^\mu
  = \half T^{\mu\nu} \delta_\scB g_{\mu\nu}
  + \half J^{\mu\nu} \delta_\scB b_{\mu\nu}
  + K^\Phi \delta_\scB \Phi
  + \Delta~~,~~
  \Delta \geq 0~~~,
\end{equation}
where the different quantities involved may also include non-hydrostatic
contributions, and can be viewed as generalisation of the requirement of a
hydrostatic effective action. A set of physical constitutive relations
$T^{\mu\nu}$, $J^{\mu\nu}$, $K^\Phi$ associated with a hydrodynamic system are
required to be accompanied by a free energy current $N^\mu$ and a quadratic form
$\Delta$ such that \cref{eq:adiabaticity} is satisfied for all fluid
configurations. Below we will show how the adiabaticity equation can be used to
obtain constraints on the hydrodynamic constitutive relations.

Traditionally, one takes a slightly different route with the second law of
thermodynamics as the starting point in order to arrive at these
constraints~\cite{landau2013fluid}. Switching off the extra gapless modes $\Phi$
for the moment, one requires that for every set of physical constitutive
relations $T^{\mu\nu}$, $J^{\mu\nu}$, the hydrodynamic system in question must
admit an associated entropy current $S^\mu$ whose divergence is positive
semi-definite, $\nabla_\mu S^\mu \geq 0$, on the solutions of the conservation
equations. Given such an entropy current, it is always possible to go off-shell
and write down an equivalent statement for the second law by introducing
arbitrary linear combinations of the conservation
equations~\cite{Loganayagam:2011mu} (see also~\cite{Jain:2016rlz})
\begin{equation}\label{eq:offshell_2ndLaw}
  \nabla_\mu S^\mu + \mathcal{A}_\nu \lb \nabla_\mu T^{\mu\nu}
  - \half H^{\nu\rho\sigma} J_{\rho\sigma} \rb
  + \mathcal{B}_\nu \nabla_\mu J^{\mu\nu} = \Delta \geq 0~~.
\end{equation}
Here $\mathcal{A}_\mu$ and $\mathcal{B}_\mu$ are arbitrary multipliers composed of
the hydrodynamic and background fields, and introduced as to satisfy this equation
offshell. Recall that the hydrodynamic fields $u^\mu$, $T$, and $\mu_\mu$ were
some arbitrary set of fields chosen to describe the system, and like in any
field theory, can admit arbitrary field redefinitions. We can use this freedom
to set $\mathcal{A}_\mu = u_\mu/T$ and $\mathcal{B}_\mu = \mu_\mu/T$. Having
done that, and using the relation between free-energy and entropy currents, i.e.
\begin{equation}
  N^\mu = S^\mu
  + \frac{1}{T} T^{\mu\nu} u_\nu
  + \frac{1}{T} J^{\mu\nu} \mu_\nu~~,
\end{equation}
it is easy to see that the offshell second law of thermodynamics
\bref{eq:offshell_2ndLaw} reduces to the adiabaticity equation
\bref{eq:adiabaticity}. Hence the constraints imposed by the second law of
thermodynamics are equivalent to the ones imposed by the adiabaticity
equation. The latter, however, turns out to be functionally advantageous to
implement. An entirely analogous argument follows in the presence of additional
gapless modes~\cite{Jain:2018jxj}.

\subsection{Constitutive relations up to first order}
\label{sec:firstOrderGen}

In the bulk of this paper, we will derive the constitutive relations allowed by
the adiabaticity equation \bref{eq:adiabaticity} up to one-derivative order for
several cases of interest. As shall be explained in the later sections, for all
of these cases, the adiabaticity equation \eqref{eq:adiabaticity} can be reduced
to a simpler version
\begin{equation}\label{adiabaticity_simple} 
  \nabla_\mu N^\mu
  = \half T^{\mu\nu} \delta_\scB g_{\mu\nu}
  + \half J^{\mu\nu} \delta_\scB b_{\mu\nu}
  + \Delta~~,~~
  \Delta \geq 0~~,
\end{equation}
where the $\delta_\scB \Phi$ term has been removed by going onshell and using
the available field redefinition freedom. It is possible to broadly classify the
constitutive relations satisfying \cref{adiabaticity_simple} into hydrostatic,
i.e. constitutive relations that remain independent in a hydrostatic
configuration, and non-hydrostatic, i.e. constitutive relations that vanish in a
hydrostatic configuration.

\begin{table}[t]
    \renewcommand{\arraystretch}{1.4}
  \centering
  \begin{tabular}{cc} 
    \toprule
    \multicolumn{2}{c}{Ordinary one-form fluids} \\
    \midrule
    $\mathcal{N}$ (1)
    & $\epsilon^{\mu\nu\rho\sigma}u_\mu H_{\nu\rho\sigma}$ \\
    $\eta^{(\mu\nu)(\rho\sigma)}$ (2)
    & $P^{\mu\nu} P^{\rho\sigma}$, $P^{\rho(\mu}P^{\nu)\sigma}$ \\
    $\chi^{(\mu\nu)[\rho\sigma]}$, $\chi'^{[\rho\sigma](\mu\nu)}$ (2$\times$0)
    & -- \\
    $\sigma^{[\mu\nu][\rho\sigma]}$ (2)
    & $u^{[\mu}P^{\nu][\rho} u^{\sigma]}$, $P^{\rho[\mu}P^{\nu]\sigma}$ \\
    \toprule
    \multicolumn{2}{c}{One-form string fluids} \\
    \midrule
    $\mathcal{N}$ (5)
    & $h^\mu \dow_\mu T$, $h^\mu \dow_\mu \frac{\varpi}{T}$,
      $\epsilon^{\mu\nu\rho\sigma}u_\mu H_{\nu\rho\sigma}$,
      $\epsilon^{\mu\nu\rho\sigma} u_\mu h_\nu \dow_\rho u_\sigma$,
      $\epsilon^{\mu\nu\rho\sigma} u_\mu h_\nu \dow_\rho h_\sigma$ \\
    $\eta^{(\mu\nu)(\rho\sigma)}$ (8)
    & $h^\mu h^\nu h^\rho h^\sigma$, $h^\mu h^\nu \Delta^{\rho\sigma}$,
      $\Delta^{\mu\nu} h^\rho h^\sigma$,
      $h^{(\mu} \Delta^{\nu)(\rho} h^{\sigma)}$,
      $\Delta^{\mu\nu}\Delta^{\rho\sigma}$,
      $\Delta^{\rho(\mu}\Delta^{\nu)\sigma}$, \\
    &
      $h^{(\mu} \epsilon^{\nu)(\rho} h^{\sigma)}$, 
      $\epsilon^{\rho(\mu} \Delta^{\nu)\sigma}$ \\
    $\chi^{(\mu\nu)[\rho\sigma]}$, $\chi'^{[\rho\sigma](\mu\nu)}$ (2$\times$4)
    & $h^{(\mu}\Delta^{\nu)[\rho}h^{\sigma]}$,
      $h^{(\mu}\epsilon^{\nu)[\rho}h^{\sigma]}$,
      $h^\mu h^\nu \epsilon^{\rho\sigma}$,
      $\Delta^{\mu\nu} \epsilon^{\rho\sigma}$ \\
    $\sigma^{[\mu\nu][\rho\sigma]}$ (3)
    & $h^{[\mu}\Delta^{\nu][\rho}h^{\sigma]}$,
      $h^{[\mu}\epsilon^{\nu][\rho}h^{\sigma]}$,
      $\epsilon^{\mu\nu}\epsilon^{\rho\sigma}$ \\
    \toprule
    \multicolumn{2}{c}{One-form superfluids} \\
    \midrule
    $\mathcal{N}$ (22)
    & $h_a^\mu \dow_\mu T$,
      $\epsilon^{\mu\nu\rho\sigma} u_\mu h_{a\nu} \dow_\rho u_\sigma$,
      $\epsilon^{\mu\nu\rho\sigma} u_\mu h_{a\nu} \dow_\rho
      \zeta_\sigma$, $\epsilon^{\mu\nu\rho\sigma} u_\mu h_{a\nu} \dow_\rho
      \bar\zeta_\sigma$, \\
    & $h^\mu_a h^\nu_b \nabla_{(\mu} \zeta_{\nu)}$,
      $h^\mu_a h^\nu_b \nabla_{(\mu} \bar\zeta_{\nu)}$ \\
    $\eta^{(\mu\nu)(\rho\sigma)}$ (36)
    & $h^{(\mu}_a h^{\nu)}_b h^{(\rho}_c h^{\sigma)}_d$ \\
    $\chi^{(\mu\nu)[\rho\sigma]}$, $\chi'^{[\rho\sigma](\mu\nu)}$ (2$\times$36)
    & $h^{(\mu}_a h^{\nu)}_b h^{[\rho}_c h^{\sigma]}_d$,
    $h^{(\mu}_a h^{\nu)}_b h^{[\rho}_c u^{\sigma]}$ \\
    $\sigma^{[\mu\nu][\rho\sigma]}$ (36)
    & $h^{[\mu}_a h^{\nu]}_b h^{[\rho}_c h^{\sigma]}_d$,
      $u^{[\mu} h^{\nu]}_a h^{[\rho}_b h^{\sigma]}_c$,
      $h^{[\mu}_a h^{\nu]}_b h^{[\rho}_c u^{\sigma]}$,
      $u^{[\mu} h^{\nu]}_a h^{[\rho}_b u^{\sigma]}$ \\
    \bottomrule
  \end{tabular}
  \caption{One-derivative hydrostatic and non-hydrostatic structures for various
    phases of one-form hydrodynamic constitutive relations. Here
    $P^{\mu\nu} = g^{\mu\nu} + u^\mu u^\nu$. For string fluids, $h^\mu$ is the
    string director field while $\varpi$ is the string chemical potential;
    $\Delta^{\mu\nu} = g^{\mu\nu} + u^\mu u^\nu - h^\mu h^\nu$,
    $\epsilon^{\mu\nu} = \epsilon^{\mu\nu\rho\sigma}u_\rho h_\sigma$. For one
    form superfluids, $\xi_{\mu\nu}$ is the superfluid velocity;
    $\zeta_\mu = \xi_{\mu\nu} u^\nu$,
    $\bar \zeta^\mu = \half \epsilon^{\mu\nu\rho\sigma} u_\nu
    \xi_{\rho\sigma}$. On the other hand, $h^\mu_a$ is a set of orthonormal
    vectors made out of $\zeta^\mu$, $\bar\zeta^\mu$, and
    $\epsilon^{\mu\nu\rho\sigma}u_\nu \zeta_\rho \bar\zeta_\sigma$. These
    constitutive relations are \emph{not} generically written in Landau frame,
    but have been expressed in certain frames convenient for each case
    respectively.}
  \label{tab:structures}
\end{table}

The hydrostatic constitutive relations are characterised by a hydrostatic free
energy current
$N^\mu_{\text{hs}} = \mathcal{N} \beta^\mu + \Theta^\mu_{\mathcal{N}}$, where
$\mathcal{N}$ is made out of all the independent hydrostatic scalars, while
$\Theta^\mu_{\mathcal{N}}$ is a non-hydrostatic vector\footnote{Generically,
  $N^\mu_{\text{hs}}$ can also include a hydrostatic part transverse to
  $u^\mu$. Known as Class H$_{\text V}$ constitutive relations or transcendental
  anomalies, these contributions are completely fixed up to a finite number of
  constants~\cite{Haehl:2015pja}. For the cases considered here, such terms turn
  out to be independent of the one-form symmetry sector altogether, and hence
  have been switched off for simplicity.} defined via
\begin{equation}\label{hsConstiGeneral}
  \nabla_\mu \lb \mathcal{N} \beta^\mu \rb
  = \frac{1}{\sqrt{-g}} \delta_\scB \lb \sqrt{-g}\, \mathcal{N} \rb
  = \half \lb \mathcal{N} g^{\mu\nu} + 2\frac{\delta\mathcal{N}}{\delta
    g_{\mu\nu}} \rb \delta_\scB g_{\mu\nu}
  + 2\frac{\delta\mathcal{N}}{\delta b_{\mu\nu}} \delta_\scB b_{\mu\nu}
  - \nabla_\mu \Theta^\mu_{\mathcal{N}}~~.
\end{equation}
Comparing with \cref{adiabaticity_simple}, it is possible to read out the hydrostatic
constitutive relations as
\begin{equation}
  T^{\mu\nu}_{\text{hs}} = \mathcal{N} g^{\mu\nu} + 2\frac{\delta\mathcal{N}}{\delta
    g_{\mu\nu}}~~,~~
  J^{\mu\nu}_{\text{hs}} = 2\frac{\delta\mathcal{N}}{\delta b_{\mu\nu}}~~.
\end{equation}

In turn, the non-hydrostatic constitutive relations up to first order are simply
given as the most generic linear combinations of $\delta_\scB g_{\mu\nu}$ and
$\delta_\scB b_{\mu\nu}$. To wit
\begin{equation}\label{nhsConstiGeneral}
  \begin{pmatrix}
    T^{\mu\nu}_{\text{nhs}} \\
    J^{\mu\nu}_{\text{nhs}}
  \end{pmatrix}
  =
  - T \begin{pmatrix}
    \eta^{(\mu\nu)(\rho\sigma)} & \chi^{(\mu\nu)[\rho\sigma]} \\
    \chi'^{[\mu\nu](\rho\sigma)} & \sigma^{[\mu\nu][\rho\sigma]}
  \end{pmatrix}
  \begin{pmatrix}
    \half \delta_\scB g_{\rho\sigma} \\
    \half \delta_\scB b_{\rho\sigma}
  \end{pmatrix}~~.
\end{equation}
Here $\eta^{(\mu\nu)(\rho\sigma)}$, $\chi^{(\mu\nu)[\rho\sigma]}$,
$\chi'^{(\mu\nu)[\rho\sigma]}$ and $\sigma^{[\mu\nu][\rho\sigma]}$ are the most
general zero-derivative structures, with associated arbitrary transport
coefficients, composed of the hydrodynamic fields identified in the previous
section. In particular, there are no zero-derivative non-hydrostatic
constitutive relations.  Inserting \cref{nhsConstiGeneral} into
\cref{adiabaticity_simple} it can be inferred that they satisfy
\cref{adiabaticity_simple} with $N^\mu_{\text{nhs}} = 0$ and
\begin{equation}\label{DeltaGeneral}
  \Delta = T
  \begin{pmatrix}
    \half \delta_\scB g_{\mu\nu} &
    \half \delta_\scB b_{\mu\nu}
  \end{pmatrix}
  \begin{pmatrix}
    \eta^{(\mu\nu)(\rho\sigma)} & \chi^{(\mu\nu)[\rho\sigma]} \\
    \chi'^{[\mu\nu](\rho\sigma)} & \sigma^{[\mu\nu][\rho\sigma]}
  \end{pmatrix}
  \begin{pmatrix}
    \half \delta_\scB g_{\rho\sigma} \\
    \half \delta_\scB b_{\rho\sigma}
  \end{pmatrix}
  \geq 0~~.
\end{equation}
It follows that the symmetric part of the non-hydrostatic transport coefficient
matrix
\begin{equation}\label{symTransMatrix}
  \half \begin{pmatrix}
    \eta^{(\mu\nu)(\rho\sigma)} + \eta^{(\rho\sigma)(\mu\nu)} &
    \chi^{(\mu\nu)[\rho\sigma]} + \chi'^{[\rho\sigma](\mu\nu)} \\
    \chi'^{(\mu\nu)[\rho\sigma]} + \chi'^{[\rho\sigma](\mu\nu)} &
    \sigma^{[\mu\nu][\rho\sigma]} + \sigma^{[\rho\sigma][\mu\nu]}
  \end{pmatrix}
  \geq 0~~,
\end{equation}
is a positive semi-definite matrix. This requirement imposes certain inequality
constraints on the transport properties of the hydrodynamic theories that we
will study. A summary of the allowed tensor structures in various phases of
one-form hydrodynamics is presented in \cref{tab:structures}.

A priori, the hydrodynamic fields $u^\mu$, $T$, and $\mu_\mu$ are arbitrary
degrees of freedom chosen to describe the hydrodynamic fluctuations. In
equilibrium, these are unambiguously identified with the timelike isometry
$\scK$, but in a generic out-of-equilibrium state, they can admit arbitrary
non-hydrostatic field redefinitions. We can use this freedom to our advantage and
simplify the non-hydrostatic constitutive relations by making a choice of
``hydrodynamic frame''. The most common of such frames is the Landau frame, which fixes
the field redefinition in $u^\mu$ and $T$ by choosing
$T^{\mu\nu}_{\text{nhs}} u_\nu = 0$. The redefinition freedom in $\mu_\mu$ can
be similarly used to set $J^{\mu\nu}_{\text{nhs}} u_\nu = 0$. This leads to
\begin{equation}
  \eta^{(\mu\nu)(\rho\sigma)} u_\mu = \chi^{(\mu\nu)[\rho\sigma]} u_\mu
  = \chi'^{[\mu\nu](\rho\sigma)} u_\mu = \sigma^{[\mu\nu][\rho\sigma]} u_\mu = 0~~.
\end{equation}
To complete the quadratic form $\Delta$ in this frame, we need to further
eliminate $u^\mu \delta_\scB g_{\mu\nu}$ and $u^\mu \delta_\scB b_{\mu\nu}$ from
the non-hydrostatic constitutive relations \bref{nhsConstiGeneral}, which can be
generically done using the conservation equations \bref{0formCons}. Therefore
\begin{equation}
  \eta^{(\mu\nu)(\rho\sigma)} u_\rho = \chi^{(\mu\nu)[\rho\sigma]} u_\rho 
  = \chi'^{[\mu\nu](\rho\sigma)} u_\rho = \sigma^{[\mu\nu][\rho\sigma]} u_\rho = 0~~.
\end{equation}
Hence, all indices in $\eta^{(\mu\nu)(\rho\sigma)}$,
$\chi^{(\mu\nu)[\rho\sigma]}$, $\chi'^{(\mu\nu)[\rho\sigma]}$, and
$\sigma^{[\mu\nu][\rho\sigma]}$ can be taken to be projected orthogonally to the
fluid velocity. We will not restrict ourselves to this frame
choice throughout this work. Instead, we will make a judicious choice of basis
based on the hydrodynamic system under consideration, defaulting to the Landau frame
when no such natural choice is available.

\section{Ordinary one-form fluids} \label{sec:ordinary}

The main topic of interest of this work is one-form superfluids. However, before
delving into the intricacies of one-form superfluid dynamics, it is instructive
to consider ordinary one-form hydrodynamics first. Even though it is
comparatively simpler than the examples that will be studied in later sections,
this section provides the first formulation of one-form fluids in which the
one-form symmetry is unbroken.

At ideal order, this system is trivial because there are no zero-derivative
gauge-invariants that can be constructed from the ideal order hydrodynamic
fields $\mu_\mu$ and $b_{\mu\nu}$ identified in sec.~\ref{sec:setup}.
Consequently, at ideal order one-form fluids are characterised by the same
constitutive relations as ordinary neutral fluids. Precisely
\begin{equation}
  T^{\mu\nu} = \lb \epsilon + p \rb u^{\mu}u^{\nu} + p\, g^{\mu\nu}
  + \mathcal{O}(\dow)~~,~~
  J^{\mu\nu} = \mathcal{O}(\dow)~~,
\end{equation}
along with the thermodynamic relations
\begin{equation}
  "dd p = s\, "dd T~~,~~
  \epsilon + p = s\, T~~,~~
  "dd \epsilon = T\, "dd s~~.
\end{equation}
These constitute relations can be derived from their corresponding hydrostatic free energy density $\mathcal{N} = p(T)$, using \eqref{hsConstiGeneral}, such
that the free energy current is given by $N^\mu = p/T\, u^\mu$. The coefficients
$\epsilon$, $p$, and $s$ are identified as the energy density, isotropic
pressure, and entropy density of the fluid respectively. The first-order
equations of motion simply imply that
\begin{equation}
  u^\mu \nabla_\mu \epsilon + (\epsilon + p) \nabla_\mu u^\mu = 0~~,~~
  \frac{1}{T} P^{\mu\nu}\dow_\nu T + u^\nu \nabla_\nu u^\mu = 0~~,
\end{equation}
which can be collectively used to eliminate $u^\mu \delta_{\scB} g_{\mu\nu}$ from the first-order non-hydrostatic constitutive relations.

At one-derivative order, signatures of one-form symmetry begin to appear.  In
the hydrostatic sector there is only one gauge-invariant contribution to the
hydrostatic free energy density $\mathcal{N}$ at first order, which is given by
\begin{equation}
  \mathcal{N} = p(T)
  - \frac{\alpha(T)}{6} \epsilon^{\mu\nu\rho\sigma} u_\mu H_{\nu\rho\sigma}~~.
\end{equation}
The transport coefficient $\alpha$ is unconstrained by the adiabaticity
equation (second law).  The variation of this corrected free energy density,
according to \cref{hsConstiGeneral}, leads to the hydrostatic constitutive
relations
\begin{align}
  T^{\mu\nu}_{\text{hs}}
  &= \lb \epsilon + p \rb u^{\mu}u^{\nu} + p\, g^{\mu\nu}
    - \frac16 \epsilon^{\alpha\beta\rho\sigma} u_\alpha H_{\beta\rho\sigma}
    \frac{\dow (T\alpha)}{\dow T} u^{\mu}u^{\nu}
    - \frac\alpha3 u^{(\mu} \epsilon^{\nu)\lambda\rho\sigma}
      H_{\lambda\rho\sigma} + \mathcal{O}(\dow^2)~~,  \nn\\
  J^{\mu\nu}_{\text{hs}}
  &=
    \nabla_\sigma \lb \alpha\epsilon^{\mu\nu\rho\sigma} u_\rho\rb
     + \mathcal{O}(\dow^2)~~, \nn\\
  N^\mu_{\text{hs}}
  &= \frac{\alpha}{6T}
    \epsilon^{\mu\nu\rho\sigma} H_{\nu\rho\sigma}
    - \alpha \epsilon^{\mu\nu\rho\sigma} u_\nu  \dow_\rho \bfrac{\mu_\sigma}{T}
    + \mathcal{O}(\dow^2) ~~.
\end{align}
Note that all the dependence on $\mu_\mu$ comes via the antisymmetric derivative
$\dow_{[\mu}(\mu_{\nu]}/T)$, which is gauge-invariant.  The most general
non-hydrostatic corrections, in turn, can be decomposed along and transverse to
$u^\mu$ according to
\begin{align}
  T^{\mu\nu}_{\text{nhs}}
  &= \delta\epsilon\, u^\mu u^\nu
  + \delta f\, P^{\mu\nu}
  + 2 u^{(\mu} k^{\nu)}
  + t^{\mu\nu}~~, \nn\\
  J^{\mu\nu}_{\text{nhs}}
  &= 2 n^{[\mu} u^{\nu]} + s^{\mu\nu}~~.
\end{align}
Here all the tensor structures are transverse to $u^\mu$, while $t^{\mu\nu}$ is
symmetric-traceless and $s^{\mu\nu}$ is anti-symmetric. It is possible to use the
hydrodynamic redefinition freedom in $u^\mu$ and $T$ to set
$\delta\epsilon = k^\mu = 0$. There is also a redefinition freedom in $\mu_\mu$
but since $\mu_\mu$ does not appear in the ideal order constitutive relations,
this redefinition cannot be used to eliminate any first-order
structures. Additionaly, the first order equations of motion can be used to
remove $u^\mu \delta_\scB g_{\mu\nu}$ from set of independent first-order structures.
Finally, this leads to the following form for the first-order non-hydrostatic corrections
\begin{align}
  \delta f &= - \frac{\zeta T}{2} P^{\mu\nu} \delta_\scB g_{\mu\nu}
  = - \zeta \nabla_\mu u^\mu~~, \nn\\
  t^{\mu\nu}
  &= - \eta\, T P^{\rho\langle\mu} P^{\nu\rangle\sigma} "d_{"scB} g_{\rho\sigma}
  = - 2\eta\, P^{\mu\rho} P^{\nu\sigma} \lb \nabla_{(\rho} u_{\sigma)}
  - \frac{1}{3} P_{\rho\sigma} \nabla_\lambda u^\lambda \rb
  \equiv - \eta  \sigma^{\mu\nu}~~, \nn\\
  n^\mu &= - T \lambda P^{\mu\rho} u^\sigma \delta_{\scB} b_{\rho\sigma}
  = - 2 \lambda P^{\mu\rho} u^\sigma T \dow_{[\rho}\frac{\mu_{\sigma]}}{T}~~, \nn\\
  s^{\mu\nu} &= - T \sigma P^{\mu\rho} P^{\nu\sigma} \delta_{\scB} b_{\rho\sigma}
  = - \sigma P^{\mu\rho} P^{\nu\sigma}
  \lb 2T \dow_{[\rho}\frac{\mu_{\sigma]}}{T} + u^\lambda H_{\lambda\rho\sigma} \rb~~.
\end{align}
Introducing these into the quadratic form in \cref{DeltaGeneral}, the
non-negativity of $\Delta$ requires that all the non-hydrostatic transport
coefficients are non-negative
\begin{equation}
  \eta, \zeta, \lambda, \sigma \geq 0~~.
\end{equation}
Thus, all in all, the most generic constitutive
relations of a one-form ordinary fluid up to one-derivative order are given as
\begin{align}
  T^{\mu\nu}
  &= \lb \epsilon + p \rb u^{\mu}u^{\nu} + p\, g^{\mu\nu}
    - \zeta\, \nabla_\lambda u^\lambda P^{\mu\nu}
    - \eta\, \sigma^{\mu\nu} \nn\\
  &\qquad
    - \frac16 \epsilon^{\alpha\beta\rho\sigma} u_\alpha H_{\beta\rho\sigma}
    \frac{\dow (T\alpha)}{\dow T} u^{\mu}u^{\nu}
    - \frac\alpha3 u^{(\mu} \epsilon^{\nu)\lambda\rho\sigma}
    H_{\lambda\rho\sigma}
    + \mathcal{O}(\dow^2)~~,  \nn\\
  J^{\mu\nu}
  &=
    \nabla_\sigma \lb \alpha\epsilon^{\mu\nu\rho\sigma} u_\rho\rb
    + 2 \lambda u^{[\mu} P^{\nu]\rho} u^\sigma T
    \dow_{[\rho}\frac{\mu_{\sigma]}}{T}
    - \sigma P^{\mu\rho} P^{\nu\sigma}
  \lb 2T \dow_{[\rho}\frac{\mu_{\sigma]}}{T} + u^\lambda H_{\lambda\rho\sigma} \rb
     + \mathcal{O}(\dow^2)~~,
\end{align}
and satisfy the adiabaticity equation \eqref{adiabaticity_simple} with the free-energy current
\begin{equation}\label{freeE-unbroken}
  N^\mu
  =
  \frac{p}{T} u^\mu
  + \frac{\alpha}{6T} \epsilon^{\mu\nu\rho\sigma} H_{\nu\rho\sigma}
  - \alpha \epsilon^{\mu\nu\rho\sigma} u_\nu  \dow_\rho \bfrac{\mu_\sigma}{T}
  + \mathcal{O}(\dow^2)~~.
\end{equation}
Out of the 5 transport coefficients appearing at one derivative order, 4 are
dissipative and sign definite, while the remaining one does not cause
dissipation and is left to be sign-indefinite.

In a global thermal state, characterised by a timelike isometry
$\scK = (\mathbf k^\mu, \Lambda^{\mathbf k}_\mu)$, the dynamical fields
arrange in an equilibrium configuration
\begin{equation}
  \beta^\mu = \mathbf k^\mu~~,~~
  \Lambda^\beta_\mu = \Lambda^{\mathbf k}_\mu~~,\qquad
  u^\mu = \frac{\mathbf k^\mu}{\mathbf k}~~,~~
  T = \frac{1}{\mathbf k}~~,~~
  \mu_\mu
  = \frac{\Lambda^{\mathbf k}_\mu + \mathbf k^\nu b_{\nu\mu}}{\mathbf k},
\end{equation}
where $\mathbf k = \sqrt{-\mathbf k^\mu \mathbf k_\mu}$. If we choose a basis
$(t,x^i)$ such that $\mathbf k^\mu = \delta^\mu_t /T_0$, the hydrostatic
effective action generating the respective constitute relations can be read out
using \cref{freeE-unbroken} and \cref{eq:hydroN} leading to
\begin{equation}
  S^{\text{hs}}[g_{\mu\nu},b_{\mu\nu}]
  = \frac{1}{T_0}\int \df^3 x \sqrt{-g}\, \lB p(T)
  - \frac{\alpha(T)}{6}
  \epsilon^{\mu\nu\rho\sigma} u_\mu H_{\nu\rho\sigma} \rB.
\end{equation}

In the next section, it will be shown how the one-form symmetry can be broken and how this breaking can lead to other fields which can modify the ideal order constitutive relations.

\section{One-form superfluids} \label{sec:super}

In the previous section, hydrodynamics in the presence of an unbroken one-form
symmetry was studied. In this section, this study is extended to include
hydrodynamics with a spontaneously broken one-form symmetry by introducing a
gapless vector Goldstone mode $\varphi_\mu$ into the generic analysis of
\cref{sec:setup}. It is observed that this theory is self-dual in the absence of
external two-form sources, which is highly reminiscent of the electromagnetic
duality of sourceless Maxwell's equations. In addition to the equation of state
at ideal order, it is found that the one-form superfluid is characterised by a
total of 166 transport coefficients at one-derivative order and hence is not
extremely useful from a phenomenological standpoint. However, the various
interesting limits/sectors of the theory are highlighted, for which the spectrum
of transport coefficients is considerably more manageable. These limits will be
investigated in detail in \cref{sec:stringfluids,sec:electric}. The hydrodynamic
theory developed here finds a direct application in describing various phases of
plasma. In a certain limit, which we refer to as string fluids, one-form superfluid
dynamics provides a dual and conceptually cleaner formulation of
\emph{magnetohydrodynamics} describing plasmas with Debye screened electric
fields. In another limit, it describes plasmas without free charges, which we
refer to as \emph{bound-charge plasmas}. The details of these applications will
be given in \cref{sec:hotEM}.

\subsection{Hydrodynamics with spontaneously broken one-form
  symmetry} \label{sec:1formSF-formulation} In this section, the Josephson
condition for one-form superfluids is derived along with the ideal order
constitute relations and first-order corrections.  The hydrostatic effective
action for one-form superfluids is also given.
  
\subsubsection{Vector Goldstone and the Josephson equation}
\label{sec:Josephson}

In the theory of zero-form superfluid dynamics, the spontaneous breaking of the
global $\rmU(1)$ symmetry gives rise to a scalar Goldstone mode
$\phi$. Analogously, the Goldstone mode corresponding to a broken global
$\rmU(1)$ one-form symmetry is the one-form gauge field $\varphi_\mu$
\cite{Lake:2018dqm} that under an infinitesimal symmetry transformation
$\scX = (\chi^\mu,\Lambda^\chi_\mu)$ transforms as
\begin{equation} \label{eq:introvarphimu}
  \delta_\scX \varphi_\mu = \lie_\chi \varphi_\mu - \Lambda^\chi_\mu~~.
\end{equation}
It is useful to introduce the covariant derivative of $\varphi_\mu$ according to
\begin{equation} \label{eq:sfv}
  \xi_{\mu\nu} = 2 \dow_{[\mu} \varphi_{\nu]} + b_{\mu\nu}~~,
\end{equation}
which is gauge-invariant and transforms covariantly under the action of $\scX$,
i.e. $\delta_\scX\xi_{\mu\nu}=\lie_\scX\xi_{\mu\nu}$. In analogy with 
zero-form superfluids, for which the superfluid velocity is given by
$\xi_\mu=\partial_\mu\phi+A_\mu$, we refer to \eqref{eq:sfv} as the two-form
``superfluid velocity''. This superfluid velocity satisfies the Bianchi identity
\begin{equation} \label{eq:bisf}
  3 \dow_{[\mu} \xi_{\nu\rho]} = H_{\mu\nu\rho}~~.
\end{equation}
The existence of $\varphi_\mu$ allows for the definition of a gauge-invariant one-form chemical potential $\mu_\mu^\varphi$ such that
\begin{equation} \label{eq:defvarphi}
  \mu^\varphi_\mu
  = \mu_\mu - T \dow_\mu \lb \beta^\nu\varphi_\nu \rb~~,
\end{equation}
where $\mu_\mu$ was introduced in \cref{eq:mumu}. In this symmetry-broken phase,
the covariant information contained in $b_{\mu\nu}$, $\mu_\mu$, and $\varphi_\mu$ can be
exchanged for $\xi_{\mu\nu}$ and $\mu^\varphi_\mu$.

As mentioned in \cref{sec:setup}, the dynamics of the Goldstone mode
$\varphi_\mu$ is governed by its own equation of motion which can be represented
as
\begin{equation} \label{eq:kvarphi}
  K^\mu=0~~.
\end{equation}
This, along with the conservation equations \bref{eq:1formcon}, make the system
of dynamical equations closed. Our ignorance of the underlying microscopic
theory does not allow for a first principle derivation of
\cref{eq:kvarphi}. However, using the offshell adiabaticity equation
\eqref{eq:adiabaticity} for the case at hand
\begin{equation}
  \nabla_\mu N^\mu
  = \half T^{\mu\nu} \delta_\scB g_{\mu\nu}
  + \half J^{\mu\nu} \delta_\scB b_{\mu\nu}
  + K^\mu \delta_\scB \varphi_\mu
  + \Delta~~,~~
  \Delta \geq 0~~~,
\end{equation}
where $\delta_\scB \varphi_\mu=\beta^\nu\xi_{\nu\mu}-\mu^\varphi_\mu/T$, it is
possible to fix the form of \cref{eq:kvarphi} as in the case of usual
superfluids \cite{Jain:2016rlz}. In particular, at zero order in derivatives
using the available hydrodynamic data, the above adiabaticity equation reduces
to $-K^{\mu}\delta_\scB \varphi_\mu+\mathcal{O}(\dow)=\Delta\ge0$, where
$\mathcal{O}(\dow)$ denotes higher derivative corrections. Therefore, it is
possible to infer that
\begin{equation}
K^{\mu}=-T\alpha^{\mu\nu}\delta_\scB \varphi_\nu+\mathcal{O}(\dow)~~,~~\Delta=T\left(\delta_\scB \varphi_\mu\right)\alpha^{\mu\nu}\left(\delta_\scB \varphi_\nu\right)+\mathcal{O}(\dow)~~,
\end{equation}
for some positive semi-definite matrix $\alpha^{\mu\nu}$. Since the Goldstone
must satisfy \cref{eq:kvarphi} onshell, the above implies the relation
\begin{equation}\label{one-form-josephson}
  \delta_\scB \varphi_\mu = \mathcal{O}(\dow)
  \quad\implies\quad
  \mu^\varphi_\mu = u^\nu \xi_{\nu\mu} + \mathcal{O}(\dow)~~,
\end{equation}
which is the one-form equivalent of the \emph{Josephson equation} in superfluids
$\mu = u^\mu \xi_\mu + \mathcal{O}(\dow)$. Thus $\mu^\varphi_\mu$ does not
account for independent degrees of freedom in one-form
hydrodynamics. Additionally, the redefinition freedom associated with $\mu_\mu$
(or correspondingly $\Lambda^\beta_\mu$) can be used to absorb the potential
derivative corrections appearing in \cref{one-form-josephson}. Hence, by
redefining $\mu_\mu$, the Josephson equation \eqref{one-form-josephson} can be
turned into an exact all-order onshell statement
\begin{equation} \label{eq:exactjoseph}
  \delta_\scB \varphi_\mu = 0
  \quad\implies\quad
  \mu_\mu^\varphi =  u^\nu \xi_{\nu\mu}~~,
\end{equation}
and eliminate $\mu_\mu^\varphi$ entirely from the hydrodynamic
description. Thus, the energy-momentum conservation equation in
\eqref{eq:1formcon} provides dynamics for $u^\mu$ and $T$, while the one-form
conservation governs the dynamics of $\varphi_\mu$.\footnote{To see this, note
  that when all the $\mu_\mu$ dependence has been eliminated from the
  hydrodynamic description, the entire dependence on $b_{\mu\nu}$ in the
  hydrodynamic constitutive relations comes via $\xi_{\mu\nu}$. Since this is
  also the source of all $\varphi_\mu$ dependence, for theories admitting an
  effective action,
  $K^\mu = 2\nabla_\nu(\delta S/\delta \xi_{\mu\nu}) = 2 \nabla_\nu(\delta
  S/\delta b_{\mu\nu}) = \nabla_\nu J^{\mu\nu}$. In essence, the Josephson
  equation, that used to originally be the equation of motion for $\varphi_\mu$,
  has now been used to algebraically eliminate $\mu_\mu$. Therefore, the
  one-form charge conservation now serves as an equation of motion for
  $\varphi_\mu$.} On the other hand, the adiabaticity equation reduces to its
simple form in \eqref{adiabaticity_simple} as promised earlier. While the final
system appears to be similar to its symmetry-unbroken counterpart, it should be
noted that the constitutive relations in this case involve $\varphi_\mu$ instead
of $\mu_\mu$.

\subsubsection{Ideal one-form superfluids} \label{sec:idealFluids}

Having identified the independent set of hydrodynamic variables, it is
straightforward to derive the most general constitutive relations at ideal
order. Since we are working with four spacetime dimensions throughout this work,
it is useful to introduce an independent set of vectors
\begin{equation}\label{xi-decomposition}
  \zeta_\mu = \xi_{\mu\nu} u^\nu~~,~~
  \bar\zeta^\mu = \half \epsilon^{\mu\nu\rho\sigma} u_\nu \xi_{\rho\sigma}~~,
\end{equation}
satisfying $u_\mu\zeta^\mu=u_\mu\bar\zeta^\mu=0$, which can be thought as
electric and magnetic fields associated with $\xi_{\mu\nu}$. Here we have
introduced the completely antisymmetric Levi-Civita tensor
$\epsilon_{\mu\nu\rho\sigma}$ with conventions $\epsilon_{0123} = \sqrt{-g}$. In
turn, \cref{xi-decomposition} can be used to decompose the superfluid velocity
from \cref{eq:sfv} as
\begin{equation}
 \xi_{\mu\nu} = 2 u_{[\mu} \zeta_{\nu]} - \epsilon_{\mu\nu\rho\sigma} u^\rho
  \bar\zeta^\sigma~~,
\end{equation}
and to rewrite the Josephon equation \bref{eq:exactjoseph} as $\mu_\mu^\varphi =
-\zeta_\mu$. 

Unlike the ordinary one-form fluids studied in \cref{sec:ordinary}, one-form
superfluids exhibit signatures of one-form symmetry at ideal order itself. Using
the decomposition in \cref{xi-decomposition}, the most generic form of the
hydrostatic free energy density can be shown to take the form
\begin{equation}\label{sf-ideal-N}
  \mathcal{N}
  = P(T,\zeta^2,\bar\zeta^2,\zeta\cdot\bar\zeta)
  + \mathcal{O}(\dow)~~.
\end{equation}
Performing a variation of the functional arguments with respect to the
hydrodynamic variables $"scB$ leads to
\begin{align}
  "d_{"scB} T
  &= \frac{T}{2} u^{\mu}u^{\nu} "d_{"scB} g_{\mu\nu}, \nn\\
  "d_{"scB}\zeta^{2}
  &= \lb \zeta^{2} u^{\mu}u^{\nu} - \zeta^{\mu}\zeta^{\nu}
    \rb "d_{"scB} g_{\mu\nu}
    + 2\zeta^{[\mu}u^{\nu]} "d_{"scB} b_{\mu\nu}~~, \nn\\
  "d_{"scB}\bar\zeta^{2}
  &= 
    \lb - \bar{\zeta}^{2} P^{\mu\nu}
    + \bar\zeta^{\mu}\bar\zeta^{\nu}
    + 2 u^{(\mu} \epsilon^{\nu)\rho\sigma\tau}
    u_{\rho} \bar\zeta_{\sigma} \zeta_{\tau} \rb
    "d_{"scB} g_{\mu\nu}
    - \epsilon^{\mu\nu\rho\sigma}
    u_{\rho}\bar\zeta_{\sigma} "d_{"scB} b_{\mu\nu}~~, \nn\\
  "d_{"scB}(\zeta\cdot\bar\zeta)
  &= - \half (\zeta\cdot\bar\zeta) g^{\mu\nu} \delta_{"scB} g_{\mu\nu}
    - \frac{1}{4} \epsilon^{\mu\nu\rho\sigma}\xi_{\rho\sigma}"d_{"scB}b_{\mu\nu}~~.
\end{align}
Using \cref{hsConstiGeneral}, the one-form ideal superfluid constitutive
relations, free energy, and entropy currents are obtained as
\begin{align}\label{ideal_super_consti}
  T^{\mu\nu}
  &= \epsilon\, u^{\mu} u^{\nu}
    + \lb P - \bar q\,\bar\zeta^{2}
  - q_\times \, (\zeta\cdot\bar\zeta) \rb P^{\mu\nu}
    -  q\, \zeta^{\mu}\zeta^{\nu}
    + \bar q\, \lb \bar\zeta^{\mu}\bar\zeta^{\nu}
    + 2 u^{(\mu} \epsilon^{\nu)\rho\sigma\tau} u_{\rho}\bar\zeta_{\sigma}\zeta_{\tau}
    \rb~~, \nn\\
  J^{\mu\nu}
  &= - 2 u^{[\mu} \lb  q\,\zeta^{\nu]} + q_\times\,\bar{\zeta}^{\nu]} \rb
    - \epsilon^{\mu\nu\rho\sigma} u_{\rho}
    \lb \bar q\, \bar\zeta_{\sigma} + q_\times\, \zeta_{\sigma} \rb~~, \nn\\
  N^\mu
  &= \frac{P}{T} u^\mu~~, \nn\\
  S^\mu
  &= N^{\mu} - \beta_{\nu} T^{\mu\nu} + \frac1T
  \zeta_{\nu}J^{\mu\nu}
   = s u^\mu~~,
\end{align}
where the thermodynamic relations
\begin{equation} \label{eq:thermosfideal}
  "dd P = s\, "dd T + \half  q\, "dd \zeta^{2}
  + \half \bar q\, "dd \bar\zeta^{2}
  + q_\times \,"dd(\zeta\cdot\bar\zeta)~~,\qquad
  \epsilon + P  = s\,T +  q\,\zeta^{2} + q_\times \, (\zeta\cdot\bar\zeta)~~,
\end{equation}
were derived and used to simplify \cref{ideal_super_consti}.  From here we can
identify $P$ appearing in the free energy density as the thermodynamic
pressure. On the other hand, $\epsilon$ and $s$ stand for the energy and entropy
densities, in addition to the two superfluid densities $q$ and $\bar q$, and a
cross-density $q_\times$.\footnote{These thermodynamic relations can take a more
  appealing form if we define
  \begin{equation}
    \varpi = |\zeta|~~,~~
    \bar\varpi = |\bar\zeta|~~,~~
    \rho =  q |\zeta|~~,~~
    \bar\rho = \bar q |\bar\zeta|~~,~~
    \lambda = \zeta\cdot\bar\zeta~~,~~
    \rho_\times = q_\times~~,~~
    p = P - \bar q\,\bar\zeta^{2}
    - q_\times \, (\zeta\cdot\bar\zeta),
  \end{equation}
  which leads to
  \begin{equation}
    \epsilon + p = sT + \rho\varpi - \bar\rho \bar\varpi~~,~~
    "dd p = s\, "dd T
    + \rho \, "dd \varpi
    - \bar\varpi\, "dd \bar\rho
    - \lambda\, "dd \rho_\times~~.
  \end{equation}
  However, in the subsequent sections, limits for which $\zeta_\mu$ or
  $\bar\zeta^\mu$ is taken to be of higher-order in derivatives will be
  explored. In those situations, these definitions are ill-defined.}
\Cref{ideal_super_consti,eq:thermosfideal} imply that one-form superfluids are
completely characterised by their equation of state
$ P = P(T,\zeta^2,\bar\zeta^2, \zeta\cdot\bar\zeta)$.

\subsubsection{One derivative corrections} \label{sec:onederivative}

Having derived the constitutive relations for an ideal one-form superfluid, it
is possible to tackle the marginally more complicated first-order derivative
corrections.  This complication originates from the fact that there are 3 ideal
order mutually orthogonal spatial vectors in one-form superfluids
\begin{equation}\label{eq:1formSFVectors}
  h^\mu_1 = \frac{\zeta^\mu}{|\zeta|}~~,~~
  h^\mu_2 = \frac{\bar\zeta^\mu - (\zeta\cdot\bar\zeta)\zeta^\mu/\zeta^2}
  {\sqrt{\bar\zeta^2 - (\zeta\cdot\bar\zeta)^2/\zeta^2}}~~,~~
  h^\mu_3 = \frac{\epsilon^{\mu\nu\rho\sigma} u_\nu \zeta_\rho \bar\zeta_\sigma}
  {\sqrt{\zeta^2\bar\zeta^2 - (\zeta\cdot\bar\zeta)^2}}~~,
\end{equation}
thereby completely breaking the $\SO(3)$ rotational symmetry and providing a
decomposition for the metric
\begin{equation}
  g^{\mu\nu} = - u^\mu u^\nu + \delta^{ab} h^\mu_a h^\nu_b~~,~~h^\mu_a h^\nu_b = \delta_{ab}~~.
\end{equation}
In terms of these, the corrections to the hydrostatic free energy density
\bref{sf-ideal-N} can be written as
\begin{multline}\label{first_super_N}
  \mathcal{N}
  =
  P
  + f_1^a\, h^\mu_a \dow_\mu T
  + f_2^a\, \epsilon^{\mu\nu\rho\sigma} u_\mu h_{a\nu} \dow_{\rho} u_{\sigma}
  + f_3^a\, \epsilon^{\mu\nu\rho\sigma} u_\mu h_{a\nu} \dow_{\rho} \zeta_{\sigma}
  + f_4^a\, \epsilon^{\mu\nu\rho\sigma} u_\mu h_{a\nu} \dow_{\rho}
  \bar\zeta_{\sigma} \\
  + 2 f_5^{ab}\, h^\mu_a h^\nu_b \nabla_{(\mu} \zeta_{\nu)}
  + 2 f_6^{ab}\, h^\mu_a h^\nu_b \nabla_{(\mu} \bar\zeta_{\nu)}
  + \mathcal{O}(\dow^2)~~.
\end{multline}
Here $f_i$ are the hydrostatic transport coefficients, with $f_5^{ab}$ and
$f_6^{ab}$ being symmetric and traceless. The respective trace parts lead to
total derivative terms which do not lead to independent constitutive relations
upon taking a variation. We have not considered any corrections involving
$H_{\mu\nu\rho}$ explicitly, as they can be related to
$3\dow_{[\mu}\xi_{\nu\rho]}$ using the Bianchi identity.
Thus, in total, there are 22 transport coefficients in the hydrostatic
sector.\footnote{In principle, we can remove some terms from the free energy
  density using the $\varphi_\mu$ equation of motion. The respective
  contributions to the constitutive relations can be absorbed by redefining
  $\varphi_\mu$. We have not analysed these issues here.} As in the ideal order
case, it is possible to use \cref{hsConstiGeneral} in order to read out the
respective constitutive relations at first order in derivatives but we do not
perform this exercise here.

Using the approach detailed in sec.~\ref{sec:firstOrderGen}, we can derive the
constitutive relations in the non-hydrostatic sector. It is convenient to
parametrise the stress tensor and charge current as
\begin{align}
  T^{\mu\nu}_{\text{nhs}}
  &= \delta\epsilon\, u^\mu u^\nu
    + 2 \delta k^a\, u^{(\mu} h^{\nu)}_a
    + \delta t^{ab}\, h^{(\mu}_a h^{\nu)}_b~~, \nn\\
  J^{\mu\nu}_{\text{nhs}}
  &= 2\delta q^a\, u^{[\mu} h^{\nu]}_a
    + \delta s^{ab}\, h^{[\mu}_a h^{\nu]}_b~~.
\end{align}
The terms involving $\delta\epsilon$ and $\delta k^a$ can be set to zero using
the field redefinition freedom inherent to $T$ and $u^\mu$. The field
redefinition freedom inherent to $\mu_\mu$ has been exhausted when turning the
Josephson equation into an exact all-order statement \eqref{eq:exactjoseph},
thus $\delta q^a$ is generically non-zero. These considerations lead to the set
of non-hydrostatic constitutive relations
\begin{equation}
  \begin{pmatrix}
    \delta q^a \\
    \delta t^{ab} \\
    \delta s^{ab}
  \end{pmatrix}
  =
  -T
  \begin{pmatrix}
    \lambda_1^{ac}
    & \lambda_2^{a(cd)}
    & \lambda_2^{a[cd]} \\
    \lambda'^{a(cd)}_2
    & \eta^{(ab)(cd)}
    & \chi^{(ab)[cd]} \\
    \lambda'^{a[cd]}_2
    & \chi'^{[ab](cd)}
    & \sigma^{[ab][cd]}
  \end{pmatrix}
  \begin{pmatrix}
    u^\mu h^\nu_c \delta_\scB b_{\mu\nu} \\
    \half h^\mu_c h^\nu_d \delta_\scB g_{\mu\nu} \\
    \half h^\mu_c h^\nu_d \delta_\scB b_{\mu\nu}
  \end{pmatrix}~~,
\end{equation}
where $\lambda_1$, $\lambda_2$, $\lambda_2'$, $\eta$, $\chi$, $\chi'$, and
$\sigma$ are matrices of transport coefficients. There is a total of
$12\times 12 = 144$ non-hydrostatic transport coefficients. Positive
semi-definiteness of $\Delta$ requires that the symmetric part of the transport
coefficient matrix must have all its eigenvalues non-negative. This gives 12
inequality constraints in the non-hydrostatic sector. Onsager's relations may
impose further restrictions on the non-hydrostatic transport which we have not
considered in this analysis.

\subsubsection{Hydrostatic effective action}

At ideal order, the exact same constitutive relations \bref{ideal_super_consti}
along with the thermodynamic relations \eqref{eq:thermosfideal} can be obtained
from a hydrostatic effective action.  Under the assumption that the background
manifold admits a timelike isometry
$\scK = (\textbf{k}^{\mu}, \Lambda^{\textbf{k}}_\mu)$, we can infer the
equilibrium configuration for the hydrodynamic fields using \cref{eq:mumu} as
\begin{equation}
  \beta^\mu = \textbf{k}^{\mu}~~,~~
  \Lambda_\mu^{\beta} = \Lambda^{\mathbf k}_\mu~~,\qquad
  u^\mu = \frac{\textbf{k}^\mu}{\textbf{k}}~~,~~
  T = \frac{1}{\textbf{k}}~~,~~
  \mu_\mu = \frac{\Lambda^{\mathbf k}_\mu + \textbf{k}^\nu b_{\nu\mu}}{\mathbf k}~~,
\end{equation}
where $\textbf{k}= \sqrt{- \mathbf k^\mu \mathbf k_\mu}$ is the modulus of the
timelike Killing vector field $\textbf{k}^{\mu}$. In turn, the hydrostatic
effective action, using \cref{sf-ideal-N} and \cref{eq:hydroN}, reads
\begin{equation}
  S_{\text{hs}}[g_{\mu\nu},b_{\mu\nu};\varphi_\mu]
  = \int d^4x\sqrt{-g}\, P(T,\zeta^2,\bar\zeta^2,\zeta\cdot\bar\zeta)~~.
\end{equation}
Using \eqref{eqbConsti} we can readily obtain the currents
\bref{ideal_super_consti}. Additionally, by varying the effective action with
respect to $\varphi_\mu$ (see \cref{eq:hydroK}) yields the equation of motion
for equilibrium configurations of $\varphi_\mu$, specifically
\begin{equation} \label{eq:varphieq}
\beta^\nu\nabla_\mu\left(qT\zeta^\mu\right)=-\epsilon^{\nu\mu\rho\sigma}\nabla_\mu\left(\bar q u_\rho \bar\zeta_\sigma\right)-\frac{1}{2}\epsilon^{\nu\mu\rho\sigma}\xi_{\rho\sigma}\partial_\mu q_\times-\frac{1}{6}q_\times\epsilon^{\nu\mu\rho\sigma}H_{\mu\rho\sigma}~~,
\end{equation}
where the reader may be reminded of the defining relation in \cref{eq:defvarphi}
which leads to
$\zeta_\mu = T\partial_\mu\left(\beta^{\nu}\varphi_\nu\right) -
\mu_\mu$. Similarly at one-derivative order, the hydrostatic effective action
obtains corrections due to \cref{first_super_N}
\begin{multline}
  S_{\text{hs}}[g_{\mu\nu},b_{\mu\nu};\varphi_\mu]
  = \int d^4x\sqrt{-g}\, \bigg[ P
  + f_1^a\, h^\mu_a \dow_\mu T
  + f_2^a\, \epsilon^{\mu\nu\rho\sigma} u_\mu h_{a\nu} \dow_{\rho} u_{\sigma}
  + f_3^a\, \epsilon^{\mu\nu\rho\sigma} u_\mu h_{a\nu} \dow_{\rho}
  \zeta_{\sigma} \\
  + f_4^a\, \epsilon^{\mu\nu\rho\sigma} u_\mu h_{a\nu} \dow_{\rho}
  \bar\zeta_{\sigma}
  + 2 f_5^{ab}\, h^\mu_a h^\nu_b \nabla_{(\mu} \zeta_{\nu)}
  + 2 f_6^{ab}\, h^\mu_a h^\nu_b \nabla_{(\mu} \bar\zeta_{\nu)} \bigg]~~,
\end{multline}
which can be used to derive the hydrostatic constitutive relations and
$\varphi_\mu$ profiles.

It should be noted that no assumptions were made on the background metric and
gauge fields other than the existence of a timelike isometry. As shall be
explained in sec.~\ref{sec:stringfluids}, upon taking appropriate limits, this action describes
all equilibrium configurations in string fluids, which includes those of
\cite{Caldarelli:2010xz, Grozdanov:2016tdf, Armas:2018ibg} as special cases.

\subsection{Special limits of one-form superfluids}
\label{sec:limits}

An effective theory with 166 arbitrary transport coefficients (and 12 inequalities) at first order in derivatives is perhaps not the most useful effective theory. However, it is possible to identify limits of this general theory with a tractable number of transport coefficients and interesting applications, which are now described:
\begin{itemize}
\item \textbf{Electromagnetism:} The simplest example encompassed by this
  general theory is that of electromagnetic fields living alongside a neutral
  ideal fluid. By simply turning off the coupling between electromagnetic and
  fluid degrees of freedom, and setting
  $F_{\mu\nu}=2\partial_{[\mu}A_{\nu]}=\xi_{\mu\nu}$, the gauge field
  $\varphi_\mu$ is directly identified with the electromagnetic photon
  $A_\mu$. The identification $F_{\mu\nu}=\left(\star J\right)_{\mu\nu}$ yields
  the same theory, with the two being related by electromagnetic duality. This
  case will be described in more detail in sec.~\ref{sec:emzero}.

\item \textbf{String fluid limit.} An interesting limit of one-form superfluids,
  which will be studied in sec.~\ref{sec:stringfluids}, is the limit in which
  $\zeta^{\mu}$ enters the constitutive relations while $\bar \zeta^\mu$ is
  simply removed from the theory. As shall be explained in
  sec.~\ref{sec:stringfluids}, this limit can be understood as a partial
  breaking of the one-form symmetry along $\beta^\mu$ in which only the timelike
  component of the Goldstone mode $\varphi=\beta^\mu\varphi_\mu$ (in terms of
  which $\zeta_\mu$ is defined - see \cref{eq:defvarphi} and
  \cref{eq:exactjoseph}) enters the constitutive relations. This limit, which
  will be shown to be exactly equivalent to magnetohydrodynamics in which the
  electric fields are Debye screened, is characterised by 23 independent
  transport coefficients.

\item \textbf{Electric limit.} The electric limit is attained by considering the
  hierarchy of scales $\zeta^\mu=\mathcal{O}\left(1\right)$ and
  $\bar\zeta^\mu=\mathcal{O}\left(\dow\right)$, implying that electric fields
  $\zeta^\mu$ can be arbitrary but magnetic fields $\bar\zeta^\mu$ are weak. In
  this context, the one-form symmetry is completely broken.
  This limit is equivalent to the hydrodynamics of magnetically dominated
  bound-charge plasmas, i.e. plasmas that do not contain free charge carriers
  and have electric fields derivative suppressed. We will return to this in
  detail in sec.~\ref{sec:electric}.

\end{itemize}
In general, the formalism of one-form superfluids finds applications in many phases of (hot) electromagnetism. A more detailed description and derivation of these connections is given in sec.~\ref{sec:hotEM}.

\subsection{Self-duality of one-form superfluids} \label{sec:selfdual}
To summarise, the theory of one-form superfluid dynamics developed in the previous sections is governed by the following set of
equations\footnote{The Hodge duality operation is defined as
  ${(\star \omega)}^{\mu_1\ldots\mu_{d-k}} = \frac{1}{k!}
  \epsilon^{\nu_1\ldots\nu_k \mu_1\ldots\mu_{d-k}}\omega_{\nu_1\ldots\nu_k}$.}
\begin{alignat}{3}\label{master_equations}
  \text{Energy-momentum conservation}: &\quad \N_{"m} T^{"m"n} &&= \frac32
  \N^{[\nu} \xi^{"r"s]} J_{"r"s}- \xi^{\nu\rho} \nabla^{\sigma} J_{\sigma\rho}~~,\nn\\
  \text{$\varphi_{\mu}$ equation of motion}:
  &\quad \nabla_{\mu}J^{\mu\nu} &&= 0~~, \nn\\
  \text{$\varphi_{\mu}$ Bianchi identity}:
  &\quad \nabla_{\mu}{\star\xi}^{\mu\nu} &&= {\star H}^{\nu}~~, \nn\\
  \text{Second law of thermodynamics}: &\quad \nabla_{\mu} N^{\mu} &&= \half
  T^{\mu\nu} "d_{"scB} g_{\mu\nu} + \frac12 J^{\mu\nu} "d_{"scB}\xi_{\mu\nu} +
  "D~~,~~ "D \geq 0~~,
\end{alignat}
where due to $\delta_\scB \varphi_\mu = 0$, the following identity holds
$"d_{"scB}\xi_{\mu\nu} = "d_{"scB} b_{\mu\nu}$.\footnote{In the first equation
  in \eqref{master_equations}, the term involving the charge current divergence
  has been included in order to make the self-duality manifest. Onshell, this
  equation is identical to \cref{eq:1formcon} upon using the Bianchi identity
  and one-form conservation equation.} When the background field strength
$H_{\mu\nu\rho}$ vanishes, it is possible to check that under the mapping
\begin{gather} 
  J^{\mu\nu} \to J_{*}^{\mu\nu} = {\star\xi}^{\mu\nu}
  = \half \epsilon^{\mu\nu\rho\sigma}\xi_{\rho\sigma}~~,~~
  \xi_{\mu\nu} \to \x^{*}_{\mu\nu} = {\star J}_{\mu\nu}
  = \half \epsilon_{\mu\nu\rho\sigma}J^{\rho\sigma}~~,\nn\\
  N^\mu \to N_{*}^{\mu} = N^{\mu} - \half \beta^{\mu} J^{\rho\sigma}
  \xi_{\rho\sigma}~~,
  \label{LegendreTransform}
\end{gather}
these equations map to themselves.\footnote{In $d$ spacetime dimensions, a
  similar Legendre transform is expected to map a $q$-form superfluid to a
  $(d-q-2)$-form superfluid.} This is the self-duality of one-form superfluid
dynamics. The operation \eqref{LegendreTransform} can be seen as a Legendre
transform in the one-form sector, so that $J^{\mu\nu}$ become background sources
while $\xi_{\mu\nu}$ are seen as the respective responses.  What used to be the
$\varphi_\mu$ equation of motion, in the Legendre transformed picture becomes
the Bianchi identity for some auxiliary gauge field $\varphi^*_{\mu}$ such that
$\xi^{*}_{\mu\nu} = 2\dow_{[\mu}\varphi^{*}_{\nu]}$. The equation of motion for
$\varphi^{*}_\mu$ is given by what previously used to be the Bianchi
identity. It is interesting to note that even though the free-energy current is
Legendre transformed according to \eqref{LegendreTransform}, the physical
entropy current in the two pictures is exactly the same, namely
\begin{equation}
  S^\mu = N^{\mu} - T^{\mu\nu}\beta_{\nu} - J^{\mu\nu} \beta^{\rho}
  \xi_{\rho\nu}
  = N_{*}^{\mu} - T^{\mu\nu}\beta_{\nu} - J_{*}^{\mu\nu} \beta^{\rho}
  \xi^{*}_{\rho\nu}~~.
\end{equation}
Thus, irrespective of the formulation being used, entropy production remains the same. Additionally, it also ensures that if the
constitutive relations in one formalism are tuned in order to satisfy the second law of thermodynamics, then the coefficients in the Legendre transformed picture 
automatically respect the second law.

The realisation of the self-duality of one-form superfluids has been phrased in abstract terms by means of \eqref{LegendreTransform}. In practice, however, the exact
map between transport coefficients in both pictures can be non-trivial. In order to illustrate this, we apply the map \eqref{LegendreTransform} to one-form superfluids at 
ideal order in sec. \ref{sec:idealFluids}. The two-form superfluid velocity in the Legendre transformed picture is given by
\begin{equation}\label{IdealOrderLegendreTransform}
  \xi^{*}_{\mu\nu}
  = {\star J}_{\mu\nu}
  = 2 u_{[\mu} \lb \bar q\, \bar\zeta_{\nu]}
  + q_\times \zeta_{\nu]} \rb
    - \epsilon_{\mu\nu\rho\sigma}u^{\rho}
    \lb  q \zeta^{\sigma}\, + q_\times \bar{\zeta}^{\sigma} \rb~~.
\end{equation}
Comparison with \cref{xi-decomposition}, where $(\zeta^\mu,\bar\zeta^\mu)$ have been replaced by their corresponding Legendre transform vectors $(\zeta^\mu_*,\bar\zeta^\mu_*)$ that ought to be determined, it is possible to infer that
\begin{gather}\label{idealOrderLegendreMap}
  \zeta_{\mu}^{*}
  = \bar q\, \bar\zeta_{\mu} + q_\times \zeta_{\mu}~~,~~
    \bar\zeta^{*}_{\mu}
  =  q\, \zeta_{\mu} + q_\times \bar{\zeta}_{\mu}~~, \nn\\
  \zeta_{\mu}
  = \frac{\bar q}{ q\bar q - q_\times^{2}} \bar\zeta^{*}_{\mu}
  - \frac{q_\times}{ q\bar q - q_\times^{2}} \zeta_{\mu}^{*}~~,~~
  \bar\zeta_{\mu} = \frac{ q}{ q\bar q - q_\times^{2}} \zeta^{*}_{\mu}
  - \frac{q_\times}{ q\bar q - q_\times^{2}} \bar\zeta_{\mu}^{*}~~.
\end{gather}
Using these and comparing with \eqref{ideal_super_consti}, it is possible to find the respective two-form current via the relation
\begin{equation}
  J^{\mu\nu}_{*}
  = {\star\xi}^{\mu\nu}
  = - 2 u^{[\mu} \lb  q_{*}\,\zeta^{\nu]}_* + q_{\times *}\,\bar{\zeta}^{\nu]}_* \rb
    - \epsilon^{\mu\nu\rho\sigma} u_{\rho}
    \lb \bar q_{*}\, \bar\zeta_{\sigma}^* + q_{\times *}\, \zeta_{\sigma}^* \rb~~,
\end{equation}
where the Legendre transformed transport coefficients were identified according to
\begin{gather}
  q_{*} = - \frac{ q}{ q\bar q - q_\times^{2}}~~,~~
  \bar q_{*} = - \frac{\bar q}{ q\bar q - q_\times^{2}}~~,~~
  q_{\times *} = \frac{q_\times}{ q\bar q - q_\times^{2}}~~,~~
  \epsilon_{*} = \epsilon, \qquad
  p_{*} = p~~, \nn\\
  P_* = P -  q \zeta^{2} - \bar q \bar\zeta^{2}
  - 2q_\times (\zeta\cdot\bar\zeta)~~.
  \label{ideal_coeff_Leg}
\end{gather}
This identification brings the Legendre transformed stress tensor and charge
current to the same form as in \cref{ideal_super_consti} but with transport
coefficients and $(\zeta^\mu,\bar\zeta^\mu)$ replaced by their Legendre
transformed counterparts. It is worth noticing that the transformation
\eqref{ideal_coeff_Leg} is not defined if $q\bar q- q_\times^2=0$.

\subsection{Application to hot electromagnetism}
\label{sec:emzero}

As a simple application of one-form superfluid dynamics, consider a neutral
fluid subjected to dynamical electromagnetic fields. This is the simplest
example of a hot electromagnetic plasma, which we consider in detail in
\cref{sec:hotEM}, where the electromagnetic fields are completely decoupled from
the fluid degrees of freedom. The dynamics of this system is governed by the
energy-momentum conservation and the familiar Maxwell's equations
\begin{equation} \label{eq:eomEM}
  \nabla_\mu T^{\mu\nu}=0~~,~~
  \nabla_\mu F^{\mu\nu}=0~~,~~
  \nabla_{[\rho}F_{\mu\nu]}=0~~.
\end{equation}
The third equation (Bianchi identity) in \eqref{eq:eomEM} is solved by
introducing the photon $A_\mu$ such that
$F_{\mu\nu}=2\partial_{[\mu}A_{\nu]}$. The energy-momentum tensor of this theory
receives contributions from both the fluid component as well as the electromagnetic
fields
\begin{align}\label{decoupledEM}
  T^{\mu\nu}
  &= \epsilon_{\text m}(T) u^\mu u^\nu
  + p_{\text m}(T) (g^{\mu\nu} + u^\mu u^\nu)
  + F^{\mu}{}_{\rho}F^{\nu\rho}
  - \frac{1}{4} g^{\mu\nu} F_{\rho\sigma}F^{\rho\sigma} \nn\\
  &= \epsilon_{\text m}(T) u^\mu u^\nu
    + p_{\text m}(T) (g^{\mu\nu} + u^\mu u^\nu) \nn\\
  &\qquad
    + \frac{1}{2} (E^2 + B^2) u^\mu u^\nu 
    + \frac{1}{2} (E^2 + B^2) P^{\mu\nu} 
    - E^\mu E^\nu
    - \lb B^\mu B^\nu
    + 2 u^{(\mu} \epsilon^{\nu)\rho\sigma\tau} u_\rho B_\sigma E_\tau \rb,
\end{align}
where we have defined the electric fields $E^{\mu} = F^{\mu\nu}u_{\nu}$ and
magnetic fields
$B^{\mu} = \half \epsilon^{\mu\nu\rho\sigma}u_{\nu}F_{\rho\sigma}$.  The
electromagnetic part trivially satisfies the conservation equations for the
photon configurations that satisfy the Maxwell's equations (second equation in
\eqref{eq:eomEM}), while the conservation of the fluid part governs the dynamics
of $u^\mu$ and $T$.


This setup can be equivalently described by ideal one-form superfluid
dynamics. To this aim, we perform the identification $F_{\mu\nu}=\xi_{\mu\nu}$,
which implies
\begin{equation}
  E^\mu = \zeta^\mu~~,~~
  B^{\mu} = \bar\zeta^\mu~~.
\end{equation}
Comparing the energy-momentum tensor in \cref{decoupledEM} with
\cref{ideal_super_consti} we can read out that
\begin{gather}\label{ideal_coeff_EM}
  P = p_{\text m}(T) + \half (\zeta^2 - \bar\zeta^2)~~,~~
  q = -\bar q = 1~~,~~
  q_\times = 0~~,~~
  \epsilon = \epsilon_{\text m}(T)
  + \frac{1}{2} \lb \zeta^{2} + \bar\zeta^{2}\rb~~.
\end{gather}
It follows that the two-form current
$J^{\mu\nu} = - \xi^{\mu\nu} = - F^{\mu\nu}$. Having made the identification,
the equations of one-form superfluid dynamics in \eqref{master_equations} with
$H_{\mu\nu\rho}=0$ map directly to \eqref{eq:eomEM}. The respective
energy-momentum tensors and Bianchi identities map to each other, while the
equation of motion for $\varphi_\mu$ is equivalent to Maxwell's
equations. Therefore, the one-form Goldstone $\varphi_\mu$ can be identified
with the photon $A_\mu$. The associated hydrostatic free-energy density for this
one-form superfluid is given by
\begin{equation} \label{eq:emN}
  \mathcal{N} = P = p_{\text m}(T) - \frac{1}{4} \xi_{\mu\nu}\xi^{\mu\nu}
  = p_{\text m}(T) - \frac{1}{4} F_{\mu\nu} F^{\mu\nu} ~~.
\end{equation}
This is precisely the Lagrangian density for electromagnetism minimally coupled
to a neutral fluid, where the vacuum permeability has been set to unity.

Due to the self-duality of one-form superfluids
\eqref{IdealOrderLegendreTransform} at $H_{\mu\nu\rho} = 0$, we can also make
the identification $F_{\mu\nu}=\xi_{\mu\nu}^*=\star J_{\mu\nu}$. The electric
and magnetic fields now reverse their roles
\begin{equation}
  E^\mu=-\bar\zeta^\mu~~,~~B^\mu=\zeta^\mu~~,
\end{equation}
while the mapping for transport coefficients remains the same (see
\cref{ideal_coeff_Leg}). This dual description is essentially the consequence of
electromagnetic duality of vacuum Maxwell's equations under $E^{\mu} \to B^\mu$
and $B^\mu \to - E^\mu$. In this case, the Bianchi identity in
\eqref{master_equations} maps to Maxwell equations in \eqref{eq:eomEM}, while
the equation of motion for $\varphi_\mu$ maps to the electromagnetic Bianchi
identity. In this picture the vector Goldstone $\varphi_\mu$ can be understood
as an auxiliary ``magnetic photon''. The energy momentum tensor
\bref{decoupledEM} and the Lagrangian density \eqref{eq:emN}, when defined with
respect to the Legendre transformed $P_*$ in \cref{ideal_coeff_Leg}, remain
invariant.

The relations between one-form superfluids at finite temperature and hot
electromagnetism will be considered in more generality in \cref{sec:hotEM}. In
any case, the relations established here should provide confidence to the reader
that one-form superfluids can be used to construct effective theories where the
electromagnetic degrees of freedom interact with the mechanical and thermal
degrees of freedom of relativistic matter.

\section{String fluids}
\label{sec:stringfluids} 

In this section a theory of parity-violating string fluids is formulated up to first order in derivatives, extending and completing earlier formulations \cite{Schubring:2014iwa, Grozdanov:2016tdf, Hernandez:2017mch}. This theory can be formulated by partially breaking the one-form symmetry along the fluid velocity $\beta^\mu$, yielding a scalar Goldstone mode $\varphi$, or by a direct limit of one-form superfluids as discussed in sec.~\ref{sec:limits}. Both these directions will be described in this section. String fluids provide a dual formulation of MHD that is cast only in terms of symmetries, eliminating the need of introducing the non-propagating degrees of freedom $\mu$ (chemical potential) and $E^\mu$ (electric fields) in traditional treatments of MHD \cite{Armas:2018atq}. The exact relation between the two formulations will be described in detail in sec.~\ref{sec:hotEM}.

\subsection{Partial breaking of one-form symmetry}

String fluids can be obtained directly from one-form fluids discussed in
sec.~\ref{sec:ordinary} where the one-form symmetry is spontaneously broken in
the direction of the fluid flow. In practice, it implies that the theory admits
a scalar Goldstone $\varphi$ in the hydrodynamic regime along with the usual
hydrodynamic fields $u^\mu$, $T$, and $\mu_\mu$ introduced in
sec.~\ref{sec:setup}. Under a symmetry transformation $\scX$, $\varphi$
transforms as
\begin{equation} \label{eq:introvarphi}
  \delta_\scX \varphi = \lie_\chi \varphi - \beta^\mu \Lambda_\mu^\chi~~.
\end{equation}
This new mode allows for the introduction of a new gauge-invariant vector combination $\varpi h^\mu$ that captures the covariant derivatives of $\varphi$, namely
\begin{equation} \label{eq:newvector}
  \varpi h_\mu
  = \mu_\mu - T \dow_\mu \varphi~~.
\end{equation}
Here $h^\mu h_\mu = 1$ and we have isolated the norm $\varpi$ of the vector. It
can be verified that, at this stage, $u^\mu$ and $h^\mu$ are not necessarily
orthogonal, instead their inner product satisfies
$u^\mu h_\mu = - T^2/\varpi \delta_\scB \varphi$. The hydrodynamic systems built
using these degrees of freedom are referred to as \emph{string fluids}. In
particular, the vector $h_\mu$ characterises the direction of the strings while
$\varpi$ is interpreted as a string chemical potential.

Following a similar procedure as in sec.~\ref{sec:1formSF-formulation} we can 
determine the Josephson equation for string fluids. The Goldstone mode
$\varphi$ is accompanied by its equation of motion $K=0$, which can be used to
write down the offshell adiabaticity equation \eqref{eq:adiabaticity} in the
form
\begin{equation}
  \nabla_\mu N^\mu
  = \half T^{\mu\nu} \delta_\scB g_{\mu\nu}
  + \half J^{\mu\nu} \delta_\scB b_{\mu\nu}
  + K \delta_\scB \varphi
  + \Delta~~,~~
  \Delta \geq 0~~.
\end{equation}
Using the available hydrodynamic data, at ideal order this equation becomes $- K \delta_\scB \varphi = \Delta \geq 0$, implying that
\begin{equation}
  K = - \alpha \delta_\scB \varphi + \mathcal{O}(\dow)~~,~~
  \Delta = \alpha (\delta_\scB \varphi)^2 + \mathcal{O}(\dow)~~,~~
  \alpha \geq 0~~,
\end{equation}
where $\alpha$ is some transport coefficient. Imposing the $\varphi$ equation of motion $K=0$, it follows that
$\delta_\scB \varphi = \mathcal{O}(\dow)$, which in turn implies the Josephson equation for string fluids
$u^\mu h_\mu = \mathcal{O}(\dow)$. Analogous to sec.~\ref{sec:1formSF-formulation}, it is possible to use the redefinition freedom associated with $\mu_\mu$ to absorb potential derivative corrections and to turn it into the exact statement
\begin{equation} \label{eq:stringjos}
  \delta_\scB \varphi = 0
  \quad\implies\quad
  u^\mu h_\mu = 0~~.
\end{equation}
Thus, the string direction can generically be chosen to be transverse to the
fluid flow.  Therefore, the independent dynamical fields in string fluids, just
like the previous considerations of~\cite{Grozdanov:2016tdf}, are $u^\mu$, $T$,
$\varpi$, and $h_\mu$ with $u^\mu u_\mu = -1$, $h^\mu h_\mu = 1$, and
$u^\mu h_\mu = 0$. The dynamics for $u^\mu$ and $T$ is governed by the
energy-momentum conservation in \cref{eq:1formcon}, while that for $\varpi$ and
$h_\mu$ by the components of the one-form conservation transverse to
$u^\mu$. The component of the one-form conservation along $u^\mu$, on the other
hand, acts as a constraint on the allowed field configurations on an initial
Cauchy slice. In our picture, this constraint is seen as determining the
configurations of the scalar Goldstone $\varphi$.\footnote{To see this, note
  that there are two sources of $b_{\mu\nu}$ dependence in string fluids:
  $H_{\mu\nu\rho}$ and $\varpi h_\mu$. Therefore, for theories admitting an
  effective action, we can infer that
  $J^{\mu\nu} = - 3 \nabla_\lambda (\delta S/\delta H_{\lambda\mu\nu}) + 2
  u^{[\mu} (\delta S/\delta(\varpi h_{\nu]})) $. On the other hand, all the
  $\varphi$ dependence comes from $\varpi h_\mu$ leading to
  $K = \nabla_\mu (T \delta S/\delta(\varpi h_\mu)) = (\delta S/\delta(\varpi
  h_{\mu})) T^2 u^\nu \delta_\scB g_{\mu\nu} + T u_\nu \nabla_\mu
  J^{\mu\nu}$. We have used that $u^\mu (\delta S/\delta(\varpi h_{\mu})) =
  0$. Therefore, after the time component of $\mu_\mu$ has been algebraically
  eliminated using the $\varphi$ equation of motion, the time component of the
  one-form conservation equation serves as the equation of motion for
  $\varphi$.} Additionally, once \cref{eq:stringjos} is imposed, the
adiabaticity equation \eqref{eq:adiabaticity} reduces to
\eqref{adiabaticity_simple}.

\subsection{Ideal string fluids}
At ideal order, string fluids are characterised by the free energy density
$\mathcal{N} = p(T,\varpi)$. The $\delta_\scB$ variations of $T$ and $\varpi$ read
\begin{equation}
  \delta_\scB T = \frac{T}{2} u^\mu u^\nu \delta_\scB g_{\mu\nu}~~,~~
  \delta_\scB \varpi
  = \frac{\varpi}{2} \lb u^\mu u^\nu
  - h^{\mu} h^{\nu} \rb \delta_\scB g_{\mu\nu}
  + u^{[\mu}h^{\nu]}\delta_{"scB}\xi_{\mu\nu}~~,
\end{equation}
and can be used, together with \cref{nhsConstiGeneral}, to derive the respective constitutive
relations. Specifically, these read
\begin{align}\label{ideal_string_consti}
  T^{\mu\nu} &= (\epsilon+p)\, u^{\mu} u^{\nu} + p\,g^{\mu\nu}
               - \varpi \rho\, h^{\mu}h^{\nu} + \mathcal{O}(\dow)~~, \nn\\
  J^{\mu\nu} &= 2 \rho\, u^{[\mu}h^{\nu]} + \mathcal{O}(\dow)~~,
\end{align}
where the thermodynamic relations
\begin{equation}\label{string_thermodynamics}
  "dd p = s\, "dd T + \rho\, "dd \varpi~~,~~
  \epsilon + p = s\,T +  \rho\,\varpi~~,
\end{equation}
were defined and led to the identification of $p$ as pressure, $\epsilon$ as
energy density, $\rho$ as string density and $s$ as entropy density. The
associated free energy and entropy currents are given as
\begin{equation}
  N^{\mu} = \frac{p}{T} u^{\mu}~~,~~
  S^{\mu} = s\, u^{\mu}~~.
\end{equation}
Since $\Delta$ at ideal order vanishes, ideal string fluids are non-dissipative.

It is instructive to work out the ideal order equations of motion governing the
dynamics of the string fluid hydrodynamic fields. In particular, the components
of the energy-momentum conservation imply
\begin{subequations}\label{EOM_string_first_EM}
  \begin{align}
    \N_{"m} T^{"m"n}
    = \half H^{\nu\rho\sigma} J_{\rho\sigma}
    &+ 2\zeta^{[\nu} u^{\rho]} \nabla^\mu J_{\mu\rho} \nn\\
    \implies& \delta_\scB s
              + \frac{s}{2} P^{\mu\nu} \delta_\scB g_{\mu\nu}
              = \mathcal{O}(\dow^2)~~, \label{EOM1} \\
            & u^\mu h^{\nu} \delta_\scB g_{\mu\nu}
              = \mathcal{O}(\dow^2)~~, \label{EOM2} \\
            & (\epsilon+p) u^\mu \Delta^{\rho\nu} \delta_\scB g_{\mu\nu}
              - \rho h^{\mu} \Delta^{\rho\nu} \delta_\scB b_{\mu\nu}
              = \mathcal{O}(\dow^2)~~, \label{EOM3}
  \end{align}
\end{subequations}
while those of the one-form current conservation reduce to
\begin{subequations}\label{EOM_string_first_J}
  \begin{align}
    \N_{"m} J^{"m"n} = 0
    \implies& \delta_\scB \rho
              + \frac{\rho}{2} \Delta^{\mu\nu} \delta_\scB g_{\mu\nu}
              = \mathcal{O}(\dow^2)~~, \label{EOM4} \\
            & \Delta^{\rho\mu} u^\nu \delta_\scB b_{\mu\nu}
              + \varpi \Delta^{\rho\mu} h^\nu \delta_\scB g_{\mu\nu}
              = \mathcal{O}(\dow^2)~~, \label{EOM5} \\
            & \frac{1}{T} \N_{"m} \lb T \rho h^{\mu} \rb
              - \rho T u^\mu h^{\nu} \delta_\scB g_{\mu\nu}
              = \mathcal{O}(\dow^2)~~.  \label{EOMconstraint}
  \end{align}
\end{subequations}
Here $\Delta^{\mu\nu} = g^{\mu\nu} + u^\mu u^\nu - h^\mu h^\nu$ and
\begin{equation}
  \delta_\scB g_{\mu\nu} = 2 \nabla_{(\mu} \bfrac{u_{\nu)}}{T}~~,~~
  \delta_\scB b_{\mu\nu} = 2 \dow_{[\mu} \bfrac{\varpi h_{\nu]}}{T} 
  + \frac{u^{\sigma}}{T} H_{\sigma\mu\nu}~~,
\end{equation}
were used to simplify the expressions. \Cref{EOM1,EOM2,EOM3,EOM4,EOM5} can be
used to eliminate $u^\mu \delta_\scB g_{\mu\nu}$ and
$u^\mu \delta_\scB b_{\mu\nu}$ from the set of independent first order
non-hydrostatic tensors. On the other hand, \cref{EOMconstraint}, upon using
\cref{EOM2}, gives a constraint equation for $\varphi$ configurations on an
initial Cauchy slice
\begin{equation} \label{eq:stringvareom}
\N_{"m} \lb T\rho h^{\mu} \rb = 0~~,
\end{equation}
which is the no-monopole constraint of \cite{Gammie:2003rj}. Additionally, the second equation in \eqref{EOM_string_first_J} is the induction equation of \cite{Gammie:2003rj}.

As already explained in \cite{Armas:2018atq}, the introduction of $\varphi$ in the formulation of string fluid dynamics allows for a well-defined hydrostatic effective action \eqref{eq:hydroN}, where $N_{\text{hs}}^\mu=(p/T)u^{\mu}$ and from which \eqref{eq:stringvareom} arises as the variation with respect to $\varphi$ (see \cref{eq:hydroK}).

\subsubsection{Strings fluids as a limit of one-form superfluids}
\label{stringLimitofSF}

As mentioned in sec.~\ref{sec:limits}, string fluids as described above can be obtained as a limit of one-form superfluids introduced in sec.~\ref{sec:1formSF-formulation}. This limit is obtained by removing any dependence on $\bar\zeta^\mu$ from the one-form superfluid theory, in which case the Bianchi identity \eqref{eq:bisf} looses its meaning. Comparing \eqref{eq:introvarphi} with \eqref{eq:introvarphimu}, it is straightforwardly inferred that the Goldstone scalar $\varphi$ is the component of the Goldstone vector $\varphi_\mu$ along $\beta_\mu$, i.e. $\varphi=\beta^\mu\varphi_\mu$. The complete equivalence is made by comparing \eqref{eq:newvector} with \eqref{eq:defvarphi} under the light of the Josephson equations \eqref{eq:exactjoseph} and \eqref{eq:stringjos} leading to the identification
\begin{equation}
  \zeta_\mu = - \varpi h_\mu~~,~~q = \frac{\rho}{\varpi}~~,
\end{equation}
while the conditions $\bar q = q_\times = 0$ arise due to the removal of any dependence on $\bar\zeta$ from the constitutive relations of sec.~\ref{sec:idealFluids}, thus recovering \bref{ideal_string_consti} from
\cref{ideal_super_consti}. Additionally, \cref{eq:stringvareom} can be obtained from the equilibrium equation \eqref{eq:varphieq} for $\varphi_\mu$.

\subsection{One derivative corrections to string fluids}
\label{sec:stringfluidcorrections}

Having established the ideal order
constitutive relations, it is possible to continue the hydrodynamic expansion to
one higher order. The results will be a sector of the transport coefficients
given in sec.~\ref{sec:onederivative}, which also includes parity-violating
terms, hence providing an extension of earlier literature
\cite{Schubring:2014iwa, Grozdanov:2016tdf, Hernandez:2017mch}.

\subsubsection{Hydrostatic corrections}

Hydrostatic corrections to ideal string fluids are composed of the first order
scalars that can appear in the hydrostatic free energy $\mathcal{N}$ and are
non-vanishing in equilibrium. At first order, it is possible to identify a total
of 5 transport coefficients
\begin{equation}\label{stringN}
  \mathcal{N} = p
  - \frac{\alpha}{6} \epsilon^{\mu\nu\rho\sigma}u_\mu H_{\nu\rho\sigma}
  - \beta \epsilon^{\mu\nu\rho\sigma} u_\mu h_\nu \dow_\rho u_\sigma
  - \tilde\beta_1 h^\mu \dow_\mu T
  - \tilde\beta_2 h^\mu \dow_\mu \frac{\varpi}{T}
  - \tilde\beta_3 \epsilon^{\mu\nu\rho\sigma} u_\mu h_\nu \dow_\rho h_\sigma~~.
\end{equation}
Since boundary transport is not being considered, total derivative scalars such
as $\nabla_\mu h^\mu$ can be removed from the independent set. Additionally, the
equilibrium condition \eqref{eq:stringvareom} allows us to set $\tilde\beta_2=0$
and hence only 4 scalars are independent. However, allowing for a non-zero
$\tilde\beta_2$ will ease comparison with earlier literature in
\cref{sec:hotEM}. The terms coupling to $\alpha$ and $\beta$ are CP-even while
those coupling to $\tilde\beta_1$, $\tilde\beta_2$, $\tilde\beta_3$ are
CP-odd.\footnote{The discrete parity symmetry P acts on various quantities as
  usual, while the quantities odd under the one-form charge conjugation C are
  $b_{\mu\nu}$, $H_{\mu\nu\rho}$, $\xi_{\mu\nu}$, $\zeta_\mu$, $\bar\zeta^\mu$,
  $h_\mu$, and $J^{\mu\nu}$. We are interested in CP-odd terms in string fluids
  because when relating string fluids to magnetohydrodynamics in section
  \ref{sec:MHDd}, these correspond to P-odd terms in magnetohydrodynamics. We
  have deferred a more exhaustive discussion of the action of discrete
  symmetries, including CPT, to \cref{app:CPT}.} The distinguished notation
$\alpha$ for the first transport coefficient is due to the fact that it will
play a crucial role later in the mapping to magnetohydrodynamics in
\cref{sec:hotEM}. Performing the $\delta_\scB$ variation of all the
one-derivative terms in \cref{stringN} and using \cref{hsConstiGeneral}, the
contributions of each term to the constitutive relations and free energy current
can be obtained, and are given in app.~\ref{app:hydrostatic}.

\subsubsection{Non-hydrostatic corrections} \label{sec:hydrostring}
In order to derive non-hydrostatic constitutive relations, it is useful to decompose the currents in this sector of the theory along and transverse to $u^\mu$ and $h^\mu$, such that
\begin{align}\label{string-nhs-corrections}
  T^{\mu\nu}_{\text{nhs}}
  &= \delta\epsilon ~u^{\mu}u^{\nu} + \delta f \Delta^{\mu\nu}
  + \delta\tau ~h^{\mu}h^{\nu} + 2\ell^{(\mu}h^{\nu)} + 2 k^{(\mu}u^{\nu)} +
    t^{\mu\nu}~~, \nn\\
  J^{\mu\nu}_{\text{nhs}}
  &= 2 \delta\rho ~u^{[\mu}h^{\nu]} + 2m^{[\mu} h^{\nu]} + 2 n^{[\mu} u^{\nu]} +
    \delta s\, \epsilon^{\mu\nu}~~,
\end{align}
where $\epsilon^{\mu\nu} = \epsilon^{\mu\nu\rho\sigma}u_\rho h_\sigma$ is a
parity-odd contribution. In particular, any antisymmetric tensor transverse to
$u^{\mu}$ and $h^{\nu}$ in 4 spacetime dimensions only has one degree of freedom
and is always proportional to $\epsilon^{\mu\nu}$. Choosing to work in the Landau
frame following the discussion in sec.~\ref{sec:firstOrderGen}, and eliminating
$u^\mu\delta_\scB g_{\mu\nu}$ and $u^\mu \delta_\scB b_{\mu\nu}$ using the first
order equations of motion, the non-hydrostatic constitutive relations can be
represented as
\begin{gather}
  \begin{pmatrix}
    \delta f \\ \delta\tau \\ \delta s
  \end{pmatrix}
  =
  - \frac{T}{2} \begin{pmatrix}
    \zeta_{\perp} & \zeta_{\times} & \textcolor{blue}{\tilde\kappa_{1}}  \\
    \zeta'_{\times} & \zeta_{\parallel} & \textcolor{blue}{\tilde\kappa_{2}}  \\
    \textcolor{blue}{\tilde\kappa'_{1}} &
    \textcolor{blue}{\tilde\kappa'_{2}} & r_{\parallel}
  \end{pmatrix}
  \begin{pmatrix}
    \Delta^{\mu\nu}"d_{"scB}g_{\mu\nu} \\
    h^{\mu}h^{\nu} "d_{"scB}g_{\mu\nu} \\
    \epsilon^{\mu\nu} "d_{"scB}\xi_{\mu\nu}
  \end{pmatrix}~~, \nn\\
  \begin{pmatrix}
    \ell^{\mu} \\ m^{\mu}
  \end{pmatrix}
  =
  - T \begin{pmatrix}
    \eta_{\parallel} & \textcolor{blue}{r_{\times}} & \tilde\eta_{\parallel} &
    \textcolor{blue}{\tilde r_{\times}} \\
    \textcolor{blue}{r'_{\times}} & r_{\perp} &
    \textcolor{blue}{\tilde r'_{\times}} & \tilde r_{\perp}
  \end{pmatrix}
  \begin{pmatrix}
    \Delta^{\mu\sigma}h^{\nu}"d_{"scB}g_{\sigma\nu} \\
    \Delta^{\mu\sigma}h^{\nu}"d_{"scB}\xi_{\sigma\nu} \\
    \epsilon^{\mu\sigma}h^{\nu}"d_{"scB}g_{\sigma\nu} \\
    \epsilon^{\mu\sigma}h^{\nu}"d_{"scB}\xi_{\sigma\nu}
  \end{pmatrix}~~, \nn\\
  t^{\mu\nu} = - \eta_{\perp} T\Delta^{\rho\langle\mu}\Delta^{\nu\rangle\sigma}
  "d_{"scB} g_{\rho\sigma}
  + \tilde\eta_{\perp} T\epsilon^{\rho\langle\mu}\Delta^{\nu\rangle\sigma}
  "d_{"scB} g_{\rho\sigma}~~.
\end{gather}
The redefinition freedom in $u^\mu$ and $T$ has been used to set
$\delta\epsilon = k^\mu = 0$, whereas the residual freedom in $\mu_\mu$ after
setting $u^\mu h_\mu = 0$ is used to set $\delta \rho = n^\mu = 0$. Here we have
introduced 19 non-hydrostatic transport coefficients, which are functions of $T$
and $\varpi$. In tab.~\ref{table:cptstring}, the transformation properties of
these coefficients under CP transformations is summarised.
\begin{table}[t]
\begin{center}
\begin{tabular}{|c|c|}
  \hline
  \textbf{CP}
  & \textbf{Transport coefficients} \\
  \hline
  CP-even
  & $\zeta_\perp~,~
    \zeta_\times~,~
    \zeta'_\times~,~
    \zeta_\parallel~,~
    r_\parallel~,~
    \eta_\parallel~,~
    \tilde\eta_\parallel~,~
    r_\perp~,~
    \tilde r_\perp~,~
    \eta_\perp~,~
    \tilde\eta_\perp$ \\[0.2em]
  CP-odd
  & $ \textcolor{blue}{\tilde\kappa_1}~,~
    \textcolor{blue}{\tilde\kappa'_1}~,~
    \textcolor{blue}{\tilde\kappa_2}~,~
    \textcolor{blue}{\tilde\kappa'_2}~,~
    \textcolor{blue}{r_\times}~,~
    \textcolor{blue}{r'_\times}~,~
    \textcolor{blue}{\tilde r_\times}~,~
    \textcolor{blue}{\tilde r'_\times}$ \\
  \hline
\end{tabular}   
\caption {Transformation properties under CP for non-hydrostatic transport
  coefficients in string fluids.}
\label{table:cptstring}
\end{center}
\end{table}
Thus, the first 11 coefficients already identified in \cite{Hernandez:2017mch} are in the CP-even sector and the remaining new 8 coefficients (in \emph{blue}) are in the CP-odd sector and had not been previously identified in the literature. Of these 8 coefficients, 4 can be understood as new current resistivities and are related to the remaining 4 via Onsager's relations under certain assumptions as will be explained in sec.~\ref{sec:Kubostring}.

Using the adiabaticity equation \eqref{adiabaticity_simple} and after some non-trivial algebra, it is possible to derive that
\begin{multline}
  \frac{1}{T}\Delta =
  \frac14 \begin{pmatrix}
    \Delta^{\mu\nu}"d_{"scB}g_{\mu\nu} \\
    h^{\mu}h^{\nu} "d_{"scB}g_{\mu\nu} \\
    \epsilon^{\mu\nu} "d_{"scB}\xi_{\mu\nu}
  \end{pmatrix}^{"rmT}
  \begin{pmatrix}
    \zeta_{\perp}
    & \half(\zeta_{\times}+\zeta'_{\times})
    & \half(\textcolor{blue}{\tilde\kappa_{1}}+\textcolor{blue}{\tilde\kappa'_{1}}) \\
    \half(\zeta_{\times}+\zeta'_{\times})
    & \zeta_{\parallel}
    & \half(\textcolor{blue}{\tilde\kappa_{2}} + \textcolor{blue}{\tilde\kappa'_{2}}) \\
    \half(\textcolor{blue}{\tilde\kappa_{1}} + \textcolor{blue}{\tilde\kappa'_{1}})
    & \half(\textcolor{blue}{\tilde\kappa_{2}} + \textcolor{blue}{\tilde\kappa'_{2}})
    & r_{\perp}
  \end{pmatrix}
  \begin{pmatrix}
    \Delta^{\mu\nu}"d_{"scB}g_{\mu\nu} \\
    h^{\mu}h^{\nu} "d_{"scB}g_{\mu\nu} \\
    \epsilon^{\mu\nu} "d_{"scB}\xi_{\mu\nu}
  \end{pmatrix} \\
  + \half \begin{pmatrix}
    \Delta^{\mu\sigma} h^{\nu}"d_{"scB}g_{\sigma\nu}
    + i \epsilon^{\mu\sigma} h^{\nu}"d_{"scB}\xi_{\sigma\nu}\\
    \Delta^{\mu\sigma} h^{\nu}"d_{"scB}\xi_{\sigma\nu}
  \end{pmatrix}^{"rmT}
  \begin{pmatrix}
    \eta_{\parallel} & \frac{1}{2}\lb \textcolor{blue}{r_{\times}} + \textcolor{blue}{r'_{\times}}\rb \\
    \frac{1}{2}\lb \textcolor{blue}{r_{\times}} + \textcolor{blue}{r'_{\times}}\rb & \textcolor{blue}{r_{\perp}} - i^{2} \eta_{\parallel}
  \end{pmatrix}
  \begin{pmatrix}
    \Delta^{\mu\sigma} h^{\nu}"d_{"scB}g_{\sigma\nu}
    + i \epsilon^{\mu\sigma} h^{\nu}"d_{"scB}\xi_{\sigma\nu}\\
    \Delta^{\mu\sigma} h^{\nu}"d_{"scB}\xi_{\sigma\nu}
  \end{pmatrix} \\
  + \half \eta_{\perp} "d_{"scB} g_{\mu\nu} \Delta^{\rho\langle\mu}\Delta^{\nu\rangle\sigma}
  "d_{"scB} g_{\rho\sigma}~~,
\end{multline}
where
$i = \frac{1}{2\eta_{\parallel}} \lb\textcolor{blue}{\tilde r_{\times}} -
\textcolor{blue}{\tilde{r}'_{\times}}\rb$. Out of the 19 non-hydrostatic transport coefficients,
the following 8 linear combinations trivially drop out of the quadratic form
\begin{gather}
  \zeta_{\times} - \zeta'_{\times}~~,~~
  \textcolor{blue}{\tilde\kappa_{1}} - \textcolor{blue}{\tilde\kappa'_{1}}~~,~~
 \textcolor{blue}{\tilde\kappa_{2}} - \textcolor{blue}{\tilde\kappa'_{2}}~~,~~
  \textcolor{blue}{r_{\times}} - \textcolor{blue}{r'_{\times}}~~,~~
  \textcolor{blue}{\tilde r_{\times}} + \textcolor{blue}{\tilde r'_{\times}}~~,~~
  \tilde\eta_{\parallel}~~,~~
  \tilde r_{\perp}~~,~~
  \tilde\eta_{\perp}~~,
\end{gather}
and hence are left totally unconstrained. These combinations can be identified as the
non-hydrostatic non-dissipative transport coefficients, as they do not
contribute to dissipation. Finally, requiring $\Delta\geq 0$ gives 6 inequality
constraints among the remaining 11 dissipative transport coefficients. In terms
of matrices of transport coefficients they can be expressed as
\begin{equation}
  \begin{pmatrix}
    \zeta_{\perp}
    & \half(\zeta_{\times}+\zeta'_{\times})
    & \half(\textcolor{blue}{\tilde\kappa_{1}}+\textcolor{blue}{\tilde\kappa'_{1}}) \\
    \half(\zeta_{\times}+\zeta'_{\times})
    & \zeta_{\parallel}
    & \half(\textcolor{blue}{\tilde\kappa_{2}} + \textcolor{blue}{\tilde\kappa'_{2}}) \\
    \half(\textcolor{blue}{\tilde\kappa_{1}} + \textcolor{blue}{\tilde\kappa'_{1}})
    & \half(\textcolor{blue}{\tilde\kappa_{2}} + \textcolor{blue}{\tilde\kappa'_{2}})
    & r_{\perp}
  \end{pmatrix} \geq 0~~,~~
  \begin{pmatrix}
    \eta_{\parallel} & \frac{1}{2}\lb \textcolor{blue}{r_{\times}} + \textcolor{blue}{r'_{\times}}\rb \\
    \frac{1}{2}\lb \textcolor{blue}{r_{\times}} + \textcolor{blue}{r'_{\times}}\rb & r_{\perp} - i^{2} \eta_{\parallel}
  \end{pmatrix} \geq 0~~,~~
  \eta_{\perp} \geq 0~~,
\end{equation}
whereby positive semi-definiteness of a matrix is understood as the requirement that all its eigenvalues
are non-negative. In total, therefore, the number of non-hydrostatic transport
coefficients can be summarised as in tab.~\ref{table:typestring}.
\begin{table}[h]
  \centering
  \begin{tabular}[h]{|l|c|c|}
  \hline 
    & \textbf{CP-even} & \textbf{CP-odd} \\
    \hline

    Non-dissipative non-hydrostatic & 4 & 4 \\
    
    Dissipative & 7 & 4 \\
    \hline
  \end{tabular}
  \caption {Classes of non-hydrostatic transport coefficients in string fluids}
\label{table:typestring}
\end{table}
Under certain assumptions, not all of these 19 transport coefficients are
independent as it will be shown via Kubo formulae and Onsager's relations.

\subsection{Kubo formulae}
\label{sec:Kubostring}

Using the variational background method of~\cite{Son:2007vk} it is possible to
derive Kubo formulae for string fluids, which are of particular interest for
evaluating transport coefficients in holographic setups. In what follows, the
hydrostatic corrections of sec.~\ref{sec:hydrostring} have been ignored and only
the non-hydrostatic have been taken into account.\footnote{In the particular
  holographic setup of \cite{Grozdanov:2017kyl}, the 4 independent hydrostatic
  transport coefficients of sec.~\ref{sec:hydrostring} vanished.} It is
convenient to split the background coordinates $x^\mu$ into the set $(t,x^i,z)$
and to consider a simple equilibrium configuration in a flat background
spacetime with vanishing $b_{\mu\nu}$ and velocity profile
$u^{\mu} = \delta^\mu{}_t$,  $h^\mu = \pm \delta^\mu{}_{z}$. In order to obtain Kubo
formulae, the one-point functions are introduced
 \begin{equation} \label{eq:one-point}
  \mathbb{T}^{\mu\nu}
  = \sqrt{-g}~\langle T^{\mu\nu}\rangle~~,~~
  \mathbb{J}^{\mu\nu}
  = \sqrt{-g}~\langle J^{\mu\nu}\rangle~~,
\end{equation}
and a small time-dependent but homogeneous in space perturbation around the
equilibrium state is performed such that $u^{\mu} \to u^{\mu} + \delta u^{\mu}$,
$h^{\mu} \to h^{\mu} + \delta h{}^{\mu}$,
$g_{\mu\nu} \to \eta_{\mu\nu} + \delta h_{\mu\nu}$, and
$b_{\lambda\sigma}\to\delta b_{\lambda\sigma}$.\footnote{Explicitly we find
  $ \delta u^t = \frac{1}{2}\delta h_{tt}~,~ \delta v_i^t = \delta u^{i} +
  \delta h_{ti}~,~ \delta h^z = -\frac{1}{2}\delta h_{zz}$ and
  $\delta\left(\nabla_{(i}u_{j)}\right) = \partial_t \delta h_{ij}/2$.}  These
perturbations should be understood as small deformations that generically take
the form $\delta b_{\lambda\sigma}=A_{\lambda\sigma}e^{-\omega t}$ for some
amplitude matrix $A_{\lambda\sigma}$. According to linear response theory, small
variations of \eqref{eq:one-point} can be written in terms of retarded Green's
functions of frequency $\omega$ such that
\begin{equation} \label{eq:Green}
  \delta \mathbb{T}^{\mu\nu}
  = \frac{1}{2}G_{TT}^{\mu\nu,\lambda\rho} \delta h_{\lambda\rho}
  + \frac{1}{2} G_{TJ}^{\mu\nu,\lambda\rho}
  \delta b_{\lambda\rho}~,~~\delta \mathbb{J}^{\mu\nu}
  = \frac{1}{2}G_{JT}^{\mu\nu,\lambda\rho}\delta h_{\lambda\rho}
  + \frac{1}{2} G_{JJ}^{\mu\nu,\lambda\rho}\delta
  b_{\lambda\rho}~~.
\end{equation}
Evaluating \eqref{eq:Green} for the specific initial equilibrium configuration
and writing it in components, it is found that
\begin{gather} \label{eq:Kubostring}
\zeta'_\times = \lim_{\omega\to0}\frac1\omega ~\mathrm{Im}~ G_{TT}^{zz,ii}~~,~~\zeta_{||}=\lim_{\omega\to0}\frac1\omega ~\mathrm{Im}~ G_{TT}^{zz,zz}~~,~~\textcolor{blue}{\tilde{\kappa}_{2}}\text{sign}(h)=\lim_{\omega\to0}\frac1\omega ~\mathrm{Im}~ G_{TJ}^{zz,ij}~~, \nn\\
\eta_{||}=\lim_{\omega\to0}\frac1\omega ~\mathrm{Im}~
G_{TT}^{zi,zi}~~,~~\tilde{\eta}_{||}\text{sign}(h)=\lim_{\omega\to0}\frac1\omega
~\mathrm{Im}~
G_{TT}^{zi,zj}~~,~~\textcolor{blue}{r_{\times}}=\lim_{\omega\to0}\frac1\omega
~\mathrm{Im}~ G_{TJ}^{zi,iz}~~, \nn\\
\textcolor{blue}{\tilde{r}_{\times}}\text{sign}(h)=\lim_{\omega\to0}\frac1\omega
~\mathrm{Im}~
G_{TJ}^{zi,jz}~~,~~
\textcolor{blue}{\tilde\kappa_1'} \text{sign}(h)
= \lim_{\omega\to0}\frac1\omega ~\mathrm{Im}~G_{JT}^{ij,kk}~~,
~~\textcolor{blue}{\tilde{\kappa}_{2}'}\text{sign}(h)=\lim_{\omega\to0}\frac1\omega
~\mathrm{Im}~ G_{JT}^{ij,zz}~~, \nn\\
r_{||}=\lim_{\omega\to0}\frac1\omega ~\mathrm{Im}~ G_{JJ}^{ij,ij}~~,~~\textcolor{blue}{r_{\times}'}=\lim_{\omega\to0}\frac1\omega ~\mathrm{Im}~ G_{JT}^{iz,iz}~~,~~r_{\perp}=\lim_{\omega\to0}\frac1\omega ~\mathrm{Im}~ G_{JJ}^{iz,iz}~~, \nn\\
\textcolor{blue}{\tilde{r}_{\times}'}\text{sign}(h)=\lim_{\omega\to0}\frac1\omega
~\mathrm{Im}~ G_{JT}^{iz,jz}~(i\ne
j)~~,~~\tilde{r}_{\perp}\text{sign}(h)=\lim_{\omega\to0}\frac1\omega
~\mathrm{Im}~ G_{JJ}^{iz,jz}~(i\ne j)~~, \nn\\
\zeta_\perp+\frac{(d-3)}{2(d-2)}\eta_\perp=\lim_{\omega\to0}\frac1\omega
~\mathrm{Im}~ G_{TT}^{ii,ii}~~,~~\zeta_\times=\lim_{\omega\to0}\frac1\omega
~\mathrm{Im}~
G_{TT}^{ii,zz}~~,~~\textcolor{blue}{\tilde{\kappa}_1}\text{sign}(h)=\lim_{\omega\to0}\frac1\omega
~\mathrm{Im}~ G_{TJ}^{ii,jk}~~, \nn\\
\eta_{\perp}=\lim_{\omega\to0}\frac1\omega ~\mathrm{Im}~ G_{TT}^{ij,ij}~,~(i\ne j)~~,~~\tilde{\eta}_{\perp}\text{sign}(h)=\lim_{\omega\to0}\frac1\omega ~\mathrm{Im}~ G_{TT}^{ij,ii}~,~(i\ne j)~~.
\end{gather}

If the microscopic theory in question has some sort of discrete symmetry
$\Theta$ including time-reversal, we can use the Onsager's relations to relate
some of the transport coefficients. For operators $O_a=\{T^{ij},J^{ij}\}$, the
Onsager's relations state (see e.g. \cite{Kovtun:2012rj})
\begin{equation}
  G_{O_aO_b}(\omega, h) = i_a i_b G_{O_b O_a}(\omega, \Theta h)~~,
\end{equation}
where $i_a$ are eigenvalues of $O_a$ under $\Theta$. For $\Theta = \text{CT}$
(which is just time-reversal in the dual hot electromagnetism picture),
$i_T = i_J = 1$ and $\Theta h = -h$; see \cref{app:CPT} for more details. This
leads to the following relations among transport coefficients
\begin{equation} \label{eq:Onsagerstrings}
  \zeta_\times=\zeta_\times'~~,~~
  \textcolor{blue}{r_\times}=\textcolor{blue}{r_\times'}~~,~~
  \textcolor{blue}{\tilde r_\times}=\textcolor{blue}{\tilde r_\times'}~~,~~
  \textcolor{blue}{\tilde {\kappa}_1}=-\textcolor{blue}{\tilde {\kappa}_1'}~~,~~
    \textcolor{blue}{\tilde{\kappa}_2}=-\textcolor{blue}{\tilde{\kappa}_2'}~~.
\end{equation}
On the other hand, for $\Theta = \text{CPT}$, $i_T = i_J = 1$ and
$\Theta h = h$. In this case, the constraints are slightly different and we get
\begin{equation} \label{eq:Onsagerstrings_CPT}
  \zeta_\times=\zeta_\times'~~,~~
  \textcolor{blue}{r_\times}=\textcolor{blue}{r_\times'}~~,~~
  \textcolor{blue}{\tilde r_\times}=-\textcolor{blue}{\tilde r_\times'}~~,~~
  \textcolor{blue}{\tilde {\kappa}_1}=\textcolor{blue}{\tilde {\kappa}_1'}~~,~~
    \textcolor{blue}{\tilde{\kappa}_2}=\textcolor{blue}{\tilde{\kappa}_2'}~~.
\end{equation}
Thus, within either of these contexts, there are 4 independent hydrostatic
transport coefficients and 14 non-hydrostatic transport coefficients. Hence,
string fluids are characterised by a total of 18 transport coefficients at first
order in derivatives.

While other phenomenological realisations of string fluids are possible, here we
consider it in the context in which $J^{\mu\nu}$ has a positive eigenvalue under
time-reversal symmetry and $h^\mu$ has a negative eigenvalue. These
considerations are motivated by the mapping of string fluids to MHD as will be
discussed in sec.~\ref{sec:hotEM}. In this context, Onsager's relations for the
operators  require that

\section{Electric limit of one-form superfluids}
\label{sec:electric}

This section explores the electric limit of one-form superfluids discussed in
sec.~\ref{sec:limits}. This limit is characterised by the derivative hierarchy
$\zeta^\mu=\mathcal{O}\left(1\right)$ and
$\bar\zeta^\mu=\mathcal{O}\left(\dow\right)$, in which case, contrary to the
previous section, the Bianchi identity \eqref{eq:bisf} plays a relevant role. A
discussion on the Bianchi identity and its consequences allows the determination
of the relevant hydrodynamic structures. This is followed by the derivation of
the first order corrections in the electric limit, yielding a total of 29
transport coefficients (modulo Onsager's relations). As shall be established in
sec.~\ref{sec:boundcharge}, this limit provides a dual formulation of
magnetic-dominated bound-charge plasmas, which under particular assumptions are
directly related to MHD without free charges.

\subsection{Bianchi identity and order mixing}

The electric limit of one-form superfluids is defined as the regime where the
$\zeta_\mu$ components of $\xi_{\mu\nu}$ are treated at ideal order, while the
components $\bar\zeta^\mu$ are treated at one-derivative order. Naively, this
may appear to be qualitatively similar to string fluids where $\zeta_\mu$ was
treated at ideal order while $\bar\zeta^\mu$ was entirely removed from the
hydrodynamic description. However, there is an important distinction. In
particular, note that the Bianchi identities \bref{eq:bisf} relate certain
derivatives of $\zeta_\mu$ to those of $\bar\zeta^\mu$. In components
\begin{align} \label{eq:Bianchies}
  \epsilon^{\mu\nu\rho\sigma} u_\mu \zeta_\nu \dow_\rho u_\sigma
  - \frac16 \epsilon^{\mu\nu\rho\sigma} u_\mu H_{\nu\rho\sigma}
  &= \nabla_\mu \bar\zeta^\mu
  - \bar\zeta^\mu u^\nu \nabla_\nu u_\mu~~, \nn\\
  \epsilon^{\mu\nu\rho\sigma} u_\nu \zeta_\rho
  \lb \frac1T \dow_\sigma T + u^\lambda \nabla_\lambda u_\sigma \rb
  + \half  \epsilon^{\mu\nu\rho\sigma} u_\nu
  &\lb - 2T \dow_\rho \frac{\zeta_\sigma}{T}
  + u^\lambda H_{\lambda\rho\sigma} \rb \nn\\
  &= \bar\zeta^\mu \nabla_\nu u^\nu
  + T P^{\mu}{}_\lambda
  \lb \beta^\nu \nabla_\nu \bar\zeta^\lambda
  - \bar\zeta^\nu \nabla_\nu \beta^\lambda \rb~~.
\end{align}
In string fluids, where $\bar\zeta^\mu$ is not a dynamical field, these
equations are irrelevant. On the other hand, in the electric limit, these
equations become important. Upon setting $\bar\zeta^\mu = \mathcal{O}(\dow)$,
these read
\begin{gather}
  \frac16 \epsilon^{\mu\nu\rho\sigma} u_\mu H_{\nu\rho\sigma}
  = \epsilon^{\mu\nu\rho\sigma} u_\mu \zeta_\nu \dow_\rho u_\sigma
  + \mathcal{O}(\dow^2), \nn\\
  P^{\mu\rho} P^{\nu\sigma}  \delta_\scB b_{\rho\sigma}
  = - 2 \zeta^{[\mu} P^{\nu]\sigma} u^\rho \delta_\scB g_{\rho\sigma}
  + \mathcal{O}(\dow^2)~~.
  \label{termRemoval}
\end{gather}
Therefore, the first order terms appearing on the left hand side, which used to be
independent in string fluids, are no longer independent in the electric limit. This has an important consequence which is referred here as ``order mixing'' between
consecutive derivative orders in the electric fluid constitutive
relations. Noting that $\delta_\scB b_{\mu\nu} = \delta_\scB \xi_{\mu\nu}$, it is possible to
massage the adiabaticity equation \bref{adiabaticity_simple} into
\begin{multline}\label{adiabaticity_electric}
  \nabla_\mu N^\mu
  = \half \lb T^{\mu\nu}
  + 2 u^{(\mu} P^{\nu)\rho} \zeta^{\sigma} J_{\rho\sigma}
  \rb \delta_\scB g_{\mu\nu}
  - J^{\rho\sigma} u_\rho P^{[\nu}{}_\sigma u^{\mu]} \delta_\scB \xi_{\mu\nu} \\
  + \half J^{\tau\lambda} P_{\tau\rho} P_{\lambda\sigma}
  \lb P^{\rho\mu} P^{\sigma\nu} \delta_\scB \xi_{\mu\nu}
  + 2 \zeta^{[\rho} P^{\sigma]\nu} u^\mu \delta_\scB g_{\mu\nu} \rb
  + \Delta~~.
\end{multline}
This equation implies, in general, the appearance of $k$-derivative order terms
in $T^{\mu\nu}$ and $J^{\mu\nu}$ if $N^\mu$ was being studied at $k$-derivative
order.  Since the term in the parentheses in the second line in
\eqref{adiabaticity_electric} is two-derivative order, we could also have a
$(k-1)$-derivative contribution to $J^{\mu\nu}$. Furthermore, since
$\delta_\scB g_{\mu\nu}$ is one-derivative order, terms in the parentheses in
the first line in \eqref{adiabaticity_electric} must be $k$-derivative order,
leading to certain $(k-1)$ derivative contributions in $T^{\mu\nu}$ as well. In
turn, this could lead to the same transport coefficient appearing across
consecutive derivative orders.

In the hydrostatic sector, such order-mixing only comes from the terms in
$\mathcal{N}$ dependent on $\bar\zeta^\mu$. Generically, if attention is being focused on the $k$th
order terms in $\mathcal{N}$ and define
\begin{equation}
  \mathcal{R}^{(k-1)}_\mu = \frac{\delta\mathcal{N}_{(k)}}{\delta\bar\zeta^\mu}~~,
\end{equation}
such order-mixing contributions are given by
\begin{equation}\label{general-order-mixing}
  T^{\mu\nu}_{(k-1)} \sim - 2 u^{(\mu}
  \epsilon^{\nu)\rho\sigma\tau} u_\rho \zeta_{\sigma}  \mathcal{R}^{(k-1)}_\tau~~,~~
  J^{\mu\nu}_{(k-1)} \sim - \epsilon^{\mu\nu\rho\sigma} u_\rho
  \mathcal{R}^{(k-1)}_\sigma~~.
\end{equation}
In the non-hydrostatic sector, on the other hand, no independent transport
coefficient appear across derivative orders. However, whereas the inequality
constraints imposed by $\Delta \geq 0$ usually only apply to one-derivative dissipative
transport coefficients, in this case they can also involve transport coefficients from
two-derivative order. This will be made explicit below.

\subsection{Ideal one-form superfluids in the electric limit}

Defining $\zeta_\mu = - \varpi h_\mu$ for later convenience and suppressing
$\bar\zeta^\mu$ to one-derivative order, the ideal one-form superfluid
constitutive relations \bref{ideal_super_consti} become
\begin{align}\label{ideal_electric_consti}
  T^{\mu\nu}
  &= \epsilon\, u^{\mu} u^{\nu}
    + p\, P^{\mu\nu}
    -  \rho\varpi\, h^{\mu} h^{\nu}
    + \mathcal{O}(\dow)~~, \nn\\
  J^{\mu\nu}
  &= 2 \rho \, u^{[\mu} h^{\nu]}
    + \textcolor{blue}
    {\varpi q_\times\, \epsilon^{\mu\nu\rho\sigma} u_{\rho} h_{\sigma}}
    + \mathcal{O}(\dow)~~, \nn\\
  N^\mu
  &= \frac{p}{T} u^\mu
    + \mathcal{O}(\dow)~~,
\end{align}
where $\rho = q\varpi$ was defined. All the coefficients appearing here are now
seen as functions of $T$ and $\varpi$. Except for the $q_\times$ term highlighted
in blue, the constitutive relations of an ideal one-form superfluid in the electric
limit are precisely the same as for string fluids given in
\cref{ideal_string_consti} and satisfy the thermodynamic relations
\bref{string_thermodynamics}.

The $q_\times$ term, on the other hand, is a manifestation of the order-mixing that was 
alluded to above. Comparing its form with \cref{general-order-mixing}, it is possible to
infer that it originates from a one-derivative term
$q_\times \zeta_\mu \bar\zeta^\mu$ in the free energy density. This is, in fact,
the case as it can be verified by expanding the ideal one-form superfluid free-energy density
\bref{sf-ideal-N} up to one derivative order, obtaining
\begin{equation}
  P(T,\zeta^2,\bar\zeta^2,\zeta\cdot\bar\zeta)
  = p(T,\varpi)
  + q_\times(T,\varpi) \, \zeta\cdot\bar\zeta
  + \mathcal{O}(\dow^2)~~.
\end{equation}
Additionally, due to the presence of the $q_\times$ term, the first order equations of motion
significantly modify compared to string fluids. The components of the
energy-momentum conservation stay the same as in \cref{EOM_string_first_EM},
while those of the one-form conservation receive contributions from the $q_\times$
term. Precisely, it is found 
\begin{subequations}\label{EOM_electric_first_J}
  \begin{align}
    \N_{"m} J^{"m"n} = 0& \nn\\
    \implies& \delta_\scB \rho
              + \frac{\rho}{2} \Delta^{\mu\nu} \delta_\scB g_{\mu\nu}
              + \frac{q_\times}{6T} \epsilon^{\mu\nu\rho\sigma}h_\mu H_{\nu\rho\sigma}
              = \mathcal{O}(\dow^2)~~, \label{EOM4e} \\
            & \Delta^{\mu\rho} u^\sigma \delta_\scB b_{\rho\sigma}
              + \varpi\Delta^{\mu\rho} h^\sigma \delta_\scB g_{\rho\sigma}
              + \frac{\varpi^2}{T\rho} \epsilon^{\mu\nu} \dow_\nu q_\times
              - \frac{q_\times\varpi}{6T\rho} \Delta^\mu{}_\lambda
              \epsilon^{\lambda\nu\rho\sigma}
              H_{\nu\rho\sigma}
              = \mathcal{O}(\dow^2)~~, \label{EOM5e} \\
            & \frac{1}{T} \N_{"m} \lb T \rho h^{\mu} \rb
              - \rho T u^\mu h^{\nu} \delta_\scB g_{\mu\nu}
              + \frac{q_\times}{6} \epsilon^{\mu\nu\rho\sigma} u_\mu H_{\nu\rho\sigma}
              = \mathcal{O}(\dow^2)~~.  \label{EOMconstraint_e}
  \end{align}
\end{subequations}
These equations imply that, as in the string fluid case, it is still possible to eliminate $u^\mu \delta_\scB g_{\mu\nu}$ using the first
order energy-momentum conservation equations. However, it is no longer possible to eliminate $u^\mu
\delta_\scB b_{\mu\nu}$ in terms of other non-hydrostatic data. This has
important consequences for one-derivative non-hydrostatic corrections.

Since the study of one-derivative corrections is the subject of our attention below, it is instructive to expand the
ideal one-form superfluid constitutive relations \bref{ideal_super_consti} to
one-derivative order. This expansion gives rise to
\begin{align}\label{ideal_electric_consti_1}
  T^{\mu\nu}
  &= \epsilon\, u^{\mu} u^{\nu}
    + p\, P^{\mu\nu}
    -  \rho\varpi\, h^{\mu} h^{\nu}
    + \zeta\cdot\bar\zeta \lb T \frac{\dow q_\times}{\dow T}
    + \varpi \frac{\dow q_\times}{\dow \varpi} \rb u^{\mu}u^{\nu}
    - \zeta\cdot\bar\zeta\,
    \varpi\frac{\dow q_\times}{\dow \varpi} h^{\mu}h^{\nu} \nn\\
  &\qquad
    - \textcolor{blue}
    {2\bar q\, u^{(\mu} \epsilon^{\nu)\rho\sigma\tau}
    u_{\rho}\zeta_{\sigma}\bar\zeta_{\tau}}
    + \mathcal{O}(\dow^2)~~, \nn\\
  J^{\mu\nu}
  &= 2\rho\, u^{[\mu} h^{\nu]}
    + \zeta\cdot\bar\zeta\,
    \frac{\dow q_\times}{\dow \varpi} 2 u^{[\mu}h^{\nu]}
    - 2q_\times u^{[\mu} \bar\zeta^{\nu]}
    - q_\times \epsilon^{\mu\nu\rho\sigma} u_\rho \zeta_\sigma \nn\\
  &\qquad
    - \textcolor{blue}{\epsilon^{\mu\nu\rho\sigma} u_{\rho}
    \lb \bar q\, \bar\zeta_{\sigma}
    + q'_\times(\zeta\cdot\bar\zeta)\, \zeta_{\sigma} \rb}
    + \mathcal{O}(\dow^2)~~, \nn\\
  N^\mu
  &= \frac{p}{T} u^\mu
    + \frac{q_\times}{T} \zeta\cdot\bar\zeta\, u^\mu
    + \mathcal{O}(\dow^2)~~.
\end{align}
The contributions from the one-derivative order term $q_\times$ are now complete,
while two new order-mixing contributions, $q'_\times$ and $\bar q$, from
two-derivative order appear. Their origin can be traced back to the free energy
density \bref{sf-ideal-N} expanded up to two-derivative order as
\begin{equation}\label{P-expanded-2}
  P(T,\zeta^2,\bar\zeta^2,\zeta\cdot\bar\zeta)
  = p(T,\varpi)
  + q_\times(T,\varpi) \zeta\cdot\bar\zeta
  + \half q'_\times (T,\varpi) (\zeta\cdot\bar\zeta)^2
  + \half \bar q(T,\varpi) \bar\zeta^2
  + \mathcal{O}(\dow^3)~~.
\end{equation}
It is clear from these considerations that order mixing significantly increases the difficulty of studying these hydrodynamic systems, nevertheless it is possible to keep track of it precisely and to obtain constitutive relations in a hydrodynamic expansion.

\subsection{One-derivative corrections}

\subsubsection{Hydrostatic corrections}
\label{sec:electric-hs-corrections}

Above it was shown that taking electric limit of ideal one-form superfluids generates
some one-derivative corrections to the respective constitutive
relations. However, the constitutive relations can also receive more generic
one-derivative corrections allowed by the adiabaticity equation \eqref{adiabaticity_simple}. 
Consider first the order mixing terms, whose general expression was given in 
\cref{general-order-mixing}. The most generic two-derivative terms in the
hydrostatic free energy density involving $\bar\zeta^\mu$ can be represented as
\begin{equation}
  \mathcal{N}_{(2)}
  = \half \lb q'_\times \zeta_\mu\zeta_\nu + \bar q P_{\mu\nu} \rb
  \bar\zeta^\mu \bar\zeta^\nu
  + R_\mu \bar\zeta^\mu + \ldots~~.
\end{equation}
Here the quadratic terms in $\bar\zeta^\mu$, i.e. $q'_\times$ and $\bar q$, are
the same as those obtained in \cref{P-expanded-2} in ideal one-form superfluids. The
linear terms in $\bar\zeta^\mu$ are parametrised by a generic one-derivative
vector structure $R_\mu$ which involves an explicit derivative. It will contain,
for example, terms proportional to $P_{\mu\nu} \nabla^\nu T$ and
$\epsilon_{\mu\nu\rho\sigma} u^\nu \nabla^\rho u^\sigma$ among many others. Using
\cref{general-order-mixing}, their contribution to one-derivative constitutive
relations is given as
\begin{align}\label{order-mixing-consti}
  T^{\mu\nu}_{\text{hs},\text{order-mixing}}
  &= - 2u^{(\mu} \epsilon^{\nu)\rho\sigma\tau} u_\rho \zeta_\sigma
    \lb \bar q \bar\zeta_\tau + R_\tau \rb
    + \mathcal{O}(\dow^2)~~, \nn\\
  J^{\mu\nu}_{\text{hs},\text{order-mixing}}
  &= - \epsilon^{\mu\nu\rho\sigma} u_\rho
    \lb \bar q \bar \zeta_\sigma + q'_\times (\zeta\cdot\bar\zeta) \zeta_\sigma
    + R_\sigma \rb
    + \mathcal{O}(\dow^2)~~, \nn\\
  N^{\mu}_{\text{hs},\text{order-mixing}}
  &= \mathcal{O}(\dow^2)~~.
\end{align}
Secondly, it is necessary to consider explicitly one-derivative order terms in the
hydrostatic free-energy density. It is possible to import all the terms directly from
string fluids in \cref{stringN}, except the $\alpha$ term which is no longer
independent due to the Bianchi identity \bref{termRemoval}. Taking into account
the contributions mentioned above, the total hydrostatic free energy
density for one-derivative constitutive relations reads
\begin{multline}\label{electricN}
  \mathcal{N} =
  p
  + q_\times \zeta_\mu \bar \zeta^\mu
  + \half \lb q'_\times \zeta_\mu\zeta_\nu + \bar q P_{\mu\nu} \rb
  \bar\zeta^\mu \bar\zeta^\nu
  + R_\mu \bar\zeta^\mu \\
  - \beta \epsilon^{\mu\nu\rho\sigma} u_\mu h_\nu \dow_\rho u_\sigma
  - \tilde\beta_1 h^\mu \dow_\mu T
  - \tilde\beta_2 h^\mu \dow_\mu \frac{\varpi}{T}
  - \tilde\beta_3 \epsilon^{\mu\nu\rho\sigma} u_\mu h_\nu \dow_\rho h_\sigma~~.
\end{multline}
The contributions from $p$, $q_\times$, $q'_\times$, and $\bar q$ are given in
\cref{ideal_electric_consti_1}, from $R_\mu$ in \cref{order-mixing-consti},
while those from $\beta$ and $\tilde\beta_i$ can be directly imported from
\cref{app:hydrostaticString}. As in the case of string fluids, the equation of
motion \eqref{EOMconstraint_e} together with the Bianchi identity
\eqref{termRemoval} allow to set $\tilde\beta_2=0$, thus leading to 3
independent hydrostatic transport coefficients at first order in derivatives.
This completes the analysis of first order hydrostatic corrections.

\subsubsection{Non-hydrostatic corrections}
For the non-hydrostatic contributions, it is useful to parametric the stress tensor and charge current as
\begin{align}
  T^{\mu\nu}_{\text{nhs}}
  &= \delta\epsilon ~u^{\mu}u^{\nu} + \delta f \Delta^{\mu\nu}
  + \delta\tau ~h^{\mu}h^{\nu} + 2\ell^{(\mu}h^{\nu)} + 2 k^{(\mu}u^{\nu)} +
    t^{\mu\nu}~~, \nn\\
  J^{\mu\nu}_{\text{nhs}}
  &= 2 \delta\rho ~u^{[\mu}h^{\nu]} + 2m^{[\mu} h^{\nu]} + 2 n^{[\mu} u^{\nu]} +
    \delta s\, \epsilon^{\mu\nu}~~.
\end{align}
Introducing these into \bref{adiabaticity_electric}, it is possible to massage the
adiabaticity equation into
\begin{multline}
  \nabla_\mu N^\mu
  = \lb \delta\epsilon
  - \frac{\delta\rho}{\dow\rho/\dow\varpi}
  \lb T \frac{\dow\rho}{\dow T} + \varpi \frac{\dow\rho}{\dow \varpi} \rb
  \rb
  \half u^\mu u^\nu \delta_\scB g_{\mu\nu}
  + \lb k^\mu - \varpi m^\mu \rb u^\nu \delta_\scB g_{\mu\nu} \\
  + \delta\rho
  \lb \delta_\scB\rho + \half\rho \Delta^{\mu\nu} \delta_\scB g_{\mu\nu} \rb
  + n^\mu \lb u^{\nu}"d_{"scB}b_{\mu\nu} + \varpi h^{\nu}"d_{"scB}g_{\mu\nu}
  \rb\\
  + \delta s \half \epsilon^{\mu\nu} \delta_\scB b_{\mu\nu}
  + m^\mu \lb h^{\nu}"d_{"scB}b_{\mu\nu}
  + \varpi u^{\nu}"d_{"scB} g_{\mu\nu} \rb \\
  + \lb \delta f - \frac{\rho\delta\rho}{\dow\rho/\dow\varpi} \rb
  \half \Delta^{\mu\nu} \delta_\scB g_{\mu\nu}
  + \lb \delta\tau + \varpi \delta\rho \rb
  \half h^\mu h^\nu \delta_\scB g_{\mu\nu}
  + \lb \ell^{\mu} - \varpi n^\mu \rb h^\nu \delta_\scB g_{\mu\nu} \\
  + \half t^{\mu\nu} \delta_\scB g_{\mu\nu}
  + \Delta~~.
  \label{eq:freeBCP}
\end{multline}
The rationale behind this arrangement is that the terms in the third line in \eqref{eq:freeBCP} drop
out using the Bianchi identities \bref{termRemoval}, while those in the second line in \eqref{eq:freeBCP}
drop out using the first order equations of motion \bref{EOM_electric_first_J} when
$q_\times$ is zero. However, these terms are important to complete the quadratic
form $\Delta$. It is possible to use the redefinition freedom in $u^\mu$ and $T$ to set
\begin{equation}
  \delta\epsilon
  = \frac{\delta\rho}{\dow\rho/\dow\varpi}
  \lb T \frac{\dow\rho}{\dow T} + \varpi \frac{\dow\rho}{\dow \varpi} \rb~~,
  \qquad
  k^\mu = \varpi m^\mu~~,
\end{equation}
and eliminate terms in the first line in \eqref{eq:freeBCP}. In string fluids,
it was possible to use the residual redefinition freedom in $\mu_\mu$ to set
$\delta\rho = n^\mu = 0$ as well. However, in the current context there is no
such freedom as it was already used to make the Josephson equation
\bref{eq:exactjoseph} exact. Schematically, the non-hydrostatic corrections can
be written as
\begin{gather}
  \begin{pmatrix}
    \delta \rho \\
    \delta f - \frac{\rho\delta\rho}{\dow\rho/\dow\varpi} \\
    \delta\tau + \varpi \delta\rho \\
    \delta s
  \end{pmatrix}
  =
  - \frac{T}{2} \begin{pmatrix}
    \textcolor{Plum}{\lambda_1} & \lambda_2
    & \lambda_3 & \textcolor{blue}{\lambda_4} \\
    \textcolor{Plum}{\lambda'_2} & \zeta_{\perp}
    & \zeta_{\times}
    & \textcolor{blue}{\tilde\kappa_{1}}  \\
    \textcolor{Plum}{\lambda'_3} & \zeta'_{\times}
    & \zeta_{\parallel} & \textcolor{blue}{\tilde\kappa_{2}}  \\
    \textcolor{Plum}{\lambda'_4} & \tilde\kappa'_{1} &
    \tilde\kappa'_{2} & \textcolor{blue}{r_{\parallel}}
  \end{pmatrix}
  \begin{pmatrix}
    2 \delta_\scB\rho + \rho \Delta^{\mu\nu} \delta_\scB g_{\mu\nu} \\
    \Delta^{\mu\nu}"d_{"scB}g_{\mu\nu} \\
    h^{\mu}h^{\nu} "d_{"scB}g_{\mu\nu} \\
    \epsilon^{\mu\nu} "d_{"scB}b_{\mu\nu}
  \end{pmatrix}, \nn\\
  \begin{pmatrix}
    n^\mu \\ \ell^{\mu} - \varpi n^\mu \\ m^{\mu}
  \end{pmatrix}
  =
  - T \begin{pmatrix}
    \textcolor{Plum}{\lambda_5} & \lambda_6
    & \textcolor{blue}{\lambda_7}
    & \textcolor{Plum}{\lambda_8} & \lambda_9
    & \textcolor{blue}{\lambda_{10}} \\
    \textcolor{Plum}{\lambda'_6}  & \eta_{\parallel}
    & \textcolor{blue}{r_{\times}}
    & \textcolor{Plum}{\lambda'_9}
    & \tilde\eta_{\parallel}
    & \textcolor{blue}{\tilde r_{\times}} \\
    \textcolor{Plum}{\lambda'_7} & r'_{\times}
    & \textcolor{blue}{r_{\perp}}
    & \textcolor{Plum}{\lambda'_{10}}
    & \tilde r'_{\times}
    & \textcolor{blue}{\tilde r_{\perp}}
  \end{pmatrix} 
  \begin{pmatrix}
    \Delta^{\mu\sigma}u^{\nu}"d_{"scB}b_{\sigma\nu}
    + \varpi \Delta^{\mu\sigma}h^{\nu}"d_{"scB}g_{\sigma\nu} \\
    \Delta^{\mu\sigma}h^{\nu}"d_{"scB}g_{\sigma\nu} \\
    \Delta^{\mu\sigma}h^{\nu}"d_{"scB}b_{\sigma\nu}
    + \varpi \Delta^{\mu\sigma}u^{\nu}"d_{"scB} g_{\sigma\nu} \\
    \epsilon^{\mu\sigma}u^{\nu}"d_{"scB}b_{\sigma\nu}
    + \varpi \epsilon^{\mu\sigma}h^{\nu}"d_{"scB}g_{\sigma\nu} \\
    \epsilon^{\mu\sigma}h^{\nu}"d_{"scB}g_{\sigma\nu} \\
    \epsilon^{\mu\sigma}h^{\nu}"d_{"scB}b_{\sigma\nu}
    + \varpi \epsilon^{\mu\sigma}u^{\nu}"d_{"scB} g_{\sigma\nu}
  \end{pmatrix}, \nn\\
  t^{\mu\nu} = - \eta_{\perp} T\Delta^{\rho\langle\mu}\Delta^{\nu\rangle\sigma}
  "d_{"scB} g_{\rho\sigma}
  + \tilde\eta_{\perp} T\epsilon^{\rho\langle\mu}\Delta^{\nu\rangle\sigma}
  "d_{"scB} g_{\rho\sigma}~~.
  \label{eSF_nhs_corrections}
\end{gather}
Since the tensor structures
\begin{equation}
  \epsilon^{\mu\nu} "d_{"scB}b_{\mu\nu}~~,~~
  \Delta^{\mu\sigma}h^{\nu}"d_{"scB}b_{\sigma\nu}
  + \varpi \Delta^{\mu\sigma}u^{\nu}"d_{"scB} g_{\sigma\nu}~~,~~
  \epsilon^{\mu\sigma}h^{\nu}"d_{"scB}b_{\sigma\nu}
  + \varpi \epsilon^{\mu\sigma}u^{\nu}"d_{"scB} g_{\sigma\nu}~~,
\end{equation}
are second order due to the Bianchi identities \bref{termRemoval}, the transport
coefficients highlighted in blue are actually second order, but are required for
positive definiteness of $\Delta$. The terms highlighted in purple are first
order in general but become second order when using the
first order equations of motion \bref{EOM_electric_first_J} if $q_\times =
0$. In general, the positive definiteness of $\Delta$ gives 9 inequalities among
these transport coefficients and at first order there is a total of 26
non-hydrostatic transport coefficients.  However, the application of this theory
to magnetic dominated bound-charge plasmas that is provided in
sec.~\ref{sec:boundcharge} consists of setting $q_\times=0$ and leads to, upon
appropriate identification, 8 non-hydrostatic transport coefficients, namely
$\zeta_\perp$, $\zeta_\times$, $\zeta_\times'$, $\zeta_{||}$, $\eta_{||}$,
$\tilde\zeta_{||}$, $\eta_\perp$, $\tilde\eta_\perp$.

\section{Hot electromagnetism}
\label{sec:hotEM}

Hot electromagnetism is the theory that results from the interaction of electromagnetic degrees of freedom with mechanical and thermal degrees of freedom of matter. At long wavelength and large timescales compared to the mean free path of the microscopic theories, matter can be approximated by a hot plasma and hydrodynamic theory determines the dynamical evolution of fluctuations around equilibrium. In this section, the term \emph{hot electromagnetism} is used to denote the traditional treatments of hydrodynamic regimes of plasmas where the electromagnetic gauge field $A_\mu$ is incorporated as dynamical degrees of freedom. After a brief exposure of the different types of regimes that are considered in this work, namely MHD where electric fields are Debye screened, and bound-charge plasmas where they are not, exact dualities between different limits of one-form superfluids considered in the previous sections and the these two regimes are derived.

\subsection{Heating up Maxwell's equations}
Consider an electromagnetic plasma heated up to a finite temperature. The near
equilibrium physics of such a plasma is governed by charged hydrodynamics
coupled to dynamical electromagnetic fields. The dynamics of the electromagnetic
fields $F_{\mu\nu}$ is governed by Maxwell equations in
matter\footnote{\Cref{MHD-EOM} is a modified version of the second equation in
  \eqref{eq:eomEM} that accounts for the presence of matter and couplings to
  external currents.}
\begin{subequations}\label{MHD-EOM}
  \begin{equation}\label{Maxwells-Equations}
    \nabla_\nu F^{\nu\mu} + J^\mu_{\text{matter}} + J^\mu_\ext = 0~~,
  \end{equation}
  along with the Bianchi identity
  \begin{equation}\label{Bianchi-MHD}
    \nabla_{[\mu} F_{\nu\rho]} = 0~~.
  \end{equation}
  Here $J^\mu_\ext$ denotes an identically conserved background charge current
  distribution coupled to the plasma such that $\nabla_\mu J^\mu_\ext =
  0$. Prime examples of $J^\mu_\ext$ include a lattice of ions or an auxiliary
  field theory source that facilitates the computation of correlation
  functions. $J^\mu_{\text{matter}}$ is the charge current associated with the
  matter component of the plasma and it is not required to be trivially
  conserved at finite temperature. In fact, the conservation equation
  $\nabla_\mu J^\mu_{\text{matter}} = 0$, which can be seen as the divergence of
  \cref{Maxwells-Equations}, serves as an equation of motion for the
  hydrodynamic chemical potential $\mu$. As already commented in
  sec.~\ref{sec:emzero}, the Bianchi identity \bref{Bianchi-MHD} is solved by
  introducing the dynamical photon field $A_\mu$ such that
  $F_{\mu\nu} = 2 \dow_{[\mu} A_{\nu]}$. Having done that,
  \cref{Maxwells-Equations} provides dynamics for 4 physical degrees of freedom
  in $A_\mu$ and $\mu$. In addition, the plasma is characterised by the usual
  hydrodynamic fields $u^\mu$ and $T$, whose dynamics is governed by
  energy-momentum conservation
  \begin{equation}\label{EMConservation-MHD}
    \nabla_\mu T^{\mu\nu} = F^{\nu\rho} J_\rho~~,
  \end{equation}
  where the total dynamical charge current of the plasma $J^\mu = \nabla_\nu F^{\nu\mu} + J^\mu_{\text{matter}}$ was introduced.
\end{subequations}

The constitutive relations of hot electromagnetism are written as expressions
for $T^{\mu\nu}$ and $J^\mu$ in terms of $u^\mu$, $T$, $\mu$, and $F_{\mu\nu}$,
arranged in a derivative expansion. A priori, these may be expected to be
exactly the same as ordinary charged fluids with background electromagnetic
fields.  However, since the electromagnetic fields are dynamical, they can be
relevant at ideal order in the derivative expansion, i.e.
$F_{\mu\nu} = \mathcal{O}(1)$.  This considerably modifies the actual
constitutive relations~\cite{Kovtun:2016lfw, Hernandez:2017mch}. Similar to
ordinary hydrodynamics, the constitutive relations of a plasma are also required
to satisfy the second law of thermodynamics. This requirement is formulated in
terms of the zero-form version of the adiabaticity equation
\eqref{adiabaticity_simple}, namely
\begin{equation}\label{adiabaticity-MHD}
  \nabla_\mu N^\mu
  = \half T^{\mu\nu} \delta_\scB g_{\mu\nu}
  + J^\mu \delta_\scB A_\mu + \Delta~~,~~
  \Delta \geq 0~~,
\end{equation}
which has to be satisfied for some free energy current $N^\mu$ and quadratic
form $\Delta$. Here $\delta_\scB$ denotes an infinitesimal symmetry
transformation along $\scB = (\beta^\mu, \Lambda^\beta)$ introduced in
\cref{eq:hydro0fields}, which when applied to the metric and gauge field read
\begin{equation}\label{dbVariations_MHD}
  \delta_\scB g_{\mu\nu} = 2 \nabla_{(\mu} \bfrac{u_{\nu)}}{T}~~,~~
  \delta_\scB A_\mu = \dow_\mu \frac{\mu}{T} - \frac1T E_\mu~~,
\end{equation}
where the electric $E^\mu$ and magnetic fields $B^\mu$ are defined as
\begin{equation}
  \label{eq:stEM}
  E^\mu = F^{\mu\nu}u_\nu~~,~~
  B^\mu = \half \epsilon^{\mu\nu\rho\sigma} u_\nu F_{\rho\sigma}~~, \qquad
  F_{\mu\nu} = 2 u_{[\mu]} E_{\nu} - \epsilon_{\mu\nu\rho\sigma} u^\rho B^\sigma.
\end{equation}
Provided that \cref{adiabaticity-MHD} is satisfied, the entropy current, defined
as $S^\mu = N^\mu - T^{\mu\nu} u_\nu/T - J^\mu \mu/T$, has positive
semi-definite divergence onshell (i.e. once the equations of motion are
satisfied).

\subsubsection{Ideal fluid minimally coupled to electromagnetism}
As a working example, and to aid intuition, consider the well-known model in the context of MHD \cite{davidson2001introduction} of an ideal fluid 
minimally coupled to Maxwell's electromagnetism via a conductivity term $\sigma$ in the constitutive relations
\begin{align}\label{decoupled-MHD-consti}
  T^{\mu\nu}
  &=
    F^{\mu}{}_\rho F^{\nu\rho}
    - \frac14 F_{\rho\sigma} F^{\rho\sigma} g^{\mu\nu}
    + \epsilon(T,\mu) u^\mu u^\nu
    + p(T,\mu) \lb g^{\mu\nu} + u^\mu u^\nu \rb~~, \nn\\
  J^\mu
  &= \nabla_\nu F^{\nu\mu} + q(T,\mu)\, u^\mu
  - \sigma(T,\mu) P^{\mu\nu} \lb T \dow_\nu \frac{\mu}{T} - E_\nu \rb~~,
\end{align}
where the fluid part of the currents satisfies the usual thermodynamic relations
$\df p = s\df T + q \df\mu$ and $\epsilon + p = sT + q\mu$. The energy-momentum
tensor includes the purely electromagnetic contribution given in \cref{eq:stEM} alongside the usual fluid
contributions. These relations satisfy \cref{adiabaticity-MHD} with
\begin{gather}
  N^\mu = - \frac{1}{4T} F_{\rho\sigma} F^{\rho\sigma}\, u^\mu
  + \frac1T F^{\mu\nu} \lb T \dow_\nu \frac{\mu}{T} - E_\nu \rb
  + \frac{p(T,\mu)}{T} u^\mu~~, \nn\\
  \Delta
  = \frac{\sigma(\mu,T)}{T} P^{\mu\nu} \lb T \dow_\mu \frac{\mu}{T} - E_\mu \rb
  \lb T \dow_\nu \frac{\mu}{T} - E_\nu \rb~~,
\end{gather}
provided that the conductivity obeys the positivity constraint
$\sigma(\mu,T) > 0$. In general, the constitutive relations of the plasma
\eqref{decoupled-MHD-consti} could admit further derivative corrections and
exhibit more intricate couplings between the electromagnetic and fluid sectors
as it will be described later.

Using \cref{MHD-EOM}, it is possible to work out the equations of motion for this simple
plasma model. For the purposes of the current discussion, it suffices to look at \cref{Maxwells-Equations} which leads to
\begin{equation}\label{intro-muE}
  q(T,\mu)
  = u_\mu \nabla_\nu F^{\nu\mu}
  + u_\mu J^\mu_\ext~~,~~
  \sigma(T,\mu) E^\mu
  = - P^{\mu}{}_\lambda\nabla_\nu F^{\nu\lambda}
  - P^{\mu}{}_\lambda J^\lambda_\ext
  + T\sigma(T,\mu) P^{\mu\nu} \dow_\nu \frac{\mu}{T}~~.
\end{equation}
The first equation expresses the point that, to leading order in derivatives,
the charge density of the plasma organises itself according to the charge
density of the background. The second equation states that, to leading order,
the electric fields in the plasma are induced by external
currents. Additionally, these two equations algebraically determine the plasma
dynamical fields $\mu$ and $E^\mu$ in terms of the other dynamical and
background fields of the theory order by order in the derivative
expansion.\footnote{The rationale here is that if a dynamical field $f$
  satisfies an equation $f = f_0 + \mathcal{F}(\nabla,f)$, where
  $\mathcal{F}(\nabla,f)$ is at least one order in derivatives, then we can
  algebraically determine it recursively within the derivative expansion as
  $f = f_0 + \mathcal{F}(\nabla, f_0 + \mathcal{F}(\nabla, f_0 +
  \mathcal{F}(\nabla, f_0 + \mathcal{F}(\nabla,f_0 + \ldots ))))$.  } Therefore,
$\mu$ and $E^\mu$ do not in general obtain independent dynamics in the
hydrodynamic regime of hot electromagnetism. In fact, this statement continues
to hold when the most general coupling and derivative corrections are taken into
account (see sec.~\ref{sec:MHDd}). An interesting exception to this, which will
be studied below, is the case of plasmas which have
$q(T,\mu) = \sigma(T,\mu) = 0$.

\subsubsection{The magnetohydrodynamics regime} \label{sec:MHDregime}
Consider the sector of hot electromagnetism for which the background currents are
stationary to leading order, i.e. the spatial currents are derivative suppressed
$P_{\mu\nu} J^\nu_\ext = \mathcal{O}(\dow)$.\footnote{One can show that this
  requirement is frame-invariant by noting that under
  $u^\mu \to u^\mu + \delta u^\mu$, where $\delta u^\mu = \mathcal{O}(\dow)$
  such that $u_\mu \delta u^\mu = 0$, it remains invariant.} An example of such
backgrounds is the case of a fixed lattice of ions. From \cref{intro-muE}, it follows
that the electric fields in such plasmas are derivative suppressed, i.e.
$E_\mu = \mathcal{O}(\dow)$, while the magnetic fields can be arbitrarily
large. This is the hydrodynamic incarnation of \emph{Debye screening}: electric
fields are screened over large distances due to the presence of free
charges.\footnote{The usual requirement for Debye screening, found in traditional textbooks of MHD, is to take the limit $\sigma\to\infty$. From the second equation in \eqref{intro-muE}, it is obvious that this has the same effect as that attained by requiring $P_{\mu\nu} J^\nu_\ext = \mathcal{O}(\dow)$. However, this ``infinite conductivity limit'' breaks the hydrodynamic derivative expansion. For this reason, it appears that the requirement $P_{\mu\nu} J^\nu_\ext = \mathcal{O}(\dow)$ is more physically sound.} Such hydrodynamic systems are commonly referred to as
\emph{magnetohydrodynamics} (MHD) (see e.g.~\cite{Hernandez:2017mch}). Under the
MHD limit, the energy-momentum tensor in \cref{decoupled-MHD-consti} becomes
\begin{equation}
  T^{\mu\nu}
  = \lb \epsilon(T,\mu) + p(T,\mu) \rb u^\mu u^\nu
  + \lb p(T,\mu) - \frac12 B^2 \rb g^{\mu\nu}
  + B^2 \bbB^{\mu\nu}
  + \mathcal{O}(\dow)~~,
\end{equation}
where $\bbB^{\mu\nu} = P^{\mu\nu} - \hat B^\mu \hat B^\nu$,
$P^{\mu\nu}=g^{\mu\nu}+u^{\mu}u^{\nu}$ and $\hat B^\mu = B^\mu/|B|$, with $|B|$
being the modulus of $B^\mu$.  For most applications of MHD, it is useful to
consider the scenario where the background charge current is entirely derivative
suppressed, i.e.  $J^\mu_\ext = \mathcal{O}(\dow)$, making the requirement of
sub-leading external currents ``covariant''. Such models are applicable when the
background charge currents are either non-existent or negligible, as in the case
of solar physics. Thus, in addition to the electric fields being screened, such
plasmas are electrically neutral over large length scales, i.e.
$q(T,\mu) = \mathcal{O}(\dow)$. In this regime, MHD can be reformulated in terms
of a string fluid with a global one-form symmetry \cite{Armas:2018atq}. This
connection will be developed further in \cref{sec:MHD-String}.

\subsubsection{The bound-charge plasma regime}

An often unstated requirement for the MHD regime to dominate the hydrodynamics
of plasmas is that the plasmas are conducting, i.e. $\sigma(T,\mu) \neq 0$,
otherwise the second equation in \cref{intro-muE} would not impose any
restriction on electric fields, and hence they could be arbitrarily
large. Consider a fluid which does not contain any free charges. For instance, a
gas of neutral atoms which can nonetheless be polarised. In the absence of any
free charge carriers, the conductivity $\sigma(T,\mu)$ is identically zero. Over
large distances, the charge density $q(T,\mu)$ also adds up to zero. More
rigorously, these are plasmas whose constitutive relations do not depend on
$\mu$. That is, in the simple case of \cref{decoupled-MHD-consti}, $p = p(T)$
and $\sigma = 0$ leading to $\epsilon = \epsilon(T)$ and $q = 0$ by means of the
thermodynamic relations. Consequently, $\mu$ drops out from the set of
independent degrees of freedom and the charge conservation
$\nabla_\mu J^\mu = 0$, which had the role of providing dynamics to $\mu$,
becomes identically satisfied implying that
\begin{equation} \label{eq:Mmunu}
  \nabla_\mu J^\mu = 0
  \quad\implies\quad
  J^\mu = \nabla_\nu M^{\mu\nu}~~, ~~
  M^{\mu\nu} = - F^{\mu\nu} + M^{\mu\nu}_{\text{matter}}~~,
\end{equation}
where $M^{\mu\nu}_{\text{matter}}$ is the antisymmetric polarisation tensor
characteristic of the material that constitutes the plasma. The physical content
of the leading order Maxwell's equations \bref{intro-muE} is then that such a
system can only be described by hydrodynamics when the background charge current
is weak, i.e.  $J^\mu_\ext = \mathcal{O}(\dow)$. The dynamical equations
\bref{MHD-EOM} and adiabaticity equation \bref{adiabaticity-MHD} for a
\emph{bound-charge plasma} can be recast as
\begin{gather}
  \nabla_\mu T^{\mu\nu}
  = - F^{\nu\rho} J_\rho^\ext~~,~~
  \nabla_\mu M^{\mu\nu} = J^\nu_\ext~~,~~
  \epsilon^{\mu\nu\rho\sigma}\nabla_\nu F_{\rho\sigma} = 0~~, \nn\\
  \nabla_\mu N^\mu
  = \half T^{\mu\nu} \delta_\scB g_{\mu\nu}
  + \half M^{\mu\nu} \delta_\scB F_{\mu\nu}
  + \Delta~~, \qquad \Delta \geq 0~~,
  \label{bound-plasma-EOM}
\end{gather}
where $N^\mu \to N^\mu - M^{\mu\nu} \delta_\scB A_\nu$ was redefined.  Maxwell's electromagnetism in vacuum is self-dual under electromagnetic
duality. There is a version of this duality that is still respected by the bound-charge plasma. It may be verified that under the transformation
\begin{gather}
  F_{\mu\nu} \to \half \epsilon_{\mu\nu\rho\sigma} M^{\rho\sigma}~~,~~
  M^{\mu\nu} \to \half \epsilon^{\mu\nu\rho\sigma} F_{\rho\sigma}~~,~~N^\mu \to N^\mu - \half \beta^\mu M^{\mu\nu} F_{\mu\nu}~~,
\end{gather}
the equations of motion \eqref{bound-plasma-EOM} map to themselves when
$J^\mu_\ext = 0$ and with the same energy-momentum tensor $T^{\mu\nu}$. In
\cref{sec:boundcharge}, it will be shown that eqs.~\eqref{bound-plasma-EOM} are
essentially the governing equations of one-form superfluid dynamics.

\subsection{Magnetohydrodynamics} \label{sec:MHDd}

This section deals with the MHD regime of hot electromagnetic plasmas described
in sec.~\ref{sec:MHDregime}. The ideal order constitutive relations of these
plasmas are essentially the same as the constitutive relations of ordinary
charged hydrodynamics, except that magnetic fields can be arbitrary large, i.e.
$B^\mu = \mathcal{O}(1)$, and the electric fields are derivative suppressed,
i.e. $E^\mu = \mathcal{O}(\dow)$.  Though many of the results that will be
presented in this section already appeared in \cite{Hernandez:2017mch}, the
details given here provide a cleaner derivation of these results and extends the
traditional treatment of MHD to include parity-violating terms.

\subsubsection{Ideal magnetohydrodynamics}

At ideal order, MHD is characterised by a hydrostatic free energy density of the
form $\mathcal{N} = P(T,\mu,B^2)$. This free energy is the most general at ideal
order and makes no assumptions on the strength of the coupling between
electromagnetic degrees of freedom and thermal degrees of freedom.  Using the
$\delta_\scB$ variations of the free arguments with respect to the fields
\eqref{eq:hydro0fields}
\begin{gather}
  \delta_\scB T = \frac{T}{2} u^\mu u^\nu \delta_\scB g_{\mu\nu}~~,~~
  \delta_\scB \mu = \frac{\mu}{2} u^\mu u^\nu \delta_\scB g_{\mu\nu}
  + u^\mu \delta_\scB A_\mu~~, \nn\\
  \delta_\scB B^2 = \lb B^\mu B^\nu - B^2 P^{\mu\nu}
  - 2 u^{(\mu} \epsilon^{\nu)\lambda\rho\sigma} B_\lambda u_\rho E_{\sigma} \rb \delta_\scB g_{\mu\nu}
  - 2 \epsilon^{\mu\nu\rho\sigma} B_\rho u_\sigma \nabla_\nu \delta_\scB A_\mu~~,
\end{gather}
together with the zero-form version of \cref{hsConstiGeneral} (i.e. with
$b_{\mu\nu}\to A_\mu$), it is possible to infer the respective constitutive
relations, free energy, and entropy currents. These take the form
\begin{align}
  T^{\mu\nu}
  &= (\epsilon + P) u^\mu u^\nu
    + P g^{\mu\nu}
    + \varpi |B| \bbB^{\mu\nu}
    + 2\varpi u^{(\mu} \epsilon^{\nu)\lambda\rho\sigma} \hat B_\lambda u_\rho
    E_{\sigma}~~, \nn\\
  J^\mu
  &= q u^\mu - \nabla_\nu \lb \varpi \epsilon^{\mu\nu\rho\sigma} \hat B_\rho
    u_\sigma\rb~~, \nn\\
  N^\mu
  &= \frac{P}{T} u^\mu + \varpi \epsilon^{\mu\nu\rho\sigma} \hat B_\rho u_\sigma
    \lb \dow_\nu \frac{\mu}{T} - \frac1T E_\nu \rb~~, \nn\\
  S^\mu
  &= s u^\mu
    + \nabla_\nu \lb \frac{\mu\varpi}{T} \epsilon^{\mu\nu\rho\sigma} \hat B_\rho
    u_\sigma\rb~~,
\end{align}
where the thermodynamics can be expressed as 
\begin{equation}\label{MHD-thermodynamics}
  \df P = s \df T + q \df \mu - \frac{\varpi}{2|B|} \df B^2~~,~~
  \epsilon + P = sT + q\mu~~.
\end{equation}
Here $\varpi$ is being defined as $-2|B| \dow P/\dow B^2$ and will later be
identified with the string chemical potential in the dual higher-form language.
Note that from \cref{hsConstiGeneral}, the first order terms appear in the ideal
MHD constitutive relations but these can be ignored when focusing on zero
derivative order. These constitutive relations reduce to the simple model
\eqref{decoupled-MHD-consti} upon using $P(T,\mu,B^2) = - B^2/2 + p(T,\mu)$.

The equations of motion \eqref{MHD-EOM}-\eqref{EMConservation-MHD} at ideal order take the form
\begin{align}\label{MHD-firstOrderEOM}
  \nabla_\mu T^{\mu\nu} = F^{\nu\rho} J_\rho
  \implies
  &
    - u^\mu \nabla_\mu \epsilon
    - (\epsilon + P) \nabla_\mu u^\mu
    - \varpi |B| \bbB^{\mu\nu} \nabla_\mu u_\nu
    = \mathcal{O}(\dow^2)~~, \nn\\
  &
    (\epsilon + P) P^{\mu\nu} \lb \frac{1}{T} \dow_\nu T
    + u^\lambda \nabla_\lambda u_\nu \rb
    + q P^{\mu\nu} \lb T \dow_\nu \frac{\mu}{T} - E_\nu \rb \nn\\
  &\qquad
    + \epsilon^{\nu\rho\alpha\beta} u_\alpha B_\beta J^{(1)}_\rho
    = \mathcal{O}(\dow^2)~~, \nn\\
  J^\mu + J^\mu_\ext = 0
  \implies 
  & q(T,\mu,B^2) = u_\mu J^\mu_\ext + \mathcal{O}(\dow)~~, \nn\\
  & P_{\mu\nu} J^\nu_\ext = \mathcal{O}(\dow)~~.
\end{align}
where a component of the Bianchi identity \eqref{Bianchi-MHD} 
\begin{equation}
  \nabla_\mu B^\mu = B^\mu u^\nu \nabla_\nu u_\mu
  - \epsilon^{\mu\nu\rho\sigma} E_\mu u_\nu \dow_\rho u_\sigma~~,
\end{equation}
was used. In particular, note the appearance of one derivative corrections to
the charge current $J^\mu_{(1)}$ in the transverse components to $u^\mu$ of the
energy-momentum conservation. The transverse components of the Maxwell's
equations imply that the transverse components of the external current sources
$J^\mu_\ext$ must be derivative suppressed, as earlier advertised in
sec.~\ref{sec:MHDregime}. The component along the velocity, on the other hand,
implies that $q(T,\mu,B^2) = u_\mu J^\mu_\ext$ onshell at ideal order. This
equation can be formally solved for $\mu$ and leads to the inference
\begin{equation}\label{appendix-remove-mu}
  \mu = \mu_0(T,B^2,u_\mu J^\mu_\ext) + \mathcal{O}(\dow)~~.
\end{equation}
Therefore, $\mu$ is not a true independent degree of freedom of the theory. At
first order in derivatives, it will be seen that this statement also holds true for electric
fields. Thus the magnetic fields are the only true dynamical degrees of
freedom in the $\rmU(1)$ sector of magnetohydrodynamics.

\subsubsection{One derivative corrections} \label{sec:onederivativeMHD}
The ideal MHD theory described above can be extended to one derivative order in both the hydrostatic and non-hydrostatic sectors. 
The most generic hydrostatic free energy density at first order is given by
\begin{multline}\label{N-MHD}
  \mathcal{N} = P
  + M_1 B^\mu \dow_\mu \frac{B^2}{T^4}
  + M_2 \epsilon^{\mu\nu\rho\sigma} u_\mu B_\nu \dow_\rho B_\sigma \\
  - \frac{M_3}{T} B^\mu \dow_\mu T
  - M_4 \epsilon^{\mu\nu\rho\sigma} u_\mu B_\nu \dow_\rho u_\sigma
  + T M_5 B^\mu \dow_\mu \frac{\mu}{T}
  + \mathcal{O}(\dow^2)~~,
\end{multline}
where all the coefficients $M_i ~ (i=1,...,5)$ are functions of $T$, $\mu$, and
$B^2$. It is possible to vary these first order contributions so as to obtain
the respective contributions to the constitutive relations, which are detailed
in app.~\ref{app:hydrostaticMHD}. The hydrostatic free energy \eqref{N-MHD} had
been considered in \cite{Hernandez:2017mch}. However, it is noted here that due
to the Bianchi identity \eqref{Bianchi-MHD}, the term involving $M_1$ is not
independent and hence $M_1$ can be set to zero. This has led to an over-counting
of independent hydrostatic coefficients in
\cite{Hernandez:2017mch}. Nevertheless, for the purposes of comparison with
earlier literature, a non-vanishing $M_1$ coefficient is considered here.

In turn, the non-hydrostatic corrections can be obtained as in previous sections. As in earlier cases, the 
equations of motion allow to remove $u^\mu \delta_\scB g_{\mu\nu}$ and
$u^\mu \delta_\scB A_\mu$ from the independent non-hydrostatic tensor
structures. The most generic corrections can then be written as
\begin{align}\label{MHD-nhs-corrections}
  T^{\mu\nu}_{\text{nhs}}
  &= \delta\mathcal{F} \bbB^{\mu\nu}
  + \delta\mathcal{T} \hat B^\mu \hat B^\nu
  + 2 \mathcal{L}^{(\mu} \hat B^{\nu)}
  + \mathcal{T}^{\mu\nu}~~, \nn\\
  J^\mu_{\text{nhs}}
  &= \delta\mathcal{S} \hat B^\mu
  + \mathcal{M}^\mu~~,
\end{align}
where the different components of the stress tensor and charge current can be
written in terms of matrices of transport coefficients
\begin{gather}
  \begin{pmatrix}
    \delta\mathcal{F} \\
    \delta\mathcal{T} \\
    \delta\mathcal{S} 
  \end{pmatrix}
  = - T
  \begin{pmatrix}
    \zeta_{11} & \zeta_{12} & \textcolor{blue}{\tilde\chi_1} \\
    \zeta'_{12} & \zeta_{22} & \textcolor{blue}{\tilde\chi_2} \\
    \textcolor{blue}{\tilde\chi'_1} & \textcolor{blue}{\tilde\chi'_2} & \sigma_\parallel
  \end{pmatrix}
  \begin{pmatrix}
    \half\bbB^{\mu\nu}\delta_\scB g_{\mu\nu} \\
    \half\hat B^\mu \hat B^\nu \delta_\scB g_{\mu\nu} \\
    \hat B^\mu \delta_\scB A_\mu
  \end{pmatrix}~~, \nn\\
  \begin{pmatrix}
    \mathcal{L}^\mu \\
    \mathcal{M}^\mu
  \end{pmatrix}
  = - T
  \begin{pmatrix}
    \eta_{11} & \textcolor{blue}{\sigma_{\times}} & \tilde\eta_{11} & \textcolor{blue}{\tilde\sigma_\times} \\
    \textcolor{blue}{\sigma'_\times} & \sigma_\perp & \textcolor{blue}{\tilde\sigma'_\times} & \tilde\sigma_\perp
  \end{pmatrix}
  \begin{pmatrix}
    \bbB^{\mu\sigma} \hat B^\nu \delta_\scB g_{\sigma\nu} \\
    \bbB^{\mu\sigma} \delta_\scB A_\sigma \\
    \epsilon^{\mu\alpha\beta\sigma} u_\alpha \hat B_\beta \hat B^\nu \delta_\scB
    g_{\sigma\nu} \\
    \epsilon^{\mu\alpha\beta\sigma} u_\alpha \hat B_\beta \delta_\scB A_{\sigma}
  \end{pmatrix} ~~,\\
  \mathcal{T}^{\mu\nu}
  = - \eta_{22} T \bbB^{\rho\langle\mu} \bbB^{\nu\rangle\sigma}\delta_\scB
  g_{\rho\sigma}
  + \tilde\eta_{22} T \epsilon^{\rho\alpha\beta\langle\mu}
  u_\alpha \hat B_\beta \bbB^{\nu\rangle\sigma} \delta_\scB g_{\rho\sigma}~~.
\end{gather}
The 8 coefficients in blue are parity-violating terms whose existence had been identified in \cite{Hernandez:2017mch} but were not studied in any detail.

\subsubsection{Maxwell's equations}
In this section it is shown that $\mu$ and $E^\mu$ are not dynamical degrees of freedom in MHD. Assembling all the contributions from the previous subsections, the most general charge current $J^\mu$ up to first order in derivatives can be written in the form
\begin{multline}\label{fullJ-MHD}
  J^\mu
  = q u^\mu
  - \nabla_\nu \lb \varpi \epsilon^{\mu\nu\rho\sigma}\hat B_\rho u_\sigma \rb
  + \lb B^\lambda \dow_\lambda \frac{B^2}{T^4} \frac{\dow M_1}{\dow \mu}
  + \epsilon^{\lambda\nu\rho\sigma} 
  u_\lambda B_\nu \dow_\rho B_\sigma \frac{\dow M_2}{\dow \mu}
  \dbrk
  - \frac{1}{T} B^\lambda \dow_\lambda T \frac{\dow M_3}{\dow \mu}
  - \epsilon^{\lambda\nu\rho\sigma} u_\lambda B_\nu \dow_\rho u_\sigma
  \frac{\dow M_4}{\dow \mu}
  + T B^\lambda \dow_\lambda \frac{\mu}{T} \frac{\dow M_5}{\dow \mu}
  - \frac1T \nabla_\lambda \lb T M_5 B^\lambda \rb
  \rb u^\mu \\
  - \lb \textcolor{blue}{\tilde\chi'_1} \hat B^\mu \bbB^{\rho\sigma}
  + \textcolor{blue}{\tilde\chi'_2} \hat B^\mu  \hat B^\rho \hat B^\sigma
  + 2 \textcolor{blue}{\sigma'_\times} \bbB^{\mu(\rho} \hat B^{\sigma)}
  + 2\textcolor{blue}{\tilde\sigma'_\times}
  \epsilon^{\mu\alpha\beta(\rho} u_\alpha \hat B_\beta \hat B^{\sigma)}
  \rb \frac{T}{2} \delta_\scB g_{\rho\sigma} \\
  - \lb \sigma_\parallel \hat B^\mu \hat B^\nu
  + \sigma_\perp \bbB^{\mu\nu}
  + \tilde\sigma_\perp \epsilon^{\mu\nu\alpha\beta} u_\alpha \hat B_\beta
  \rb T \delta_\scB A_\nu
  + \mathcal{O}(\dow^2)~~.
\end{multline}
Inserting this current into Maxwell's equations \cref{Maxwells-Equations}, the different components read
\begin{align}\label{Maxwells-firstOrder-MHD}
  q(T,\mu,B^2)
  &= u_\mu J^\mu_\ext
    - B^\lambda \dow_\lambda \frac{B^2}{T^4} \frac{\dow M_1}{\dow \mu}
    - \epsilon^{\lambda\nu\rho\sigma} 
    u_\lambda B_\nu \dow_\rho B_\sigma \frac{\dow M_2}{\dow \mu}
    + \frac{1}{T} B^\lambda \dow_\lambda T \frac{\dow M_3}{\dow \mu} \nn\\
  &\quad
    - \lb \frac{\dow M_4}{\dow \mu} + \frac{\varpi}{|B|} \rb
    B_\mu u_\nu \dow_\rho u_\sigma
    - T B^\lambda \dow_\lambda \frac{\mu}{T} \frac{\dow M_5}{\dow \mu}
    + \frac1T \nabla_\lambda \lb T M_5 B^\lambda \rb
    + \mathcal{O}(\dow^2), \nn\\
  \sigma_\parallel \hat B^\nu \delta_\scB A_\nu
  &= \half \bbE^{\mu\nu} X_{\mu\nu}
    - \lb \textcolor{blue}{\tilde\chi'_1} \bbB^{\rho\sigma}
    + \textcolor{blue}{\tilde\chi'_2} \hat B^\rho \hat B^\sigma
    \rb \half \delta_\scB g_{\rho\sigma}
    + \mathcal{O}(\dow^2)~, \nn\\
  \lb \sigma_\perp \bbB^{\mu\nu}
  + \tilde\sigma_\perp \bbE^{\mu\nu}
  \rb \delta_\scB A_\nu
  &= \bbE^{\mu\sigma} \hat B^\nu X_{\sigma\nu}
  - 2\lb \textcolor{blue}{\sigma'_\times} \bbB^{\mu(\rho} \hat B^{\sigma)}
  + \textcolor{blue}{\tilde\sigma'_\times} \bbE^{\mu(\rho} \hat B^{\sigma)}
  - \varpi \bbE^{\mu(\rho} u^{\sigma)}
    \rb  \half \delta_\scB g_{\rho\sigma}
    + \mathcal{O}(\dow^2)~~,
\end{align}
where $\bbE^{\mu\nu}$ was defined according to
$\bbE^{\mu\nu} = \epsilon^{\mu\nu\rho\sigma} u_\rho \hat B_\sigma$ along with
\begin{equation}\label{XmnDef}
  X_{\mu\nu} =
  2 \dow_{[\mu} \bfrac{\varpi \hat B_{\nu]}}{T}
  + \frac{u^\lambda}{T} \epsilon_{\lambda\mu\nu\rho} J^\rho_\ext~~.
\end{equation}
Recalling that $T\delta_\scB A_\mu = T\dow_\mu(\mu/T) - E_\mu$, these equations
can be used to algebraically determine $\mu$ and $E_\mu$ in
MHD. Below, it is shown precisely how this can be accomplished.

Introducing $J^\mu_{(1)}$, i.e. one-derivative corrections appearing in the charge current \eqref{fullJ-MHD},
into the first order equations of motion \bref{MHD-firstOrderEOM} and eliminating $P^{\mu\nu}\delta_\scB A_\nu$ using
\cref{Maxwells-firstOrder-MHD}, it is possible to derive the onshell relation
\begin{equation}\label{theFancyEOM-MHD}
  P^{\mu(\rho} u^{\sigma)} \delta_\scB g_{\rho\sigma}
  = - \lb \frac{|B|}{\epsilon + P + \varpi |B|} \rb
  \bbB^{\mu{[\rho}} \hat B^{\sigma]} X_{\rho\sigma}
  + \mathcal{O}(\dow^2)~~,
\end{equation}
which will be useful in solving for $\mu$ and $E^\mu$. For the remainder of this subsection, it is assumed that
$u_\mu J^\mu_\ext = \mathcal{O}(\dow)$ for simplicity, leading to all the
components of the background currents to be derivative suppressed. Under this
assumption, \cref{Maxwells-firstOrder-MHD} can be solved for $\mu$ and $E_\mu$ within the derivative expansion leading to
\begin{align}\label{TheFinal_MuE_elimination}
  \mu
  &= \mu_0(T,B^2)
    + \frac{1}{\dow q/\dow \mu}
    \lB u_\mu J^\mu_\ext
    - B^\lambda \dow_\lambda \frac{B^2}{T^4} \frac{\dow M_1}{\dow \mu}
    - \epsilon^{\lambda\nu\rho\sigma} 
    u_\lambda B_\nu \dow_\rho B_\sigma \frac{\dow M_2}{\dow \mu}
    + \frac{1}{T} B^\lambda \dow_\lambda T \frac{\dow M_3}{\dow \mu} \right. \nn\\
  &\qquad
    \left. - \lb \frac{\dow M_4}{\dow \mu} + \frac{\varpi}{|B|} \rb
    \epsilon^{\lambda\nu\rho\sigma} B_\lambda u_\nu \dow_\rho u_\sigma
    - T B^\lambda \dow_\lambda \frac{\mu}{T} \frac{\dow M_5}{\dow \mu}
    + \frac1T \nabla_\lambda \lb T M_5 B^\lambda \rb
    \rB_{\mu = \mu_0}
    + \mathcal{O}(\dow^2)~~,  \nn\\
  %
  E^\mu
  &=
    T P^{\mu\nu} \dow_\nu \frac{\mu}{T}
    - \frac{T}{2\sigma_\parallel} \hat B^\mu \bbE^{\rho\sigma} X_{\rho\sigma}
    + \frac{T}{\sigma_\parallel} \hat B^\mu \lb \textcolor{blue}{\tilde\chi'_1} \bbB^{\rho\sigma}
    + \textcolor{blue}{\tilde\chi'_2} \hat B^\rho \hat B^\sigma
  \rb \half \delta_\scB g_{\rho\sigma} \nn\\
  &\qquad - T
    \lb \frac{\epsilon+P}{\epsilon + P + \varpi |B|} \rb
    \lb \frac{\sigma_\perp}{\sigma^2_\perp + \tilde\sigma_\perp^2 }
    \bbE^{\mu\rho} \hat B^\sigma
    + \frac{\tilde\sigma_\perp}{\sigma^2_\perp + \tilde\sigma_\perp^2 }
    \bbB^{\mu\rho} \hat B^\sigma \rb X_{\rho\sigma} \nn\\
  &\qquad
    - 2T \lb
    \frac{\tilde\sigma_\perp \textcolor{blue}{\sigma'_\times} - \sigma_\perp \textcolor{blue}{\tilde\sigma'_\times}}
    {\sigma^2_\perp + \tilde\sigma_\perp^2 }
    \bbE^{\mu(\rho} \hat B^{\sigma)} 
    - \frac{\tilde\sigma_\perp \textcolor{blue}{\tilde\sigma'_\times} + \sigma_\perp\textcolor{blue}{\sigma'_\times}}
    {\sigma^2_\perp + \tilde\sigma_\perp^2 }
    \bbB^{\mu(\rho} \hat B^{\sigma)}
    \rb  \half \delta_\scB g_{\rho\sigma}
    + \mathcal{O}(\dow^2)~~,
\end{align}
where \cref{theFancyEOM-MHD} was used to derive the second equation above and
 $\mu_0(T,B^2)$ was defined as the root of the equation
\begin{equation}\label{define-mu0}
  q(T,\mu_0(T,B^2),B^2)
  = \frac{\dow P(T,\mu,B^2)}{\dow\mu} \bigg|_{\mu = \mu_0(T,B^2)}
  = 0~~.
\end{equation}
Therefore, within the MHD derivative expansion, Maxwell's equations can be used
to explicitly eliminate the chemical potential and the electric fields from the
hydrodynamic description. As it will be shown in sec.~\ref{sec:MHD-String}, this
elimination is the backbone for recasting MHD as the string fluid of
sec.~\ref{sec:stringfluids}.

\subsubsection{Kubo formulae and Onsager's relations}

Analogously to sec.~\ref{sec:Kubostring}, Kubo formulae can be obtained by
perturbing around an initial equilibrium configuration. In the context of MHD,
the relevant operators are $O_a=\{T^{\mu\nu}, F_{\mu\nu}\}$, whose one point
functions are defined as
\begin{equation} \label{eq:one-pointMHD}
  \mathbb{T}^{\mu\nu}
  = \sqrt{-g}~\langle T^{\mu\nu}\rangle~~,~~
  \mathbb{F}_{\mu\nu}
  = \sqrt{-g}~\langle F_{\mu\nu}\rangle~~.
\end{equation}
In order to obtain Kubo formulas in MHD, perturbations of the background metric $g_{\mu\nu}$ and the external currents $J^{\mu}_{\text{ext}}$ are performed. Thus, solving for the electric field as in \eqref{TheFinal_MuE_elimination} is required, at least at the linearised level. The retarded Green's functions, for small time-dependent and spatially homogeneous perturbations $\delta h_{\lambda\rho}$ and $\delta J_{ext}^{\mu}$ are defined as in \cite{Hernandez:2017mch}
\begin{equation}
\delta \mathbb{T}^{\mu\nu}=\frac{1}{2}G_{TT}^{\mu\nu,\lambda\rho}(\omega)\delta h_{\lambda\rho}-i \omega {{G_{TF}}^{\mu\nu}}_{0\lambda}\delta J^{\lambda}_{\text{ext}}~~,~~\delta \mathbb{F}_{\mu\nu}=\frac{1}{2}{{G_{FT}}_{\mu\nu}}^{\lambda\rho}\delta h_{\lambda\rho}-i\omega {G_{FF}}_{\mu\nu,0\lambda}\delta J^{\lambda}_{\text{ext}}~~.
\end{equation} 
Considering an equilibrium configuration with $u^\mu=\delta^{\mu}_t$,
$\mu=\mu_0=0$, and magnetic field aligned in the z-direction with magnitude
$B^z=B_0$, it is straightforward to derive the Kubo formulae
\begin{equation} \label{eq:KuboMHD}
\begin{split}
&\frac{\textcolor{blue}{\tilde\chi_1'}}{\sigma_{||}}\text{sign}(B_0)=\lim_{\omega\to0}\frac1\omega ~\mathrm{Im}~ G_{FT}^{tz,xx}~~,~~\frac{\textcolor{blue}{\tilde\chi_2'}}{\sigma_{||}}\text{sign}(B_0)=\lim_{\omega\to0}\frac1\omega ~\mathrm{Im}~ G_{FT}^{tz,zz}~~, \\
&-\left(\frac{\textcolor{blue}{\sigma_\times'}\sigma_\perp}{\sigma_\perp^2+\tilde\sigma^2}+\frac{\textcolor{blue}{\tilde\sigma_\times'}\tilde\sigma}{\sigma_\perp^2+\tilde\sigma^2}\right)\text{sign}(B_0)=\lim_{\omega\to0}\frac1\omega ~\mathrm{Im}~ G_{FT}^{xt,xz}~~,\\
&-\left(\frac{\textcolor{blue}{\tilde\sigma_\times'}\sigma_\perp}{\sigma_\perp^2+\tilde\sigma^2}-\frac{\textcolor{blue}{\sigma_\times'}\tilde\sigma}{\sigma_\perp^2+\tilde\sigma^2}\right)=\lim_{\omega\to0}\frac1\omega ~\mathrm{Im}~ G_{FT}^{xt,yz}~~,\\
&-\frac{\textcolor{blue}{\tilde\chi_1}}{\sigma_{||}}\text{sign}(B_0)=\lim_{\omega\to0}\frac1\omega ~\mathrm{Im}~ G_{TF}^{xx,tz}~~,~~-\frac{\textcolor{blue}{\tilde\chi_2}}{\sigma_{||}}\text{sign}(B_0)=\lim_{\omega\to0}\frac1\omega ~\mathrm{Im}~ G_{TF}^{zz,tz}~~, \\
&\left(\frac{\textcolor{blue}{\sigma_\times}\sigma_\perp}{\sigma_\perp^2+\tilde\sigma^2}+\frac{\textcolor{blue}{\tilde \sigma_\times}\tilde\sigma}{\sigma_\perp^2+\tilde\sigma^2}\right)\text{sign}(B_0)=\lim_{\omega\to0}\frac1\omega ~\mathrm{Im}~ G_{TF}^{xz,xt}~~,\\
&\left(\frac{\textcolor{blue}{\tilde \sigma_\times}\sigma_\perp}{\sigma_\perp^2+\tilde\sigma^2}-\frac{\textcolor{blue}{\sigma_\times}\tilde\sigma}{\sigma_\perp^2+\tilde\sigma^2}\right)=\lim_{\omega\to0}\frac1\omega ~\mathrm{Im}~ G_{TF}^{yz,xt}~~,\\
&\zeta_{22}-\frac{\textcolor{blue}{\tilde\chi_2\tilde\chi_2'}}{\sigma_{||}}=\lim_{\omega\to0}\frac1\omega ~\mathrm{Im}~ G_{TT}^{zz,zz}~~,~~\zeta_{12}'-\frac{\textcolor{blue}{\tilde\chi_2\tilde\chi_1'}}{\sigma_{||}}=\lim_{\omega\to0}\frac1\omega ~\mathrm{Im}~ G_{TT}^{zz,ii}~~, \\
&\zeta_{12}-\frac{\textcolor{blue}{\tilde\chi_1\tilde\chi_2'}}{\sigma_{||}}=\lim_{\omega\to0}\frac1\omega ~\mathrm{Im}~ G_{TT}^{ii,zz} ~~,~~\zeta_{11}-\frac{\textcolor{blue}{\tilde\chi_1\tilde\chi_1'}}{\sigma_{||}}=\lim_{\omega\to0}\frac1\omega ~\mathrm{Im}~ G_{TT}^{ii,ii}~~, \\
&\eta_{11}-\frac{\sigma_\perp\left(\textcolor{blue}{\sigma_\times}\textcolor{blue}{\sigma_\times'}-\textcolor{blue}{\tilde\sigma_\times}\textcolor{blue}{\tilde\sigma_\times'}\right)+\tilde\sigma\left(\textcolor{blue}{\sigma_\times}\textcolor{blue}{\tilde\sigma_\times'}+\textcolor{blue}{\tilde\sigma_\times}\textcolor{blue}{\sigma_\times'}\right)}{\sigma_\perp^2+\tilde\sigma^2}=\lim_{\omega\to0}\frac1\omega ~\mathrm{Im}~ G_{TT}^{zi,zi}~~, \\
&\tilde\eta_{11}-\frac{\sigma_\perp\left(\textcolor{blue}{\tilde\sigma_\times}\textcolor{blue}{\sigma_\times'}+\textcolor{blue}{\sigma_\times}\textcolor{blue}{\tilde\sigma_\times'}\right)+\tilde\sigma\left(\textcolor{blue}{\tilde\sigma_\times}\textcolor{blue}{\tilde\sigma_\times'}-\textcolor{blue}{\sigma_\times\sigma_\times'}\right)}{\sigma_\perp^2+\tilde\sigma^2}=\lim_{\omega\to0}\frac1\omega ~\mathrm{Im}~ G_{TT}^{zi,zj}~(i\ne j)~~, \\
\end{split}
\end{equation}
while the remaining $G_{FF}$ correlators were given in
\cite{Hernandez:2017mch}. In evaluating the above, the contributions arising
from the hydrostatic coefficients $M_i$ were ignored and the assumption
$\epsilon+P\gg\varpi |B|$ was made for the sake of simplicity.

For the case at hand, if the microscopic theory has a discrete symmetry
$\Theta$, the Onsager's relations require that
\begin{equation}
  G_{O_a O_b}(\omega,B_0)
  = i_a i_b G_{O_b O_a}(\omega,\Theta B_0)~~,
\end{equation}
where $i_a$ is the eigenvalue of $O_a$ under $\Theta$. See details in
\cref{app:CPT}. If $\Theta$ is simply time-reversal, we find the constraints
from Onsager's relations to be
\begin{equation} \label{eq:OnsagerMHD}
  \zeta_{12}=\zeta_{12}'~~,~~
  \textcolor{blue}{\sigma_\times}=\textcolor{blue}{\sigma_\times'}~~,~~
  \textcolor{blue}{\tilde \sigma_\times}=\textcolor{blue}{\tilde \sigma_\times'}~~,~~
  \textcolor{blue}{\tilde\chi_1}=\textcolor{blue}{\tilde\chi_1'}~~,~~
  \textcolor{blue}{\tilde\chi_2}=\textcolor{blue}{\tilde\chi_2'}~~,
\end{equation}
which in turn means that parity-violating MHD is characterised by 4 hydrostatic
transport coefficients and 14 non-hydrostatic transport coefficients. This is
the exact same number as for string fluids in sec.~\ref{sec:stringfluids}. In
the next section, it will be shown how \eqref{eq:KuboMHD} can be used to map to
transport coefficients in string fluids.

\subsection{Magnetohydrodynamics as string fluids}
\label{sec:MHD-String}

\subsubsection{The algorithm of mapping}

We now show that magnetohydrodynamics, as formulated above, can be equivalently
formulated as a string fluid discussed in \cref{sec:stringfluids}, when the
external current $J^\mu_\ext$ is derivative suppressed. To begin with, note that
after using Maxwell's equations to eliminate $\mu$ and $E_\mu$ in
\eqref{TheFinal_MuE_elimination}, the MHD constitutive relations can be
schematically represented as
\begin{gather}\label{consti-MHD-map}
  T^{\mu\nu}[u^\mu,T,B^\mu,g_{\mu\nu},J^\mu_\ext]~~,~~
  F_{\mu\nu}[u^\mu,T,B^\mu,g_{\mu\nu},J^\mu_\ext]
  = 2 u_{[\mu} E_{\nu]}[u^\mu,T,B^\mu,g_{\mu\nu},J^\mu_\ext]
  - \epsilon_{\mu\nu\rho\sigma} u^\rho B^\sigma~~, \nn\\
  \mu[u^\mu,T,B^\mu,g_{\mu\nu},J^\mu_\ext]~~.
\end{gather}
They satisfy the adiabaticity equation \cref{adiabaticity-MHD} with $J^\mu$
replaced with $-J^\mu_\ext$.  The dynamical evolution of $u^\mu$ and $T$ is
governed by the energy-momentum conservation \bref{EMConservation-MHD}, while
the evolution of $B^\mu$ is governed by the Bianchi identity
\bref{Bianchi-MHD}. Note that the constitutive relations for $\mu$ do not enter
the dynamical equations and hence are not relevant for the hydrodynamic
description.

In order to establish a connection between MHD and string fluids, it is appropriate to follow the insight of \cite{Schubring:2014iwa} and note that MHD
admits the following two-form current
\begin{equation}\label{2formCurrentDef-MHD}
  J^{\mu\nu} = \half \epsilon^{\mu\nu\rho\sigma} F_{\rho\sigma}~~,
\end{equation}
which is conserved due to the Bianchi identity \bref{Bianchi-MHD}. Physically,
the integration of this current over any codimension-2 surface counts the number
of magnetic fields lines crossing a given area element. The absence of magnetic
monopoles in Maxwell's electromagnetism implies that these magnetic field lines
are conserved. Furthermore, since the external current $J^\mu_\ext$ satisfies
$\nabla_\mu J^\mu_\ext = 0$, it can be locally re-expressed as
\begin{equation}\label{eq:mapping}
  J^\mu_\ext = \frac16 \epsilon^{\mu\nu\rho\sigma} H_{\nu\rho\sigma}~~,~~
  H_{\nu\rho\sigma} = 3 \dow_{[\nu} b_{\rho\sigma]}~~.
\end{equation}
In this language, the background charge current $J^\mu_\ext$ is traded for a
two-form background gauge field $b_{\mu\nu}$, which admits a one-form gauge
transformation $b_{\mu\nu} \to b_{\mu\nu} + 2 \dow_{[\mu} \Lambda_{\nu]}$. In
this section, we assume that $J^\mu_\ext = \mathcal{O}(\dow)$, leading to
$b_{\mu\nu} = \mathcal{O}(1)$, which is sufficient for most applications of
MHD.\footnote{This assumption does not allow us to describe MHD with
  non-vanishing charge density $q$. However, in most applications of MHD, like
  in solar physics, the plasma is assumed to be electrically neutral at the
  hydrodynamical length scales~\cite{davidson2001introduction}.} Armed
with the mappings \eqref{2formCurrentDef-MHD} and \eqref{eq:mapping}, it can be
verified that the MHD dynamical equations \bref{Bianchi-MHD} and
\bref{EMConservation-MHD} arrange themselves into
\begin{equation}\label{MHD-EOM-dual}
  \nabla_\mu T^{\mu\nu} = \half H^{\nu\rho\sigma} J_{\rho\sigma}~~,~~
  \nabla_\mu J^{\mu\nu} = 0~~,
\end{equation}
while the constitutive relations \bref{consti-MHD-map} can now be represented as
\begin{equation}\label{MHD-consti-dual}
  T^{\mu\nu}[u^\mu,T,B^\mu,g_{\mu\nu},b_{\mu\nu}]~~,~~ 
  J^{\mu\nu}[u^\mu,T,B^\mu,g_{\mu\nu},b_{\mu\nu}]
  = 2 u^{[\mu} B^{\nu]}
  + \epsilon^{\mu\nu\rho\sigma}
  u_\rho E_\sigma[u^\mu,T,B^\mu,g_{\mu\nu},b_{\mu\nu}]~~.
\end{equation}
In \cref{MHD-consti-dual}, the constitutive relations for $\mu$ have been
ignored since, as stressed earlier, they do not contribute to the dynamical
equations. \Cref{MHD-EOM-dual,MHD-consti-dual} are precisely those encountered
in the context of string fluids in \cref{sec:stringfluids}. \Cref{MHD-EOM-dual}
constitute the dynamical equations of one-form hydrodynamics given in
\cref{eq:1formcon}, while \cref{MHD-consti-dual} are the respective constitutive
relations upon identifying $B^\mu = \rho(T,\varpi) h^\mu + \mathcal{O}(\dow)$.

In order to establish an exact equivalence between the two formulations, it is
necessary to ensure that the constraints that follow from the adiabaticity
equation, or the second law of thermodynamics, are the same in both the
formulations. Consider the following map between the free energy currents
\begin{equation}\label{freeEnergy-mapping}
  N^\mu_{\text{string}}
  = N^\mu_{\text{MHD}}
  + \frac{\varpi}{2T} \epsilon^{\mu\nu\rho\sigma} h_\nu F_{\rho\sigma}
  + \frac{\mu}{T} J^\mu_\ext
  + \nabla_\nu \lb \frac{\mu\varpi}{T} \epsilon^{\mu\nu\rho\sigma} u_\rho h_\sigma \rb~~,
\end{equation}
where $N^\mu_{\text{string}}$ denotes the free energy current in string fluids and $N^\mu_{\text{MHD}}$ the free energy current in MHD.
The last term in \cref{freeEnergy-mapping} has a trivially vanishing divergence and has only been included for convenience.  It is easily checked that
\begin{align}
  \nabla_\mu N^\mu_{\text{string}}
  &= \nabla_\mu N^\mu_{\text{MHD}}
    + J^{\mu\nu} \dow_\mu \bfrac{\varpi h_\nu}{T} 
    + J^\mu_\ext \dow_\mu \frac{\mu}{T}
    + \Delta \nn\\
  &= \nabla_\mu N^\mu_{\text{MHD}}
    + \half J^{\mu\nu} \delta_\scB b_{\mu\nu}
    + J^\mu_\ext \delta_\scB A_\mu
    + \Delta
    \nn\\
  &= \half T^{\mu\nu} \delta_\scB g_{\mu\nu}
    + \half J^{\mu\nu} \delta_\scB b_{\mu\nu}
    + \Delta~~,
\end{align}
and thus we recover the string fluid adiabaticity equation
\bref{adiabaticity_simple}. This establishes that MHD with $J^\mu_\ext = \mathcal{O}(\dow)$
is entirely equivalent to one-form string fluids.

\subsubsection{Mapping of transport coefficients up to first order}
\label{sec:MHD-String-mapping}

The above discussion established the map between MHD and string fluids in quite
abstract terms. However, the explicit mapping between the transport coefficients
at first order in derivatives is highly non-trivial. This is the purpose of this
section. To begin with, it is necessary to derive the exact map between the
magnetic field $B^\mu$ in MHD and the string fluid fields $h^\mu$ and $\varpi$
at first order in derivatives.  As we have already shown below
\cref{MHD-consti-dual}, at ideal order this is just
$B^\mu = \rho(T,\varpi) h^\mu + \mathcal{O}(\dow)$. The first order derivative
corrections to string fluids in sec.~\ref{sec:stringfluidcorrections} together
with \eqref{2formCurrentDef-MHD} and the definition of magnetic fields in
\eqref{eq:stEM} allow to determine
\begin{align}\label{the-B-map}
  B^{\mu}
  &= \rho h^\mu
    - h^{\mu} \lB \frac16 \epsilon^{\alpha\beta\rho\sigma} u_\alpha H_{\beta\rho\sigma}
    \frac{\dow \alpha}{\dow \varpi}
    - \alpha\epsilon^{\lambda\nu\rho\sigma} u_\lambda h_\nu \dow_\rho u_\sigma
    + \epsilon^{\alpha\beta\rho\sigma} u_\alpha h_\beta \dow_\rho u_\sigma
    \frac{\dow\beta}{\dow \varpi} \right.\nn\\
  &\qquad\qquad
    \left.
    + h^\lambda \dow_\lambda T \frac{\dow\tilde\beta_1}{\dow \varpi}
    + h^\lambda \dow_\lambda \frac{\varpi}{T} \frac{\dow\tilde\beta_2}{\dow
    \varpi}
    - \nabla_\lambda \lb \tilde\beta_2 h^\lambda \rb \frac{1}{T}
    + \epsilon^{\alpha\beta\rho\sigma} u_\alpha h_\beta \dow_\rho h_\sigma
    \frac{\dow\tilde\beta_3}{\dow \varpi}
    \rB \nn\\
  &\qquad
    - \alpha \Delta^\mu{}_\nu \epsilon^{\nu\lambda\rho\sigma} u_\lambda \dow_\rho u_\sigma
    + \frac{\beta}{\varpi} \Delta^{\mu}{}_\nu \epsilon^{\nu\lambda\rho\sigma}
    u_\lambda \dow_\rho u_\sigma 
    - \frac{\tilde\beta_1}{\varpi} \Delta^{\mu\nu} \dow_{\nu} T 
    - \frac{\tilde\beta_2}{\varpi} \Delta^{\mu\nu} \dow_\nu \frac{\varpi}{T}
    \nn\\
  &\qquad
    + \Delta^{\mu}{}_\nu \epsilon^{\nu\lambda\rho\sigma}
      \lb \frac{T\tilde\beta_3}{\varpi^2}
      u_\lambda \dow_\rho \frac{\varpi h_\sigma}{T}
      + \frac1T\nabla_\rho \lb \frac{T\tilde\beta_3}{\varpi}
    u_\lambda h_\sigma \rb \rb
    + \mathcal{O}(\dow^2)~~.
\end{align}
Due to our choice of frame in the non-hydrostatic sector of string fluids, note
that the first order corrections to $B^\mu$ arise only due to hydrostatic
corrections. It is useful to note that $X_{\mu\nu}$ defined in \cref{XmnDef}
maps to
\begin{equation}\label{XmnMap}
  X_{\mu\nu} = \delta_\scB b_{\mu\nu} + \mathcal{O}(\dow^2)~~.
\end{equation}
Substituting $B^\mu$ in the constitutive relations
\eqref{MHD-consti-dual} allows us to determine the mapping between transport
coefficients.  Consider first the hydrostatic sector of the two formulations. It
is useful to re-express \cref{freeEnergy-mapping} as
\begin{align}\label{freeN-map}
  N^\mu_{\text{string}}
  &= N^\mu_{\text{MHD}}
    - \beta^\mu \lb \frac16 \mu
    \epsilon^{\lambda\nu\rho\sigma} u_\lambda H_{\nu\rho\sigma}
    + \half\epsilon^{\lambda\nu\rho\sigma}
    \varpi u_\lambda h_\nu F_{\rho\sigma} 
    + \mu \varpi \epsilon^{\lambda\nu\rho\sigma} u_\lambda 
     h_\nu \dow_\rho u_\sigma \rb
     \nn\\
  &\qquad
    + \epsilon^{\mu\nu\rho\sigma} u_\nu \lb
    \varpi 
      h_\rho \delta_\scB A_\sigma
    - \frac12 \mu 
    \delta_\scB b_{\rho\sigma}
    + \mu \varpi 
    h_\rho u^\lambda \delta_\scB g_{\lambda\sigma} \rb~~.
\end{align}
Given that all the transverse components are purely non-hydrostatic, it is
straightforward to derive the mapping for the hydrostatic free-energy density
between the two formulations
\begin{equation}\label{freeE-map}
  \mathcal{N}_{\text{string}}
  = \mathcal{N}_{\text{MHD}}
  - \frac16 \mu
  \epsilon^{\lambda\nu\rho\sigma} u_\lambda H_{\nu\rho\sigma}
  -  \varpi J^{\mu\nu} u_\mu h_\nu
  - \mu \varpi \epsilon^{\lambda\nu\rho\sigma} u_\lambda 
  h_\nu \dow_\rho u_\sigma~~,
\end{equation}
where $\mathcal{N}_{\text{string}}$ is given in \cref{stringN} and
$\mathcal{N}_{\text{MHD}}$ in \bref{N-MHD}.  Introducing \cref{the-B-map} into
$\mathcal{N}_{\text{MHD}}$ in \eqref{freeE-map}, it is possible to infer at
ideal order that
\begin{equation}
  p(T,\varpi)
  = P(T,\mu_0(T,\rho^2),\rho^2)
  - 2\rho^2
  \frac{\dow P(T,\mu_0(T,\rho^2),\rho^2)}{\dow \rho^2}~~,
\end{equation}
where, on the right hand side, we understand that $\rho = \rho(T,\varpi)$. Given that $\rho(T,\varpi) =
\dow p(T,\varpi)/\dow \varpi$ we can find that\footnote{The partial derivatives
  of $\varpi(T,\rho^2)$ and $\rho(T,\varpi)$ are related by
\begin{equation}
  \frac{\dow \varpi(T,\rho^2)}{\dow T}
  = - \frac{\dow\rho(T,\varpi)/\dow T}{\dow\rho(T,\varpi)/\dow\varpi}~~,~~
  \frac{\dow \varpi(T,\rho^2)}{\dow \rho^2}
  = \frac{1}{2\rho} \frac{1}{\dow\rho(T,\varpi)/\dow\varpi}~~.
\end{equation}}
\begin{equation}
  \rho
  = \frac{\dow}{\dow \rho^2} \lB
  P(T,\mu_0(T,\rho^2),\rho^2)
  - 2\rho^2
  \frac{\dow P(T,\mu_0(T,\rho^2),\rho^2)}{\dow \rho^2}
  \rB \frac{1}{\dow\varpi(T,\rho^2)/\dow\rho^2}~~,
\end{equation}
which can be solved by
\begin{equation}
  \varpi(T,\rho^2)
  = - 2\rho \frac{\dow P(T,\mu_0(T,\rho^2),\rho^2)}{\dow \rho^2}~~,
\end{equation}
yielding the functional definition of $\varpi$ in terms of the MHD thermodynamic
potentials.  Extending the free-energy density mapping \eqref{freeE-map} to one
derivative order leads to the determination of the map between hydrostatic
transport coefficients
\begin{gather}
  \alpha = \mu_0~~,~~
  \beta = M_{4} \rho + \mu_0 \varpi~~, \nn\\
  \tilde{\beta}_1
  =
  \frac{M_{3}\rho}{T}
  - 2\rho^2 \lb \frac{M_{1}}{T^4} 
  + M_{5} \frac{\dow\mu_0}{\dow \rho^2} \rb
  \lb \frac{\dow \rho}{\dow T} + \frac{\varpi}{T} \frac{\dow \rho}{\dow
    \varpi}\rb
  + \frac{4 M_{1}}{T^5} \rho^3
  - T M_{5} \rho \frac{\dow(\mu_0/T)}{\dow T} ~~,\nn\\
  \tilde{\beta}_2
  = - 2\rho^2 T \lb \frac{M_{1}}{T^4} 
  + M_{5}\frac{\dow\mu_0}{\dow \rho^2} \rb
  \frac{\dow \rho}{\dow \varpi}, \qquad
  \tilde{\beta}_3 = - M_{2} \rho^2~~,
  \label{eq:maphydro}
\end{gather}
where all the functions on the right hand side are evaluated at $\mu =
\mu_0$. Interestingly, the string fluid transport coefficient $\alpha$ maps to
the ideal order chemical potential solution $\mu_0$. This implies that if string
fluids are to describe MHD configurations at non-zero chemical potential, the
$\alpha$ term in \eqref{stringN} is required. This observation was lacking in
all previous studies \cite{Schubring:2014iwa, Grozdanov:2016tdf,
  Hernandez:2017mch}.  Note also that the 5 MHD transport coefficients
$M_{1,2,3,4,5}$ map to just 4 string fluid transport coefficients $\beta$ and
$\tilde\beta_{1,2,3}$. The reason is that, when working in a regime where
$J^\mu_\ext = \mathcal{O}(\dow)$, substituting
$\mu = \mu_0(T,B^2) + \mathcal{O}(\dow)$ in \cref{N-MHD} makes $M_5$ linearly
dependent on the other terms.



On the other hand, the mapping in the non-hydrostatic sector is slightly more
involved.  When deriving the mapping \eqref{eq:maphydro} in the hydrostatic
sector, it was inherently assumed that the fluid variables $T$ and $u^\mu$ are
the same in both the formulations. In the hydrostatic sector, this assumption is
well founded, as these hydrodynamical fields are fixed to the requirement of
$u^\mu/T$ aligning along the timelike isometry of the background defining the
equilibrium state. However, in full generality, the fields $T$ and $u^\mu$ can
admit a relative non-hydrostatic redefinition between the two
formulations. Since $T^{\mu\nu}_{\text{nhs}}$ in both the formulations is chosen
such that $T^{\mu\nu}_{\text{nhs}} u_\nu = 0$ (i.e. the constitutive relations
were expressed in the Landau frame), to find this relative redefinition it
suffices to compare
\begin{equation}
  T^{\mu\nu}_{\text{MHD,hs}}
  [u^\mu \to u^\mu + \delta u^\mu, T \to T + \delta T] u_\nu
  = T^{\mu\nu}_{\text{string,hs}} u_\nu
  + \mathcal{O}(\dow^2)~~.
\end{equation}
After a straight-forward, yet involved, computation it can be inferred that the relative change in the fluid velocity $\delta u^{\mu}$ reads
\begin{align} \label{eq:deltau}
  \delta u^{\mu}
  &= - \frac{\alpha}{s} h^\mu \half \epsilon^{\rho\sigma} \delta_\scB b_{\rho\sigma}
    + \frac{\sigma_\perp}{\sigma^2_\perp + \tilde\sigma_\perp^2 }
    \frac{\varpi T^2s}{(\epsilon + p)^2} \Delta^{\mu\rho} h^\sigma
    \delta_\scB b_{\rho\sigma}
    \nn\\
  &\qquad
    - \frac{T\varpi}{(\epsilon + p)} \lb \frac{\tilde\sigma_\perp }
    {\sigma^2_\perp + \tilde\sigma_\perp^2 }
    \lb \frac{Ts}{\epsilon + p} \rb
    + \frac\alpha\varpi \lb 1 - \frac{2\varpi \rho}{\epsilon + p} \rb
    \rb \epsilon^{\mu\rho} h^\sigma \delta_\scB b_{\rho\sigma} \nn\\
  &\qquad
    + \frac{2T\varpi}{\epsilon + p}
     \lb
    \frac{\tilde\sigma_\perp \sigma'_\times - \sigma_\perp \tilde\sigma'_\times}
    {\sigma^2_\perp + \tilde\sigma_\perp^2 }
    \Delta^{\mu(\rho} h^{\sigma)} 
    + \frac{\tilde\sigma_\perp \sigma'_\times + \sigma_\perp\tilde\sigma'_\times}
    {\sigma^2_\perp + \tilde\sigma_\perp^2 }\epsilon^{\mu(\rho} h^{\sigma)}
    \rb  \half \delta_\scB g_{\rho\sigma}~~,
\end{align}
while the relative change in temperature vanishes, i.e. $\delta T = 0$. In fact, given the informed choice of parametrisation of
the hydrostatic sector in the two formulations, it turns out that
\begin{equation}
  T^{\mu\nu}_{\text{MHD,hs}}
  [u^\mu \to u^\mu + \delta u^\mu, T \to T + \delta T]
  = T^{\mu\nu}_{\text{string,hs}}
  + \mathcal{O}(\dow^2)~~,
\end{equation}
holds exactly without further non-hydrostatic corrections. For the benefit of
inquisitive readers, these details have been relegated to \cref{app:MHD}. The
remaining step consists of comparing $T^{\mu\nu}_{\text{nhs}}$ in the two
formulations, along with $E_\mu$ in MHD to
$-\half \epsilon_{\mu\nu\rho\sigma} u^\nu J^{\rho\sigma}$, taking into account
the potential redefinition in $E_\mu$ induced by \eqref{eq:deltau}. In
particular, it is found that the field redefinition of $u^\mu$ non-trivially
mixes $E_\mu$ and $B^\mu$ leading to a one derivative shift in $E_\mu$ such that
\begin{equation}
  E_\mu \to E_\mu - |B| \bbE_{\mu\nu} \delta u^\nu + \mathcal{O}(\dow^2)~~.
\end{equation}
Consequently, the comparison must be performed according to
\begin{align}
  E_\mu - |B| \bbE_{\mu\nu} \delta u^\nu
  &= -\half \epsilon_{\mu\nu\rho\sigma} u^\nu J^{\rho\sigma}
  + \mathcal{O}(\dow^2)~~, \nn\\
  T^{\mu\nu}_{\text{MHD,nhs}}[E_\mu \to E_\mu - |B| \bbE_{\mu\nu} \delta u^\nu]
  &= T^{\mu\nu}_{\text{string,nhs}}~~,
\end{align}
which leads to a straightforward derivation of the map for non-hydrostatic transport coefficients
\begin{gather}
  \zeta_{\perp} = \zeta_{11} -
  \frac{\textcolor{blue}{\tilde\chi_1\tilde\chi'_1}}{\sigma_\parallel}~~,~~
  \zeta_{\times} = \zeta_{12} -
  \frac{\textcolor{blue}{\tilde\chi_1\tilde\chi'_2}}{\sigma_\parallel}~~,~~
  \zeta'_{\times} = \zeta'_{12} -
  \frac{\textcolor{blue}{\tilde\chi_2\tilde\chi'_1}}{\sigma_\parallel}~~,~~
  \zeta_{\parallel} = \zeta_{22} -
  \frac{\textcolor{blue}{\tilde\chi_2\tilde\chi'_2}}{\sigma_\parallel}~~, \nn\\
  \textcolor{blue}{\tilde\kappa_{1}} = \frac{\textcolor{blue}{\tilde\chi_1}}{\sigma_\parallel}~~,~~
  \textcolor{blue}{\tilde\kappa_{2}} = \frac{\textcolor{blue}{\tilde\chi_2}}{\sigma_\parallel}~~,~~
  \textcolor{blue}{\tilde\kappa'_{1}} = - \frac{\textcolor{blue}{\tilde\chi'_1}}{\sigma_\parallel}~~,~~
  \textcolor{blue}{\tilde\kappa'_{2}} = - \frac{\textcolor{blue}{\tilde\chi'_2}}{\sigma_\parallel}~~,~~
  r_\parallel = \frac{1}{\sigma_\parallel}~~,~~ \nn\\
  \eta_\parallel
  = \eta_{11}
  - \frac{\sigma_\perp(\textcolor{blue}{\sigma_\times\sigma'_\times}
    - \textcolor{blue}{\tilde\sigma_\times \tilde\sigma'_\times})
    + \tilde\sigma_\perp
    (\textcolor{blue}{\sigma_\times \tilde\sigma'_\times}
    + \textcolor{blue}{\tilde\sigma_\times \sigma'_\times})}
  {\sigma_\perp^2 + \tilde\sigma_\perp^2}~~,~~
  r_\perp
  = \lb \frac{sT}{\epsilon + p} \rb^2
    \frac{\sigma_\perp}{\sigma^2_\perp + \tilde\sigma_\perp^2 }~~, \nn\\
  \tilde{\eta}_\parallel
  = \tilde\eta_{11}
  - \frac{\sigma_\perp (\textcolor{blue}{\sigma_\times\tilde\sigma'_\times}
    + \textcolor{blue}{\tilde\sigma_\times \sigma'_\times})
    - \textcolor{blue}{\tilde\sigma_\perp (\sigma_\times \sigma'_\times}
    - \textcolor{blue}{\tilde\sigma_\times\tilde\sigma'_\times})}
  {\sigma_\perp^2 + \tilde\sigma_\perp^2}~~,~~
  \tilde r_\perp
  = \lb \frac{sT}{\epsilon + p} \rb^2
  \lb \frac{- \tilde\sigma_\perp }{\sigma^2_\perp + \tilde\sigma_\perp^2 }
  + \frac{2\alpha \rho}{sT} \rb~~, \nn\\
  \textcolor{blue}{r_\times}
  = \frac{sT}{\epsilon + p}
  \frac{- \sigma_\perp \textcolor{blue}{\tilde\sigma_\times} + \tilde\sigma_\perp \textcolor{blue}{\sigma_\times}}
  {\sigma^2_\perp + \tilde\sigma_\perp^2 }~~,~~
  \textcolor{blue}{r'_\times}
  = \frac{sT}{\epsilon + p}
  \frac{- \sigma_\perp \textcolor{blue}{\tilde\sigma'_\times} + \tilde\sigma_\perp \textcolor{blue}{\sigma'_\times}}
  {\sigma^2_\perp + \tilde\sigma_\perp^2 }~~, \nn\\
  \textcolor{blue}{\tilde r_\times}
  = \frac{sT}{\epsilon + p}
  \frac{\sigma_\perp\textcolor{blue}{\sigma_\times} + \tilde\sigma_\perp \textcolor{blue}{\tilde\sigma_\times}}
  {\sigma^2_\perp + \tilde\sigma_\perp^2 }~~,~~
  \textcolor{blue}{\tilde r'_\times}
  = \frac{sT}{\epsilon + p}
  \frac{\sigma_\perp\textcolor{blue}{\sigma'_\times} + \tilde\sigma_\perp \textcolor{blue}{\tilde\sigma'_\times}}
  {\sigma^2_\perp + \tilde\sigma_\perp^2 }~~, \nn\\
  \eta_\perp = \eta_{22}, \qquad
  \tilde\eta_\perp = \tilde\eta_{22}~~.
  \label{eq:mapnonhydro}
\end{gather}
This map expresses the fact that the that non-hydrostatic transport coefficients
are quite non-trivially related to each other. In addition, the map also
embodies the mapping of Onsager's relations found in \eqref{eq:Onsagerstrings}
and \eqref{eq:OnsagerMHD}. In particular, given that the Onsager relations
\eqref{eq:OnsagerMHD} hold in MHD, the relations \eqref{eq:Onsagerstrings} are
deduced from this map. Additionally, under the assumptions of $\alpha=0$ and
$\epsilon +P\gg \varpi |B|$, direct comparison of the Kubo formulae
\eqref{eq:KuboMHD} in MHD with the Kubo formulae in string fluids
\eqref{eq:Kubostring} by means of \eqref{2formCurrentDef-MHD} leads to a
particular case of the map derived above, as expected. The results in this
section conclude that MHD with $J^\mu_\ext = \mathcal{O}(\dow)$ is completely
equivalent to the hydrodynamic theory of string fluids formulated in
sec.~\ref{sec:stringfluids}.

\subsection{Bound-charge plasma and one-form superfluids}
\label{sec:boundcharge}

In this section, we formulate a new hydrodynamic theory describing bound-charge
plasmas (i.e. plasmas with only bound charges and no free charge carriers) in the
conventional language. We then argue how this theory can be equivalently
formulated in terms of one-form superfluids. Because the full details of
one-derivative corrections in one-form superfluid dynamics are quite involved,
we focus on the ideal sector. However, as an illustration of the robustness of
this formulation, we provide the first-order corrections in the electric limit
of one-form superfluids discussed in \cref{sec:electric}, and show that is maps to a
certain magnetically dominated sector of bound-charge plasmas.

\subsubsection{Ideal bound-charge plasma}

In order to obtain the constitutive relations for a bound-charge plasma, it
should be noted that the adiabaticity equation in \cref{bound-plasma-EOM} is
precisely the same in form as \cref{adiabaticity_simple} with
$\delta_\scB b_{\mu\nu} = \delta_\scB \xi_{\mu\nu}$ replaced with
$\delta_\scB F_{\mu\nu}$ and $J^{\mu\nu}$ replaced with $M^{\mu\nu}$ (defined
below \cref{eq:Mmunu}). Note, however, that this naive identification is only
true at the level of the adiabaticity equation. It does not hold true at the
level of equations of motion because $M^{\mu\nu}$ is not conserved.  A better,
albeit slightly non-trivial, relation to one-form superfluids will be proposed
in the next subsection. Regardless, this naive identification can be used to
write down the constitutive relations of bound charge plasmas. At ideal order,
following \cref{sec:idealFluids}, we find that
\begin{align}\label{ideal_BCP_consti}
  T^{\mu\nu}
  &= \epsilon\, u^{\mu} u^{\nu}
    + \lb P - \alpha_{BB} B^{2} - \alpha_{EB} (E\cdot B) \rb P^{\mu\nu}
    - \alpha_{EE} E^\mu E^\nu
    + \alpha_{BB} \lb B^{\mu} B^{\nu}
    + 2 u^{(\mu} \epsilon^{\nu)\rho\sigma\tau} u_{\rho}B_{\sigma}E_{\tau}
    \rb~~, \nn\\
  M^{\mu\nu}
  &= - 2 u^{[\mu} \lb  \alpha_{EE} E^{\nu]} + \alpha_{EB} B^{\nu]} \rb
    - \epsilon^{\mu\nu\rho\sigma} u_{\rho}
    \lb \alpha_{BB} B_{\sigma} + \alpha_{EB} E_{\sigma} \rb~~, \nn\\
  N^\mu
  &= \frac{P}{T} u^\mu~~, \nn\\
  S^\mu
  &= N^{\mu} - \beta_{\nu} T^{\mu\nu} + \frac1T
  E_{\nu} M^{\mu\nu}
   = s u^\mu~~,
\end{align}
where $P = P(T,E^2,B^2,E\cdot B)$, while the other thermodynamic functions were
defined via
\begin{gather} \label{eq:boundFluidThermo}
  "dd P = s\, "dd T + \half  \alpha_{EE} "dd E^{2}
  + \half \alpha_{BB} "dd B^{2}
  + \alpha_{EB} "dd(E\cdot B)~~, \nn\\
  \epsilon + P
  = s\,T +  \alpha_{EE} E^{2}
  + \alpha_{EB} (E\cdot B) ~~.
\end{gather}
$P$, $\epsilon$, and $s$ are identified as the thermodynamic pressure, energy,
and entropy density of the plasma. On the other hand, the coefficients
$\alpha_{EE}$, $\alpha_{EB}$, and $\alpha_{BB}$ are known as electromagnetic
susceptibilities of the plasma.  These thermodynamic relations and constitutive
relations have been derived earlier in~\cite{Kovtun:2016lfw}, though in a
slightly different way.  In the special case of an ideal fluid minimally coupled
to electromagnetic fields in \cref{decoupled-MHD-consti}, one chooses
$P(T,E^2,B^2,E\cdot B) = (E^2 - B^2)/2 + p(T)$, leading to $\alpha_{EE} = 1$,
$\alpha_{BB} = -1$ and $\alpha_{EB} = 0$.

It is also possible to work out the one-derivative corrections but they can be
trivially read out from \cref{sec:onederivative}. In particular, there are 166
first order transport coefficients, hinting towards the fact that one-form
superfluids and bound-charge plasmas are exactly equivalent theories.

\subsubsection{Reinterpretation as one-form superfluids}

In deriving the one-form superfluid constitutive relations above, we used the
naive similarity between the adiabaticity equations of bound-charge plasmas under
the identification $\xi_{\mu\nu} \to F_{\mu\nu}$ and
$J^{\mu\nu} \to M^{\mu\nu}$. However, as noted earlier, this identification does
not follow through to the dynamics of the system. In order to get the correct
dynamics, we propose the mapping with the respective Hodge duals
\begin{equation}
  J^{\mu\nu} = \half \epsilon^{\mu\nu\rho\sigma} F_{\rho\sigma}~~,~~
  \xi_{\mu\nu} = \half \epsilon_{\mu\nu\rho\sigma} M^{\rho\sigma}~~.
\end{equation}
This is a non-trivial mapping because, in bound-charge plasmas, $F_{\mu\nu}$ is
being treated as a constituent field and the constitutive relations are
expressed in terms of $M^{\mu\nu}$, while in one-form superfluids $\xi_{\mu\nu}$
is treated as a constituent field and the constitutive relations are expressed
in terms of $J^{\mu\nu}$. Nonetheless, it is possible to show that under this
identification, the defining equations of bound-charge plasmas map to those of a
one-form superfluid, provided the following map of background fields
\begin{equation}
  J^\mu_\ext = \frac16 \epsilon^{\mu\nu\rho\sigma} H_{\nu\rho\sigma}~~,
\end{equation}
and the map of the free-energy current
\begin{equation}\label{FreeEMap_1SFBCP}
  N^{\mu}_{\text{1SF}} = N^\mu_{\text{BCP}} - \half \beta^\mu M^{\rho\sigma} F_{\rho\sigma}~~.
\end{equation}
It is worth noting that this is precisely the self-duality operation of one-form
superfluid dynamics discussed in \cref{sec:selfdual}, except that
$H_{\mu\nu\rho}$ is no longer required to vanish. It instead maps to the
background currents in bound-charge plasmas.

Since the algebraic operation of the self-duality is the same as the map
proposed above, it is possible to directly read out the map between fields and
transport coefficients
\begin{equation}
  \zeta_{\mu}
  = \alpha_{BB} B_{\mu} + \alpha_{EB} E_{\mu}~~,~~
  \bar\zeta_{\mu}
  =  \alpha_{EE} E_{\mu} + \alpha_{EB} B_{\mu}~~,
\end{equation}
and
\begin{gather}
  q = - \frac{\alpha_{EE}}{ \alpha_{EE}\alpha_{BB} - \alpha_{EB}^{2}}~~,~~
  \bar q = - \frac{\alpha_{BB}}{ \alpha_{EE}\alpha_{BB} - \alpha_{EB}^{2}}~~,~~
  q_{\times} = \frac{\alpha_{EB}}{ \alpha_{EE}\alpha_{BB} - \alpha_{EB}^{2}}~~, \nn\\
  P_{\text{1SF}} = P_{\text{BCP}}
  -  \alpha_{EE} E^{2} - \alpha_{BB} B^{2}
  - 2 \alpha_{EB} (E\cdot B)~~.
\end{gather}
Note that this map is only well-defined if the determinant of magnetic
susceptibilities is non-zero, that is
\begin{equation}
  \alpha_{EE}\alpha_{BB} - \alpha_{EB}^{2}
  = \frac{1}{q\bar q - q_\times^2} \neq 0~~.
\end{equation}
In particular, as outlined in \cref{stringLimitofSF}, in the string fluid limit
of one-form superfluids, the coefficients $\bar q$ and $q_\times$ are zero,
leading to a violation of this condition. Therefore, they do not map to a
bound-charge plasma, but are instead dual to magnetohydrodynamics as discussed
in \cref{sec:MHD-String}.

\subsubsection{Magnetically dominated bound-charge plasma}
\label{sec:magDomPlasma}

As an interesting case, consider the regime of bound-charge plasmas where the
electric fields are derivative suppressed. The reason for focusing on this case
is because of its qualitative similarity to magnetohydrodynamics. Expanding
\cref{ideal_BCP_consti} to one-derivative order we find that
\begin{align}\label{MagDominatedBCP_ideal}
  T^{\mu\nu}
  &= \epsilon\, u^{\mu} u^{\nu} + P\, P^{\mu\nu}
    + E\cdot B
    \lB \lb T\frac{\dow \alpha_{EB}}{\dow T} \rb u^\mu u^\nu
    - 2 B^2 \frac{\dow\alpha_{EB}}{\dow B^2} \bbB^{\mu\nu} \rB \nn\\
  &\qquad
    + \alpha_{BB} \lb - B^2 \bbB^{\mu\nu}
    + 2 u^{(\mu} \epsilon^{\nu)\rho\sigma\tau} u_{\rho}B_{\sigma}E_{\tau}
    \rb
    + \mathcal{O}(\dow^2) ~~, \nn\\
  M^{\mu\nu}
  &= - \alpha_{BB} \epsilon^{\mu\nu\rho\sigma} u_{\rho} B_{\sigma}
    - 2 E\cdot B\, \frac{\dow \alpha_{EB}}{\dow B^2}
    \epsilon^{\mu\nu\rho\sigma} u_{\rho} B_{\sigma}
    - 2 \alpha_{EB}\, u^{[\mu} B^{\nu]}
    - \alpha_{EB} \epsilon^{\mu\nu\rho\sigma} u_{\rho} E_{\sigma} \nn\\
  &\qquad
    - \textcolor{blue}{ 2 u^{[\mu} \lb  \alpha_{EE} E^{\nu]}
    + \alpha'_{EB} B^{\nu]} E\cdot B \rb}
    + \mathcal{O}(\dow^2) ~~, \nn\\
  N^\mu
  &= \frac{P}{T} u^\mu
    + \frac{\alpha_{EB}}{T} E\cdot B\, u^\mu
    + \mathcal{O}(\dow^2)~~.
\end{align}
All the transport coefficients appearing here are functions of $T$ and $B^2$ and
satisfy the thermodynamics
\begin{equation}
  \df P = s \df T + \half \alpha_{BB} \df B^2~~, ~~
  \epsilon + P = Ts~~.
\end{equation}
In writing these, the ideal superfluid pressure was expanded according to
\begin{equation}
  P(T,E^2,B^2,E\cdot B)
  = P + \alpha_{EB} E\cdot B
  + \half \lb \alpha'_{EB} B_\mu B_\nu + \alpha_{EE} P_{\mu\nu} \rb E^\mu E^\nu
  + \mathcal{O}(\dow^3)~~.
\end{equation}

Note that there are order-mixing terms coupled to $\alpha_{EE}$ and
$\alpha'_{EB}$ in \cref{MagDominatedBCP_ideal}, highlighted in blue, arising from the
second order free-energy density affecting the one-derivative constitutive
relations. It is possible to add more such terms by introducing a term like
$S^\mu E_\mu$ in $\mathcal{N}$ such that $S^\mu$ includes all the possible
one-derivative hydrostatic structures barring $E_\mu$. Generically, such
order-mixing terms only give contributions to the polarisation tensor
\begin{equation}
  M^{\mu\nu}_{\text{hs,order-mixing}}
  = - 2 u^{[\mu}\lb \alpha'_{EB} B^{\nu]} E\cdot B
  + \alpha_{EE} E^{\nu]} + S^{\nu]} \rb~~.
\end{equation}
Including the explicitly one-derivative terms, it is further possible to write
down 4 hydrostatic derivative corrections, namely
\begin{multline}\label{N-mBCP}
  \mathcal{N}
  = P + \alpha_{EB} E\cdot B
  + \half \lb \alpha'_{EB} B_\mu B_\nu + \alpha_{EE} P_{\mu\nu} \rb E^\mu E^\nu
  + R^\mu E_\mu \\
  + M_1 B^\mu \dow_\mu \frac{B^2}{T^4}
  + M_2 \epsilon^{\mu\nu\rho\sigma} u_\mu B_\nu \dow_\rho B_\sigma
  - \frac{M_3}{T} B^\mu \dow_\mu T
  - M_4 \epsilon^{\mu\nu\rho\sigma} u_\mu B_\nu \dow_\rho u_\sigma~~.
\end{multline}
The contributions from the $M_i$ terms to the constitutive relations have been
recorded in \cref{app:hydrostaticMHD}. This completes the hydrostatic sector.

For the non-hydrostatic terms, we express the constitutive relations as
\begin{align}
  T^{\mu\nu}_{\text{nhs}}
  &= \delta\mathcal{E} ~u^{\mu}u^{\nu} + \delta\mathcal{F} \bbB^{\mu\nu}
    + \delta\mathcal{T} ~\hat B^{\mu}\hat B^{\nu}
    + 2\mathcal{L}^{(\mu}\hat B^{\nu)}
    + 2 \mathcal{K}^{(\mu}u^{\nu)}
    + \mathcal{T}^{\mu\nu}~~, \nn\\
  M^{\mu\nu}_{\text{nhs}}
  &= 2 \delta\mathcal{R} ~u^{[\mu} \hat B^{\nu]}
    + 2 \mathcal{M}^{[\mu} \hat B^{\nu]}
    + 2 \mathcal{N}^{[\mu} u^{\nu]} +
    \delta \mathcal{S}\, \bbE^{\mu\nu}~~.
\end{align}
It is possible to use the redefinition freedom in $u^\mu$ and $T$ to set
$\delta\mathcal{E}$ and $\mathcal{K}^\mu$ to zero. The residual terms can be
expanded according to
\begin{gather}
  \begin{pmatrix}
    \delta\mathcal{R} \\
    \delta\mathcal{F} \\
    \delta\mathcal{T} \\
    \delta\mathcal{S}
  \end{pmatrix}
  =
  - \frac{T}{2} \begin{pmatrix}
    \textcolor{blue}{\tau_1} & \tau_2
    & \tau_3 & \textcolor{blue}{\tau_4} \\
    \textcolor{blue}{\tau'_2}
    & \zeta_{11}
    & \zeta_{12}
    & \textcolor{blue}{\tilde\chi_{1}}  \\
    \textcolor{blue}{\tau'_3}
    & \zeta'_{12}
    & \zeta_{22}
    & \textcolor{blue}{\tilde\chi_{2}}  \\
    \textcolor{blue}{\tau'_4}
    & \tilde\chi'_{1}
    & \tilde\chi'_{2}
    & \textcolor{blue}{\sigma_{\parallel}}
  \end{pmatrix}
  \begin{pmatrix}
    2 u^\mu \hat B^\nu \delta_{\scB} F_{\mu\nu} \\
    \bbB^{\mu\nu}"d_{"scB}g_{\mu\nu} \\
    \hat B^{\mu} \hat B^{\nu} "d_{"scB}g_{\mu\nu} \\
    \bbE^{\mu\nu} "d_{"scB} F_{\mu\nu}
  \end{pmatrix}, \nn\\
  \begin{pmatrix}
    \mathcal{N}^\mu \\ \mathcal{L}^{\mu} \\ \mathcal{M}^{\mu}
  \end{pmatrix}
  =
  - T \begin{pmatrix}
    \textcolor{blue}{\tau_5}
    & \tau_6
    & \textcolor{blue}{\tau_7}
    & \textcolor{blue}{\tau_8}
    & \tau_9
    & \textcolor{blue}{\tau_{10}} \\
    \textcolor{blue}{\tau'_6}
    & \eta_{11}
    & \textcolor{blue}{\sigma_{\times}}
    & \textcolor{blue}{\tau'_9}
    & \tilde\eta_{11}
    & \textcolor{blue}{\tilde\sigma_{\times}} \\
    \textcolor{blue}{\tau'_7}
    & \sigma'_{\times}
    & \textcolor{blue}{\sigma_{\perp}}
    & \textcolor{blue}{\tau'_{10}}
    & \tilde\sigma'_{\times}
    & \textcolor{blue}{\tilde\sigma_{\perp}}
  \end{pmatrix}
  \begin{pmatrix}
    \bbB^{\mu\sigma}u^{\nu}"d_{"scB}F_{\sigma\nu} \\
    \bbB^{\mu\sigma} \hat B^{\nu}"d_{"scB}g_{\sigma\nu} \\
    \bbB^{\mu\sigma} \hat B^{\nu}"d_{"scB}F_{\sigma\nu} \\
    \bbE^{\mu\sigma}u^{\nu}"d_{"scB} F_{\sigma\nu} \\
    \bbE^{\mu\sigma}h^{\nu}"d_{"scB}g_{\sigma\nu} \\
    \bbE^{\mu\sigma}h^{\nu}"d_{"scB}b_{\sigma\nu}
  \end{pmatrix}~~, \nn\\
  \mathcal{T}^{\mu\nu} = - \eta_{22} T\bbB^{\rho\langle\mu}\bbB^{\nu\rangle\sigma}
  "d_{"scB} g_{\rho\sigma}
  + \tilde\eta_{22} T\bbE^{\rho\langle\mu}\bbB^{\nu\rangle\sigma}
  "d_{"scB} g_{\rho\sigma}~~.
  \label{mBCP_nhs_corrections}
\end{gather}
The blue terms have been considered here in order to complete the quadratic
form. However, they are actually second order contributions to the constitutive
relations. This mixing of derivative orders in positivity of the quadratic form is
a manifestation of the order mixing considerations explained in
sec.~\ref{sec:electric}.

Having discussed the one-derivative corrections to a bound-charge plasma in
the magnetically dominated limit, we now establish a map between these and 
one-form superfluids.  Identifying
$F_{\mu\nu} = - \half \epsilon_{\mu\nu\rho\sigma} J^{\rho\sigma}$ at ideal order
one can trivially find that
\begin{equation}
  \zeta^{\mu} = \alpha_{BB} B^\mu + \mathcal{O}(\dow)~~,~~
  \bar\zeta^\mu = \alpha_{EB} B^\mu + \mathcal{O}(\dow)~~.
\end{equation}
Thus, on the one-form superfluid side, a linear combination of $\zeta^\mu$ and
$\bar\zeta^\mu$ is derivative suppressed. One such limit was studied in
sec.~\ref{sec:electric}, namely the electric limit.  In order to map to this
limit, it is necessary to set $\alpha_{EB}(T,B^2) = 0$. Having done that, it is
possible to show that the theory is exactly equivalent to the electric limit of
one-form superfluids. Suppressing the details, the following map is found in the
hydrostatic sector
\begin{subequations}\label{mBCP_hs_map}
  \begin{gather}
    p(T,\varpi) = P(T,\rho^2) + \varpi\rho~~, \nn\\
    q_\times = 0~~, \qquad \bar q = -\frac{1}{\alpha_{EE}}~~,~~ q'_\times = -
    \frac{1}{\varpi^2} \lb \frac{1} {\alpha'_{EB} \rho^2 + \alpha_{EE}} -
    \frac{1}{\alpha_{EE}} \rb~~,
  \end{gather}
  together with the order mixing vectors
  \begin{equation}
    R_\mu = \lb \frac{h_\mu h_\nu}{\alpha'_{EB} \rho^2 + \alpha_{EE}}
    + \frac{\Delta_{\mu\nu}}{\alpha_{EE}} \rb S^\nu~~,
  \end{equation}
  and the pure first-order coefficients
  \begin{gather}
    \beta = M_4 \rho~~,~~ \tilde\beta_1 = - \lb \frac{2M_1\rho^2}{T^4}
    \frac{\dow \rho}{\dow T} + \frac{2M_1\rho^2\varpi}{T^5} \frac{\dow
      \rho}{\dow \varpi}
    - \frac{4M_1}{T^5} \rho^3 - \frac{M_3\rho}{T} \rb~~, \nn\\
    \tilde\beta_2 = - \frac{2M_1\rho^2}{T^3} \frac{\dow \rho}{\dow \varpi}~~,~~
    \tilde\beta_3 = - M_2 \rho^2~~.
  \end{gather}
\end{subequations}
Here $\rho = \varpi q = |B|$, $h^\mu = - \zeta^\mu/\varpi = B^\mu/|B|$ and
\begin{equation}
  \varpi = - 2 \rho \frac{P(T,\rho^2)}{\dow\rho^2}~~.
\end{equation}
For the first-order terms in the non-hydrostatic sector, the following trivial
map is obtained for the energy-momentum tensor
\begin{subequations}\label{mBCP_nhs_map}
  \begin{gather}
    \zeta_\perp = \zeta_{11}~~,~~ \zeta_\parallel = \zeta_{22}~~,~~
    \zeta_{\times} = \zeta_{12}~~,~~
    \zeta'_{\times} = \zeta'_{12}~~, \nn\\
    \eta_\parallel = \eta_{11}~~, \qquad \tilde\eta_\parallel =
    \tilde\eta_{11}~~,~~ \eta_\perp = \eta_{22}~~,~~ \tilde\eta_\perp =
    \tilde\eta_{22}~~,
    \label{eq:ttt}
  \end{gather}
  while for the polarisation tensor we have
  \begin{gather}
    \tau_2 = (\alpha_{EE} + \alpha'_{EB}B^2) \tilde\kappa'_1~~,~~ \tau_3 =
    (\alpha_{EE} + \alpha'_{EB}B^2) \tilde\kappa'_2~~,~~ \tau_6 = \alpha_{EE}
    \tilde r'_\times~~,~~
    \tau_9 = - \alpha_{EE} r'_\times~~, \nn\\
    \tilde\chi'_1 = - \frac{\lambda_2}{\delta\rho/\delta\varpi}~~,~~
    \tilde\chi'_2 = - \frac{\lambda_3}{\delta\rho/\delta\varpi}~~,~~
    \sigma'_\times = \frac{\varpi}{\rho} \lambda_9~~, \qquad \tilde\sigma'_\times
    = - \frac{\varpi}{\rho} \lambda_6~~.
  \end{gather}
\end{subequations}
Note that the first order terms in the polarisation tensor appear at second
order in the charge current. Hence, if we were interested in only the
one-derivative corrections to the charge current, as in MHD, these terms can be
ignored. Taking this into account, at first order in derivatives there are a
total of 8 transport coefficients in the non-hydrostatic sector, given in
\eqref{eq:ttt}, out of which the Onsager's relation set
$\zeta_{12}=\zeta_{12}'$.  This exactly matches the number of transport
coefficients found in MHD in \cref{sec:MHDd}, provided that the current of free
charges is removed by setting
$\tilde\kappa_1=\tilde\kappa_1'=\tilde\kappa_2=\tilde\kappa_2'=0$,
$r_{||}=r_\perp=\tilde r_\perp=0$, and
$\tilde r_\times=\tilde r_\times'=r_\times=r_\times'=0$.

\section{Outlook} \label{sec:discussion}

This paper has dealt with the formulation of new hydrodynamic theories with
generalised global symmetries capable of describing different hydrodynamic
regimes of hot electromagnetism. The precise correspondence between these two
classes of theories also required the formulation and extension of hydrodynamic
theories with dynamical gauge fields. This included the extension of MHD to the
parity-violating sector in sec.~\ref{sec:MHDd} and a new effective theory that
describes the hydrodynamic regime of non-conducting plasmas (i.e. plasmas
without free charge carriers) in sec.~\ref{sec:boundcharge}. Though four out of
five hydrodynamic theories that were formulated in this work can be seen as
different limits of one theory, the explicit construction of each of them
actually required a case-by-case analysis.

The connections between hydrodynamics with generalised global symmetries and hot
electromagnetism were made in the sector of the theory where the $\rmU(1)$
one-form symmetry is partially or entirely spontaneously broken. It was proven
that the theory of one-form superfluids in the electric limit in
sec.~\ref{sec:electric}, in which the one-form symmetry is completely broken, is
equivalent to a theory of magnetically dominated non-conducting plasmas with
bound charges in sec.~\ref{sec:boundcharge}. It was also proven that a theory of
one-form superfluids in the string fluid limit as in
sec.~\ref{sec:stringfluids}, in which the $\rmU(1)$ one-form symmetry is only
partially broken along $u^\mu$, is exactly equivalent to MHD with sub-leading
external currents (see sec.~\ref{sec:MHDd}). This equivalence has thus shown
that the $\rmU(1)$ one-form symmetry is spontaneously broken in these two
hydrodynamic regimes of hot plasmas.

These two theories described above were focused on the magnetic dominated phase of hot electromagnetism in which the magnetic fields can be arbitrary and the electric fields are weak. The opposite regime, that of electrohydrodynamics, in which the hydrodynamics of plasmas is electrically dominated, is still unexplored but could have interesting applications. This type of theories will also be described by one-form superfluids of sec.~\ref{sec:super} in a different regime or specific limits of one-form superfluids. In certain cases, the theories describing these regimes will be directly related to the theories developed here due to electromagnetic dualities or variations thereof, as discussed in sec.~\ref{sec:limits}. This suggests that the connections depicted in fig.~\ref{fig:diagram} between one-form (super)fluids and hot electromagnetism admit many other unexplored regimes and intricate relations between them. It would be interesting to understand this broader diagram more precisely by, for instance, classifying all the different hydrodynamic regimes of hot electromagnetism and to investigate whether fluids with generalised global symmetries can actually provide dual formulations for all these different hydrodynamic regimes.

The results of sec.~\ref{sec:stringfluids}, together with the map given in \ref{sec:MHD-String}, provide a formulation of MHD entirely in terms of 
conservation laws, including all possible dissipative effects. This has the potential to aid numerical simulations of MHD, as numerical codes
are better suited for working with conservation equations instead of dynamical Maxwell equations \cite{Gammie:2003rj}.
As such, the work we have presented here has the potential of aiding progress in the astrophysical context, not only by allowing
for numerical studies of dissipative effects in accretion disk physics but also by providing the necessary and sufficient conditions (see sec.~\ref{sec:stringfluids})
for having equilibrium solutions (without dissipation), which serve as starting points in numerical simulations. In particular,
besides providing a time-like Killing vector field, one must solve the no-monopole constraint \eqref{eq:stringvareom} in order to have an equilibrium solution
for a scalar Goldstone $\varphi$ (magnetic scalar potential). 
This has been used in \cite{Armas:2018atq} in order to obtain a new solution of a slowly rotating magnetised star but many other possibilities, such as new
accretion disk solutions, are yet to be explored. We also expect this formulation to be useful in the study of stability properties of accretion disk solutions and in probing
mechanisms for energy transport with analytic control. We intend to pursue some of these directions elsewhere.

Related to the exploration of the scope of hydrodynamics with generalised global
symmetries and its connections with electromagnetism, it was noted throughout
this paper that the traditional treatment of MHD, where the electromagnetic
photon is incorporated as a dynamical field in the hydrodynamic description, has
so far been formulated in greater generality than its counterpart as the string
fluid of sec.~\ref{sec:stringfluids}. The traditional MHD formulation given in
sec.~\ref{sec:MHDd}, extending that of \cite{Hernandez:2017mch}, allows for the
description of hot plasmas that are not electrically neutral at hydrodynamic
length scales, i.e. it is possible to consider a situation in which
$u_\mu J^{\mu}_{\text{ext}}=\mathcal{O}(1)$. It may be the case that the string
fluid formulation of sec.~\ref{sec:stringfluids} can be generalised in order to
incorporate the description of non-electrically neutral plasmas. For instance,
treating some of the components of $H_{\mu\nu\lambda}$ as $\mathcal{O}(1)$
instead of $\mathcal{O}(\dow)$ may provide the required generalisation. However,
at the present moment, it is not clear whether or not such a formulation exists
and whether it would be useful. Nevertheless, we plan on returning to this issue
in the future.

A theory of ordinary one-form fluids has also been developed in
sec.~\ref{sec:ordinary}. This theory, which is rather different from the theory
of string fluids of sec.~\ref{sec:stringfluids}, has unbroken one-form symmetry
and had not been considered previously in the literature. It is suggestive to
speculate that this effective description could describe yet another
hydrodynamic regime of hot plasmas in which the $\rmU(1)$ one-form symmetry is
unbroken. A back of the envelope calculation suggests that one-form fluids in
the unbroken phase do not describe MHD with weak magnetic fields, as could have
been naively expected. It would be interesting to pursue this direction further
and understand whether one-form fluids could find applications in other phases
of matter.

Fluid/gravity dualities have been used to describe earlier versions of string fluids (without the Goldstone mode $\varphi$) both in the context of Anti-de Sitter black branes charged under a two-form gauge field \cite{Grozdanov:2017kyl} and in the context of asymptotically flat supergravity black branes, obtained by a series of duality transformations \cite{Armas:2012ac, Armas:2013aka}. Pursuing this line of research further, it would be extremely interesting to construct gravity duals to both the string fluids of sec.~\ref{sec:stringfluids}, explicitly understanding what $\varphi$ relates to in the gravity dual, and to the one-form superfluids of sec.~\ref{sec:super}, identifying $\varphi_\mu$ in the gravity theory. The analogous fluid/gravity considerations in the case of zero-form superfluids \cite{Bhattacharya:2011eea, Bhattacharya:2011tra} will be useful. It is likely that gravity duals to string fluids, as formulated in sec.~\ref{sec:stringfluids}, will involve black branes charged under a two-form gauge field and with scalar hair.

The long wavelength perturbations of black branes in supergravity are governed by effective fluid theories with multiple higher-form currents \cite{Armas:2016mes}. Starting with the work of \cite{Armas:2018ibg}, it would be interesting to develop higher-form superfluid theories that could be used to study the stability of these gravitational solutions and to aid in finding new stationary black hole solutions via hydrostatic effective actions for the Goldstone modes.

Finally, it should be mentioned that the tools developed here and the viewpoint
expressed has repercussions to other hydrodynamic theories with generalised
global symmetries such as theories of viscoelasticity \cite{Grozdanov:2018ewh} and with weakly
broken symmetries \cite{Grozdanov:2018fic}. In particular, it is likely that some of these theories require the
introduction of the vector Goldstone mode $\varphi_\mu$ in order to define a
hydrostatic effective action. We leave this line of inquire to future work.

\subsubsection*{Acknowledgements}

We would like to thank J. Bhattacharya, J. Hernandez, P. Kovtun, P. Glorioso and
specially N. Iqbal for various helpful discussions.  We would also like to thank
an anonymous referee for useful comments to this manuscript. JA is partly supported by
the Netherlands Organization for Scientific Research (NWO). AJ would like to
thank Perimeter Institute and Durham University, where part of this project was
done, for hospitality. AJ is supported by the NSERC Discovery Grant program of
Canada.

\appendix

\section{Calculational details}

\label{app:hydrostatic}

\subsection{Hydrostatic corrections}

In this appendix,  the explicit expressions for the first order
hydrostatic corrections to various hydrodynamic systems studied in this work are derived.
Along with being of inherent phenomenological relevance, these corrections are
important when comparing the constitutive relations between one-form superfluids
and hot electromagnetism.

\subsubsection{String fluids and electric limit of one-form superfluids}
\label{app:hydrostaticString}

Using the free energy density for string fluids in \eqref{stringN}, performing a
$\delta_\scB$ variation of each of the terms and using \cref{hsConstiGeneral},
it is possible to work out their effect on the hydrostatic constitutive
relations. It is useful to parametrise these corrections as
\begin{align}\label{hs_string_consti}
  T^{\mu\nu}_{\text{hs}}
  &= (\epsilon+p)\, u^{\mu} u^{\nu} + p\,g^{\mu\nu}
    - \varpi \rho\, h^{\mu}h^{\nu}
    + T^{\mu\nu}_{\text{hs},\alpha}
    + T^{\mu\nu}_{\text{hs},\beta}
    + T^{\mu\nu}_{\text{hs},\tilde{\beta}_1}
    + T^{\mu\nu}_{\text{hs},\tilde{\beta}_2}
    + T^{\mu\nu}_{\text{hs},\tilde{\beta}_3}
    + \mathcal{O}(\dow^2)~~, \nn\\
  J^{\mu\nu}_{\text{hs}}
  &= 2 \rho\, u^{[\mu}h^{\nu]}
    + J^{\mu\nu}_{\text{hs},\alpha}
    + J^{\mu\nu}_{\text{hs},\beta}
    + J^{\mu\nu}_{\text{hs},\tilde{\beta}_1}
    + J^{\mu\nu}_{\text{hs},\tilde{\beta}_2}
    + J^{\mu\nu}_{\text{hs},\tilde{\beta}_3}
    + \mathcal{O}(\dow^2)~~, \nn\\
  N^\mu_{\text{hs}}
  &= \frac{p}{T} u^\mu
    + N^\mu_{\text{hs},\alpha}
    + N^\mu_{\text{hs},\beta}
    + N^\mu_{\text{hs},\tilde{\beta}_1}
    + N^\mu_{\text{hs},\tilde{\beta}_2}
    + N^\mu_{\text{hs},\tilde{\beta}_3}
    + \mathcal{O}(\dow^2)~~,
\end{align}
where all the explicit terms are given below, specifically
\begingroup
\allowdisplaybreaks
\begin{align}
  T^{\mu\nu}_{\text{hs},\alpha}
  &= - \frac16 \epsilon^{\alpha\beta\rho\sigma} u_\alpha H_{\beta\rho\sigma}
    \lB \lb \frac{\dow (T\alpha)}{\dow T}
    + \varpi \frac{\dow \alpha}{\dow \varpi} \rb u^{\mu}u^{\nu}
    - \varpi \frac{\dow \alpha}{\dow \varpi} h^{\mu}h^{\nu} \rB
    - \frac\alpha3 u^{(\mu} \epsilon^{\nu)\lambda\rho\sigma}
      H_{\lambda\rho\sigma}~~,  \nn\\
  J^{\mu\nu}_{\text{hs},\alpha}
  &=
    - \frac13 \epsilon^{\mu\nu\rho\sigma} u_\mu H_{\nu\rho\sigma}
    \frac{\dow \alpha}{\dow \varpi} u^{[\mu}h^{\nu]}
    + \nabla_\sigma \lb \alpha\epsilon^{\mu\nu\rho\sigma}
    u_\rho\rb~~, \nn\\
  N^\mu_{\text{hs},\alpha}
  &= \frac{\alpha}{6T}
    \epsilon^{\mu\nu\rho\sigma} H_{\nu\rho\sigma}
    - \alpha \epsilon^{\mu\nu\rho\sigma}
    u_\nu  \dow_\rho \bfrac{\varpi h_\sigma}{T}~~, \nn\\[1em]
  T^{\mu\nu}_{\text{hs},\beta}
  &= - \epsilon^{\alpha\beta\rho\sigma} u_\alpha h_\beta \dow_\rho u_\sigma
    \lB \lb \frac{1}{T^2} \frac{\dow (T^3\beta)}{\dow T}
    + \varpi^2 \frac{\dow (\beta/\varpi)}{\dow \varpi} \rb u^{\mu}u^{\nu}
    - \varpi^2 \frac{\dow (\beta/\varpi)}{\dow \varpi} h^{\mu}h^{\nu} \rB \nn\\
  &\qquad
    - 2 u^{(\mu} \epsilon^{\nu)\lambda\rho\sigma} \lb
    T \beta h_\lambda \dow_\rho \frac{u_\sigma}{T}
    + \frac1T \nabla_\rho \lb
    T \beta h_\lambda u_\sigma \rb \rb~~, \nn\\
  J^{\mu\nu}_{\text{hs},\beta}
  &= - \epsilon^{\alpha\beta\rho\sigma} u_\alpha h_\beta \dow_\rho u_\sigma
    2\varpi \frac{\dow (\beta/\varpi)}{\dow \varpi} u^{[\mu}h^{\nu]}
    + 2 \frac{\beta}{\varpi} u^{[\mu} \epsilon^{\nu]\lambda\rho\sigma}
    u_\lambda \dow_\rho u_\sigma~~, \nn\\
  N^\mu_{\text{hs},\beta}
  &= \frac{\beta}{T^2} \epsilon^{\mu\nu\rho\sigma}
    h_\nu \dow_\rho(T u_\sigma)~~, \nn\\[1em]
  T^{\mu\nu}_{\text{hs},\tilde\beta_1}
  &= - h^\lambda \dow_\lambda T
    \lB \lb \frac{\dow (T\tilde\beta_1)}{\dow T}
    + \varpi^2 \frac{\dow (\tilde\beta_1/\varpi)}{\dow \varpi} \rb u^{\mu}u^{\nu}
    - \varpi^2 \frac{\dow (\tilde\beta_1/\varpi)}{\dow \varpi} h^{\mu}h^{\nu} \rB \nn\\
  &\qquad
    - \tilde\beta_1 h^\lambda \dow_\lambda T g^{\mu\nu} 
    + 2 \tilde\beta_1 h^{(\mu}\nabla^{\nu)} T
    + T \nabla_\lambda \lb \tilde\beta_1 h^\lambda \rb
    u^\mu u^\nu~~,\nn\\
  J^{\mu\nu}_{\text{hs},\tilde\beta_1}
  &= - h^\lambda \dow_\lambda T
     2\varpi \frac{\dow (\tilde\beta_1/\varpi)}{\dow \varpi} u^{[\mu}h^{\nu]}
    - 2\frac{\tilde\beta_1}{\varpi} u^{[\mu} \nabla^{\nu]} T~~, \nn\\
  N^\mu_{\text{hs},\tilde\beta_1}
  &=  - \frac{2\tilde\beta_1}{T} u^{[\mu} h^{\lambda]} \dow_\lambda T~~, \nn\\[1em]
  T^{\mu\nu}_{\text{hs},\tilde\beta_2}
  &= - h^\lambda \dow_\lambda \frac{\varpi}{T}
      \lB \lb \frac{\dow (T\tilde\beta_2)}{\dow T}
    + \varpi^2 \frac{\dow (\tilde\beta_2/\varpi)}{\dow \varpi} \rb u^{\mu}u^{\nu}
    - \varpi^2 \frac{\dow (\tilde\beta_2/\varpi)}{\dow \varpi} h^{\mu}h^{\nu}
    \rB \nn\\
  &\qquad
    - \tilde\beta_2 h^\lambda \dow_\lambda \frac{\varpi}{T} g^{\mu\nu}
    + 2\tilde\beta_2 h^{(\mu}\nabla^{\nu)} \frac{\varpi}{T}
    - \nabla_\lambda \lb \tilde\beta_2 h^\lambda \rb \frac{\varpi}{T} h^{\mu}h^{\nu}~~,
    \nn\\
  J^{\mu\nu}_{\text{hs},\tilde\beta_2}
  &=
    - h^\lambda \dow_\lambda \frac{\varpi}{T}
      2\varpi \frac{\dow (\tilde\beta_2/\varpi)}{\dow \varpi} u^{[\mu}h^{\nu]}
    - 2\frac{\tilde\beta_2}{\varpi} u^{[\mu} \nabla^{\nu]} \frac{\varpi}{T}
    + 2\nabla_\lambda \lb \tilde\beta_2 h^\lambda \rb \frac{1}{T}
    u^{[\mu}h^{\nu]}~~, \nn\\
  N^\mu_{\text{hs},\tilde\beta_2}
  &=  - \frac{2\tilde\beta_2}{T} u^{[\mu} h^{\lambda]} \dow_\lambda
    \frac{\varpi}{T}~~, \nn\\[1em]
  T^{\mu\nu}_{\text{hs},\tilde\beta_3}
  &= - \epsilon^{\alpha\beta\rho\sigma} u_\alpha h_\beta \dow_\rho h_\sigma
      \lB \lb \frac{1}{T^2} \frac{\dow (T^3\tilde\beta_3)}{\dow T}
    + \varpi^3 \frac{\dow (\tilde\beta_3/\varpi^2)}{\dow \varpi} \rb u^{\mu}u^{\nu}
    - \varpi^3 \frac{\dow (\tilde\beta_3/\varpi^2)}{\dow \varpi} h^{\mu}h^{\nu}
    \rB
    \nn\\
  &\qquad
      - 2 \tilde\beta_3 u^{(\mu} \epsilon^{\nu)\lambda\rho\sigma}
      h_\lambda \dow_\rho h_\sigma~, \nn\\
  J^{\mu\nu}_{\text{hs},\tilde\beta_3}
  &= - \epsilon^{\alpha\beta\rho\sigma} u_\alpha h_\beta \dow_\rho h_\sigma
    2\varpi^2 \frac{\dow (\tilde\beta_3/\varpi^2)}{\dow \varpi} u^{[\mu}h^{\nu]}
      + 2 u^{[\mu} \epsilon^{\nu]\lambda\rho\sigma}
      \lb \frac{T\tilde\beta_3}{\varpi^2}
      u_\lambda \dow_\rho \frac{\varpi h_\sigma}{T}
      + \frac1T\nabla_\rho \lb \frac{T\tilde\beta_3}{\varpi}
    u_\lambda h_\sigma \rb \rb~, \nn\\
  N^\mu_{\text{hs},\tilde{\beta}_3}
  &= \frac{\tilde\beta_3}{T} \epsilon^{\mu\nu\rho\sigma} u_\lambda
    h_\nu \dow_\rho h_\sigma~~.
    \label{hs-correction-MHD}
\end{align}
\endgroup

The hydrostatic corrections in the electric limit of one-form
superfluids are obtained from the respective hydrostatic free energy density given in
\cref{electricN}. The contributions from all terms except $\beta$ and
$\tilde{\beta}_i$ have already been discussed in
\cref{sec:electric-hs-corrections}. The contribution from the remaining
terms is precisely the same as in \cref{hs-correction-MHD} for string
fluids.

\subsubsection{Magnetohydrodynamics and magnetically dominated bound-charge plasma}
\label{app:hydrostaticMHD}

Using the MHD free energy density \eqref{N-MHD}, performing the relevant
variations and ignoring certain second order contributions to the
energy-momentum tensor, the constitutive relations are the sum of the following
contributions
\begingroup \allowdisplaybreaks
\begin{align}
  T^{\mu\nu}_{\text{hs},M_1}
  &= \lB \lb T \frac{\dow M_1}{\dow T}
+ \mu \frac{\dow M_1}{\dow \mu} \rb u^\mu u^\nu
    - 2 B^2 \frac{\dow M_1}{\dow B^2} \bbB^{\mu\nu}
    \rB B^\lambda \dow_\lambda \frac{B^2}{T^4} 
  \nn\\
  &\qquad + 2 M_1 u^\lambda \dow_\lambda \frac{B^2}{T^4} u^{(\mu} B^{\nu)} 
    - M_1 B^\lambda \dow_\lambda \frac{B^2}{T^4} u^\mu u^\nu
    + \frac{2B^2}{T^4}
    \nabla_\lambda \lb M_1 B^\lambda \rb
    \lb \bbB^{\mu\nu} + 2 u^\mu u^\nu \rb, \nn\\
  M^{\mu\nu}_{\text{hs},M_1}
  &= - 2 |B| \bbE^{\mu\nu} \frac{\dow M_1}{\dow B^2}
    B^\lambda \dow_\lambda \frac{B^2}{T^4}
    - M_1 \epsilon^{\mu\nu\rho\sigma} u_\rho \dow_\sigma \frac{B^2}{T^4}
    + \frac{2|B|}{T^4} \nabla_\lambda \lb M_1 B^\lambda \rb \bbE^{\mu\nu}~~,
    \nn\\
  J^\mu_{\text{hs},M_1}
  &= u^\mu B^\lambda \dow_\lambda \frac{B^2}{T^4} \frac{\dow M_1}{\dow \mu}
    + \nabla_\nu M^{\mu\nu}_{\text{hs},M_1}~~,
    \nn\\
  N^\mu_{\text{hs},M_1}
  &= \frac{2M_1}{T} u^{[\mu} B^{\nu]} \dow_\nu \frac{B^2}{T^4}~~, \nn\\[1em]
  T^{\mu\nu}_{\text{hs},M_2}
  &= 
    \lB \lb \frac{1}{T^2} \frac{\dow (T^3 M_2)}{\dow T}
    + \mu \frac{\dow M_2}{\dow \mu} \rb u^\mu u^\nu
    - 2 B^2 \frac{\dow M_2}{\dow B^2} \bbB^{\mu\nu}
    \rB
    \epsilon^{\alpha\beta\rho\sigma} u_\alpha B_\beta \dow_\rho B_\sigma \nn\\
  &\qquad
    + 2u^{(\mu} \epsilon^{\nu)\lambda\rho\sigma} M_2
    B_\lambda \dow_\rho B_\sigma \nn\\
  &\qquad
    - \epsilon^{\alpha\beta\rho\sigma}
    \lb T M_2 u_\alpha \dow_\rho \frac{B_\beta}{T}
    + \frac1T \nabla_\rho \lb T M_2 u_\alpha B_\beta \rb \rb
    \lb
    2 B^{(\mu} P^{\nu)}{}_\sigma
    - B_\sigma (P^{\mu\nu} + u^\mu u^\nu) \rb, \nn\\
  M^{\mu\nu}_{\text{hs},M_2}
  &= - 2 |B| \bbE^{\mu\nu} \frac{\dow M_2}{\dow B^2}
    \epsilon^{\lambda\tau\rho\sigma} u_\lambda B_\tau \dow_\rho B_\sigma
    + \epsilon^{\mu\nu\lambda\tau} u_\lambda
  \epsilon^{ab\rho}{}_\tau
  \lb M_2  u_a \dow_\rho B_b
  + \nabla_\rho \lb M_2  u_a B_b \rb \rb~~, \nn\\
  J^\mu_{\text{hs},M_2}
  &= u^\mu\epsilon^{\lambda\nu\rho\sigma} 
    u_\lambda B_\nu \dow_\rho B_\sigma \frac{\dow M_2}{\dow \mu}
    + \nabla_\nu M^{\mu\nu}_{\text{hs},M_2}~~, \nn\\
  N^\mu_{\text{hs},M_2}
  &= - \frac{M_2}{T} \epsilon^{\mu\nu\rho\sigma} B_\nu \dow_\rho B_\sigma~~, \nn\\[1em]
  T^{\mu\nu}_{\text{hs},M_3}
  &= - \lB \lb T \frac{\dow (M_3/T)}{\dow T}
    + \frac{\mu}{T} \frac{\dow M_3}{\dow \mu} \rb u^\mu u^\nu
    - 2 \frac{B^2}{T} \frac{\dow M_3}{\dow B^2} \bbB^{\mu\nu}
    \rB
    B^\lambda \dow_\lambda T \nn\\
  &\qquad
    - 2 \frac{M_3}{T} u^\lambda \dow_\lambda T u^{(\mu} B^{\nu)} 
    + \frac{M_3}{T} B^\lambda \dow_\lambda T u^\mu u^\nu
    + T \nabla_\lambda \lb \frac{M_3}{T} B^\lambda \rb u^\mu u^\nu, \nn\\
  M^{\mu\nu}_{\text{hs},M_3}
  &= 2 |B| \bbE^{\mu\nu} \frac{\dow M_3}{\dow B^2}
    \frac{1}{T} B^\lambda \dow_\lambda T
    + \frac{M_3}{T} \epsilon^{\mu\nu\rho\sigma} u_\rho \dow_\sigma T~~, \nn\\
  J^\mu_{\text{hs},M_3}
  &= - \frac{1}{T} u^\mu B^\lambda \dow_\lambda T \frac{\dow M_3}{\dow \mu}
    + \nabla_\nu M^{\mu\nu}_{\text{hs},M_3}~~, \nn\\
  N^\mu_{\text{hs},M_3}
  &= - \frac{2M_3}{T^2} u^{[\mu} B^{\nu]} \dow_\nu T~~, \nn\\[1em]
  T^{\mu\nu}_{\text{hs},M_4}
  &= - \lB \lb \frac{1}{T^2} \frac{\dow (T^3 M_4}{\dow T}
    + \mu \frac{\dow M_4}{\dow \mu} \rb u^\mu u^\nu
    - 2 B^2 \frac{\dow M_4}{\dow B^2} \bbB^{\mu\nu}
    \rB
    \epsilon^{\alpha\beta\rho\sigma}
    u_\alpha B_\beta \dow_\rho u_\sigma \nn\\
  &\qquad
    + \epsilon^{\beta\alpha\rho\sigma} M_4
    u_\alpha \dow_\rho u_\sigma \lb 2 B^{(\mu} P^{\nu)}{}_\beta
    - B_\beta (P^{\mu\nu} + u^\mu u^\nu) \rb \nn\\
  &\qquad
    - 2u^{(\mu}\epsilon^{\nu)\lambda\rho\sigma} \lb T M_4
    B_\lambda \dow_\rho \frac{u_\sigma}{T}
    + \frac1T \nabla_\rho \lb T M_4 u_\sigma B_\lambda \rb \rb, \nn\\
  M^{\mu\nu}_{\text{hs},M_4}
  &= 2 |B| \bbE^{\mu\nu} \frac{\dow M_4}{\dow B^2}
    \epsilon^{\lambda\tau\rho\sigma} u_\lambda B_\tau \dow_\rho u_\sigma
    + 2 M_4 P^{\mu\rho} P^{\nu\sigma} \dow_{[\rho} u_{\sigma]}~~, \nn\\
  J^\mu_{\text{hs},M_4}
  &= - u^\mu \epsilon^{\lambda\nu\rho\sigma} u_\lambda B_\nu \dow_\rho u_\sigma
    \frac{\dow M_4}{\dow \mu}
    + \nabla_\nu M^{\mu\nu}_{\text{hs},M_4}~~, \nn\\
  N^\mu_{\text{hs},M_4}
  &= \frac{M_4}{T^2} \epsilon^{\mu\nu\rho\sigma} B_\nu \dow_\rho (T u_\sigma)~~,
    \nn\\[1em]
  T^{\mu\nu}_{\text{hs},M_5}
  &= \lB \lb T \frac{\dow (T M_5)}{\dow T}
    + T\mu \frac{\dow M_5}{\dow \mu} \rb u^\mu u^\nu
    - 2 T B^2 \frac{\dow M_5}{\dow B^2} \bbB^{\mu\nu}
    \rB B^\lambda \dow_\lambda \frac{\mu}{T} \nn\\
  &\qquad
    + 2 T M_5 u^\lambda \dow_\lambda \frac{\mu}{T} u^{(\mu} B^{\nu)} 
    - T M_5 B^\lambda \dow_\lambda \frac{\mu}{T} u^\mu u^\nu~~, \nn\\
  M^{\mu\nu}_{\text{hs},M_5}
  &= - 2 |B| \bbE^{\mu\nu} \frac{\dow M_5}{\dow B^2}
    T B^\lambda \dow_\lambda \frac{\mu}{T}
    - T M_5 \epsilon^{\mu\nu\rho\sigma} u_\rho \dow_\sigma \frac{\mu}{T}~~,
    \nn\\
  J^\mu_{\text{hs},M_5}
  &= T u^\mu B^\lambda \dow_\lambda \frac{\mu}{T} \frac{\dow M_5}{\dow \mu}
    - \frac1T u^\mu \nabla_\lambda \lb T M_5 B^\lambda \rb
    + \nabla_\nu M^{\mu\nu}_{\text{hs},M_5}~~, \nn\\
  N^\mu_{\text{hs},M_5}
  &= 2 M_5 u^{[\mu} B^{\nu]} \dow_\nu \frac{\mu}{T}~~.
    \label{hydrostatic-corrections-MHD}
\end{align}
\endgroup
In the case of magnetically dominated bound-charge plasmas with the hydrostatic free energy
density \bref{N-mBCP}, the contributions from $M_{1,2,3,4}$ to the respective
hydrostatic constitutive relations are just given in terms of the MHD expressions
in \cref{hydrostatic-corrections-MHD}, except that the transport coefficients
are taken to be independent of $\mu$.

\subsection{Mapping MHD to string fluids}
\label{app:MHD}

In this appendix the details of the mapping between MHD and string fluid
constitutive relations at first order in derivatives given in
\cref{sec:MHD-String-mapping} are provided.

\subsubsection{Eliminating chemical potential and electric field}

To begin with, we take the hydrostatic energy-momentum tensor for MHD
from \cref{app:hydrostaticMHD} and introduce it in the solutions for $\mu$ for $E_\mu$
given in \cref{TheFinal_MuE_elimination} coming from Maxwell's equations. These
are described in terms of 6 transport coefficients $P(T,\mu,B^2)$ and
$M_{1,2,3,4,5}(T,\mu,B^2)$. We expand $P(T,\mu,B^2)$ around $\mu = \mu_0(T,B^2)$
up to second order in derivatives
\begin{equation}
  P(T,\mu,B^2)
  = P_0(T,B^2)
  + \half P_2(T,B^2) \lb \mu - \mu_0(T,B^2) \rb^2
  + \mathcal{O}(\dow^3)~~.
\end{equation}
Here we have used the defining relation of $\mu_0$ from \cref{define-mu0}.
Representing the $\mu$ solution in \cref{TheFinal_MuE_elimination} as
$\mu = \mu_0 + \delta\mu$, up to the first order in derivatives, we can work out
\begin{align}
  P(T,\mu,B^2)
  &= P_0(T,B^2)
    + \mathcal{O}(\dow^2)~~, \nn\\
  \varpi(T,\mu,B^2)
  &= - 2 |B| \frac{\dow P_0(T,B^2)}{\dow B^2}
  + 2 |B| P_2(T,B^2) \frac{\dow \mu_0(T,B^2)}{\dow B^2} \delta\mu
  + \mathcal{O}(\dow^2)~~, \nn\\
  \epsilon(T,\mu,B^2)
  &= T \frac{\dow P_0(T,B^2)}{\dow T}
    - P_0(T,B^2)
    - T^2 \frac{\dow}{\dow T} \bfrac{\mu_0(T,B^2)}{T}
    P_2(T,B^2) \delta \mu
    + \mathcal{O}(\dow^2)~~.
\end{align}
In an analogous manner, we can expand $M_{1,2,3,4,5}(T,\mu,B^2)$ up to the first
order in derivatives
\begin{equation}
  M_i(T,\mu,B^2)
  = M_{i,0}(T,B^2)
  + M_{i,1}(T,B^2) \lb \mu - \mu_0(T,B^2) \rb
  + \mathcal{O}(\dow^2)~~,
\end{equation}
which after eliminating $\mu$ leads to
\begin{align}
  M_i(T,\mu,B^2)
  &= M_{i,0}(T,B^2) + \mathcal{O}(\dow)~~, \nn\\
  \frac{\dow M_i(T,\mu,B^2)}{\dow T}
  &= \frac{\dow M_{i,0}(T,B^2)}{\dow T}
    - M_{i,1}(T,B^2) \frac{\dow \mu_0(T,B^2)}{\dow T}
    + \mathcal{O}(\dow)~~, \nn\\
  \frac{\dow M_i(T,\mu,B^2)}{\dow B^2}
  &= \frac{\dow M_{i,0}(T,B^2)}{\dow B^2}
    - M_{i,1}(T,B^2) \frac{\dow \mu_0(T,B^2)}{\dow B^2}
    + \mathcal{O}(\dow)~~, \nn\\
  \frac{\dow M_i(T,\mu,B^2)}{\dow \mu}
  &= M_{i,1}(T,B^2) + \mathcal{O}(\dow)~~.
\end{align}
Schematically, splitting the first order hydrostatic contributions to the MHD
energy-momentum tensor as
$T^{\mu\nu}_{\text{hs},M_i} = T^{\mu\nu}_{\text{hs},M_{i,0}} +
T^{\mu\nu}_{\text{hs},M_{i,1}}$ and plugging in the $\mu$ and $E_\mu$ solution
from \cref{TheFinal_MuE_elimination}, after a straight-forward computation, we
can show that
\begin{align}\label{PMi1corrections}
  T^{\mu\nu}_{\text{hs},P}
  &+ \sum_i T^{\mu\nu}_{\text{hs},M_{i,1}} \nn\\
  &= \lb T \frac{\dow P_0}{\dow T} \rb u^\mu u^\nu
    + P_0 g^{\mu\nu}
    - 2 B^2 \frac{\dow P_0}{\dow B^2} \bbB^{\mu\nu} 
    + 4 T |B| \frac{\dow P_0}{\dow B^2} u^{(\mu}
    \bbE^{\nu)\sigma}
    \dow_\sigma\frac{\mu_0}{T} \nn\\
  &\qquad
    + \lb 2 B^2 \frac{\dow \mu_0}{\dow B^2} \bbB^{\mu\nu}
    - T^2 \frac{\dow(\mu_0/T)}{\dow T} u^\mu u^\nu \rb
    \lB u_\mu J^\mu_\ext
    + \frac1T \nabla_\lambda \lb T M_{5,0} B^\lambda \rb 
    + 2 \frac{\dow P_0}{\dow B^2} \epsilon^{\lambda\tau\rho\sigma} B_\lambda u_\tau \dow_\rho u_\sigma \rB \nn\\
  &\qquad
    - 4T |B| \frac{\dow P_0}{\dow B^2} u^{(\mu}
    \lB \frac{1}{\sigma^2_\perp + \tilde\sigma^2 }
    \lb \frac{T \dow P_0/\dow T}{T \dow P_0/\dow T - 2 B^2 \dow P_0/\dow B^2} \rb
    \lb - \sigma_\perp \bbB^{\nu)\rho} \hat B^\sigma
    + \tilde\sigma_\perp \bbE^{\nu)\rho} \hat B^\sigma \rb X_{\rho\sigma} \right. \nn\\
  &\qquad\qquad
    \left. + \frac{2}{\sigma^2_\perp + \tilde\sigma_\perp^2 } \lb
    - \lb \tilde\sigma_\perp \sigma'_\times - \sigma_\perp \tilde\sigma'_\times \rb
    \bbB^{\nu)(\rho} \hat B^{\sigma)} 
    - \lb \tilde\sigma \tilde\sigma'_\times
    + \sigma_\perp \sigma'_\times \rb \bbE^{\nu)(\rho} \hat B^{\sigma)}
    \rb  \half \delta_\scB g_{\rho\sigma} \rB~~.
\end{align}
Interestingly, the $M_{i,1}$ contributions entirely drop out.  On the other
hand, for the remaining hydrostatic contributions we have
\begingroup
\allowdisplaybreaks
\begin{align}\label{Mi0corrections}
  T^{\mu\nu}_{\text{hs},M_{1,0}}
  &= \lB T \frac{\dow M_{1,0}}{\dow T} u^\mu u^\nu
    - 2 B^2 \frac{\dow M_{1,0}}{\dow B^2} \bbB^{\mu\nu}
    \rB B^\lambda \dow_\lambda \frac{B^2}{T^4} 
    \nn\\
  &\qquad + 2 M_{1,0} u^\lambda \dow_\lambda \frac{B^2}{T^4} u^{(\mu} B^{\nu)} 
    - M_{1,0} B^\lambda \dow_\lambda \frac{B^2}{T^4} u^\mu u^\nu
    + \frac{2B^2}{T^4}
    \nabla_\lambda \lb M_{1,0} B^\lambda \rb
    \lb \bbB^{\mu\nu} + 2 u^\mu u^\nu \rb~~, \nn\\
  T^{\mu\nu}_{\text{hs},M_{2,0}}
  &= 
    \lB \frac{1}{T^2} \frac{\dow (T^3 M_{2,0})}{\dow T} u^\mu u^\nu
    - 2 B^2 \frac{\dow M_{2,0}}{\dow B^2} \bbB^{\mu\nu}
    \rB
    \epsilon^{\alpha\beta\rho\sigma} u_\alpha B_\beta \dow_\rho B_\sigma \nn\\
  &\qquad
    + 2 M_{2,0} u^{(\mu} \epsilon^{\nu)\lambda\rho\sigma}
    B_\lambda \dow_\rho B_\sigma \nn\\
  &\qquad
    - \epsilon^{\alpha\beta\rho\sigma}
    \lb T M_{2,0} u_\alpha \dow_\rho \frac{B_\beta}{T}
    + \frac1T \nabla_\rho \lb T M_{2,0} u_\alpha B_\beta \rb \rb
    \lb
    2 B^{(\mu} P^{\nu)}{}_\sigma
    - B_\sigma (P^{\mu\nu} + u^\mu u^\nu) \rb, \nn\\
  T^{\mu\nu}_{\text{hs},M_{3,0}}
  &= - \lB T \frac{\dow (M_{3,0}/T)}{\dow T} u^\mu u^\nu
    - 2 \frac{B^2}{T} \frac{\dow M_{3,0}}{\dow B^2} \bbB^{\mu\nu}
    \rB
    B^\lambda \dow_\lambda T \nn\\
  &\qquad
    - 2 \frac{M_{3,0}}{T} u^\lambda \dow_\lambda T u^{(\mu} B^{\nu)} 
    + \frac{M_{3,0}}{T} B^\lambda \dow_\lambda T u^\mu u^\nu
    + T \nabla_\lambda \lb \frac{M_{3,0}}{T} B^\lambda \rb u^\mu u^\nu, \nn\\
  T^{\mu\nu}_{\text{hs},M_{4,0}}
  &= - \lB \frac{1}{T^2} \frac{\dow (T^3 M_{4,0})}{\dow T} u^\mu u^\nu
    - 2 B^2 \frac{\dow M_{4,0}}{\dow B^2} \bbB^{\mu\nu}
    \rB
    \epsilon^{\alpha\beta\rho\sigma}
    u_\alpha B_\beta \dow_\rho u_\sigma \nn\\
  &\qquad
    + \epsilon^{\beta\alpha\rho\sigma} M_{4,0}
    u_\alpha \dow_\rho u_\sigma \lb 2 B^{(\mu} P^{\nu)}{}_\beta
    - B_\beta (P^{\mu\nu} + u^\mu u^\nu) \rb \nn\\
  &\qquad
    - 2u^{(\mu}\epsilon^{\nu)\lambda\rho\sigma} \lb T M_{4,0}
    B_\lambda \dow_\rho \frac{u_\sigma}{T}
    + \frac1T \nabla_\rho \lb T M_{4,0} u_\sigma B_\lambda \rb \rb, \nn\\
  T^{\mu\nu}_{\text{hs},M_{5,0}}
  &= \lB T \frac{\dow (T M_{5,0})}{\dow T} u^\mu u^\nu
    - 2 T B^2 \frac{\dow M_{5,0}}{\dow B^2} \bbB^{\mu\nu}
    \rB B^\lambda \dow_\lambda \frac{\mu_0}{T} \nn\\
  &\qquad
    + 2 T M_{5,0} u^\lambda \dow_\lambda \frac{\mu_0}{T} u^{(\mu} B^{\nu)} 
    - T M_{5,0} B^\lambda \dow_\lambda \frac{\mu_0}{T} u^\mu u^\nu.
\end{align}
\endgroup

\subsubsection{Velocity field redefinition}

As we suggested in \cref{sec:MHD-String-mapping}, the map between string fluids
and MHD can involve a non-trivial non-hydrostatic redefinition of $u^\mu$ and
$T$. In the following, we find that it is sufficient to perform a redefinition
of $u^\mu$ alone. With this hindsight, consider a redefinition of the fluid velocity
\begin{equation}
  u^\mu \to u^\mu + \delta u^\mu~~,
\end{equation}
such that $u_\mu\delta u^\mu = 0$, where $\delta u^\mu$ is purely
non-hydrostatic. The electromagnetic fields get a contribution from
$\delta u^\mu$ via
\begin{equation}
  B^\mu \to B^\mu + u^\mu B_\nu \delta u^\nu + \mathcal{O}(\dow^2)~~,~~
  E_\mu \to E_\mu - |B| \bbE_{\mu\nu} \delta u^\nu + \mathcal{O}(\dow^2)~~.
\end{equation}
Note that $B^2$ is invariant to first order. Interestingly, despite being itself
first order, $E_\mu$ shifts with a first-order piece. Therefore, the equation
determining the $E_\mu$ in \cref{TheFinal_MuE_elimination} modifies to
\begin{align}\label{dA_withdu}
  P^{\mu\nu} \delta_\scB A_\nu
  &= \half \frac{1}{\sigma_\parallel} \hat B^\mu \bbE^{\rho\sigma}
    \delta_\scB b_{\rho\sigma}
    - \frac{1}{\sigma_\parallel} \hat B^\mu
    \lb \tilde\chi'_1 \bbB^{\rho\sigma}
    + \tilde\chi'_2 \hat B^\rho \hat B^\sigma
  \rb \half \delta_\scB g_{\rho\sigma} \nn\\
  &\qquad +
    \lb \frac{\epsilon+P}{\epsilon + P + \varpi |B|} \rb
    \lb \frac{\sigma_\perp}{\sigma^2_\perp + \tilde\sigma^2 }
    \bbE^{\mu\rho} \hat B^\sigma
    + \frac{\tilde\sigma_\perp}{\sigma^2_\perp + \tilde\sigma^2 }
    \bbB^{\mu\rho} \hat B^\sigma \rb \delta_\scB b_{\rho\sigma} \nn\\
  &\qquad
    + \lb
    \frac{\tilde\sigma_\perp \sigma'_\times
    - \sigma_\perp \tilde\sigma'_\times}
    {\sigma^2_\perp + \tilde\sigma^2 }
    \bbE^{\mu(\rho} \hat B^{\sigma)} 
    - \frac{\tilde\sigma_\perp \tilde\sigma'_\times
    + \sigma_\perp \sigma'_\times}
    {\sigma^2_\perp + \tilde\sigma^2 }
    \bbB^{\mu(\rho} \hat B^{\sigma)}
    \rb \delta_\scB g_{\rho\sigma} \nn\\
  &\qquad - \frac{|B|}{T} \bbE^{\mu\nu} \delta u_\nu~~,
\end{align}
where we have used the mapping for $X_{\mu\nu}$ given in \cref{XmnMap}.  The
hydrostatic constitutive relations discussed in the previous subsection get
corrected by these redefinition and obtain a $\delta u^\mu$ contribution to the
energy-momentum tensor
\begin{align}
  T^{\mu\nu}_{\delta u}
  = 2 u^{(\mu}  \lB T \frac{\dow P_0}{\dow T} \hat B^{\nu)} \hat B_\lambda
  + \lb T \frac{\dow P_0}{\dow T}
  - 2 B^2 \frac{\dow P_0}{\dow B^2} \rb \bbB^{\nu)}{}_\lambda
  \rB \delta u^{\lambda}~~.
\end{align}

To find what the relative field redefinition for $u^\mu$ between MHD and string
fluid is, we need to compare the energy-momentum tensor in the two
formulations. Substituting $B^\mu$ in \cref{PMi1corrections,Mi0corrections}
using \cref{the-B-map}, and invoking the mapping between hydrostatic transport
coefficients given in \cref{eq:maphydro}, we can show that
\begin{equation}
  T^{\mu\nu}_{\text{MHD,hs}}
  + T^{\mu\nu}_{\delta u}
  = T^{\mu\nu}_{\text{string,hs}}~~,
\end{equation}
for $\delta u^\mu$ given in \cref{eq:deltau}. Here
$T^{\mu\nu}_{\text{string,hs}}$ are the hydrostatic corrections to string fluid
constitutive relations worked out in \cref{app:hydrostaticString}.

\subsubsection{Mapping of non-hydrostatic transport coefficients}

We have already mapped the hydrostatic transport coefficients between the two
formulations in \cref{eq:maphydro} using the hydrostatic free-energy density. To
find the mapping between non-hydrostatic transport coefficients, let us
substitute $\delta u^\mu$ from \cref{eq:deltau} into \cref{dA_withdu} and obtain
\begin{align}
  P^{\mu\nu} \delta_\scB A_\nu
  &=
    \frac{\alpha\rho}{\epsilon + p} \Delta^{\mu\sigma} h^\nu
  \delta_\scB b_{\sigma\nu}
    + \half \frac{1}{\sigma_\parallel} h^\mu
    \epsilon^{\rho\sigma} \delta_\scB b_{\rho\sigma}
  - \frac{1}{\sigma_\parallel} h^\mu \lb \tilde\chi'_1 \Delta^{\rho\sigma}
  + \tilde\chi'_2 h^\rho h^\sigma
  \rb \half \delta_\scB g_{\rho\sigma} \nn\\
  &\qquad
    + \frac{sT}{(\epsilon + p)} \lb
    \frac{\tilde\sigma_\perp\sigma'_\times - \sigma_\perp \tilde\sigma'_\times}
    {\sigma^2_\perp + \tilde\sigma^2 }
    \epsilon^{\mu(\rho} h^{\sigma)} 
    - \frac{\tilde\sigma_\perp \tilde\sigma'_\times + \sigma_\perp \sigma'_\times}
    {\sigma^2_\perp + \tilde\sigma^2 }
    \Delta^{\mu(\rho} h^{\sigma)}
    \rb \delta_\scB g_{\rho\sigma} \nn\\
  &\qquad
    - \lb \frac{sT}{\epsilon + p} \rb^2
    \lb \frac{- \tilde\sigma_\perp }{\sigma^2_\perp + \tilde\sigma^2 }
    + \frac{2\alpha \rho}{sT}
    \rb \Delta^{\mu\rho} h^\sigma \delta_\scB b_{\rho\sigma} \nn\\
  &\qquad
    + \lb \frac{sT}{\epsilon + p} \rb^2
    \frac{\sigma_\perp}{\sigma^2_\perp + \tilde\sigma^2 }
    \epsilon^{\mu\rho} h^\sigma
    \delta_\scB b_{\rho\sigma}~~.
\end{align}
Comparing it to the version obtained via the identification with string fluids
\begin{align}
  P^{\mu\nu} \delta_\scB A_\nu
  &= P^{\mu\nu} \dow_\nu \frac{\mu}{T} - \frac{1}{T} E^\mu \nn\\
  &= P^{\mu\nu} \dow_\nu \frac{\alpha}{T}
    + \frac{1}{2T} \epsilon^{\mu\nu\rho\sigma} u_\nu J_{\rho\sigma}
    + \mathcal{O}(\dow^2) \nn\\
  &= \frac{\alpha\rho}{\epsilon + p} \Delta^{\mu\sigma} h^\nu
  \delta_\scB b_{\sigma\nu}
    + \lb r'_\times \epsilon^{\mu\alpha} h^\beta
    - \tilde r'_\times \Delta^{\mu\alpha} h^\beta \rb \delta_\scB g_{\alpha\beta}
    + 
    \lb r_\perp \epsilon^{\mu\alpha} h^\beta
    - \tilde r_\perp \Delta^{\mu\alpha} h^\beta \rb \delta_\scB b_{\alpha\beta}
    \nn\\
  &\qquad
    + \half h^\mu
    \lb \tilde{\kappa}'_1 \Delta^{\alpha\beta}
    + \tilde{\kappa}'_2 h^\mu h^\nu \rb \delta_\scB g_{\mu\nu}
    + \half h^\mu
    r_\parallel \epsilon^{\alpha\beta} \delta_\scB b_{\alpha\beta}~~,
\end{align}
we can read out part of the non-hydrostatic map in \cref{eq:mapnonhydro}. For
the remaining part, we need to compare the non-hydrostatic energy-momentum
tensors in the two pictures. This is done trivially by taking the MHD
non-hydrostatic energy-momentum tensor, before the field redefinition, from
\cref{MHD-nhs-corrections}, substitute for the electric fields using
\cref{TheFinal_MuE_elimination}, and comparing it with the string fluid
expressions in \cref{string-nhs-corrections}. This finishes the mapping of all
the first-order transport coefficients presented in
\cref{sec:MHD-String-mapping}.

\subsection{Mapping magnetically dominated BCP to one-form superfluids}

The mapping from magnetically dominated bound-charge plasma to one-form
superfluids is considerably less involved because we do not have to eliminate
the chemical potential. Furthermore, as it turns out, we do not need to perform
a hydrodynamic field redefinition to map the two formulations. Firstly, we note
that the magnetic and electric fields in a magnetically dominated plasma are
given in terms of the electric limit of one-form superfluids discussed in
\cref{sec:electric} according to
\begin{align}
  B^{\mu}
  &= J^{\mu\nu} u_\nu \nn\\
  &= - q\, \zeta^\mu
    + \delta\rho\, h^\mu
    - n^\mu \nn\\
  &\qquad
    - h^{\mu} \lB \epsilon^{\alpha\beta\rho\sigma} u_\alpha h_\beta \dow_\rho u_\sigma
    \frac{\dow\beta}{\dow \varpi} 
    + h^\lambda \dow_\lambda T \frac{\dow\tilde\beta_1}{\dow \varpi}
    + h^\lambda \dow_\lambda \frac{\varpi}{T} \frac{\dow\tilde\beta_2}{\dow
    \varpi}
    - \nabla_\lambda \lb \tilde\beta_2 h^\lambda \rb \frac{1}{T}
    + \epsilon^{\alpha\beta\rho\sigma} u_\alpha h_\beta \dow_\rho h_\sigma
    \frac{\dow\tilde\beta_3}{\dow \varpi}
    \rB \nn\\
  &\qquad
    + \frac{\beta}{\varpi} \Delta^{\mu}{}_\nu \epsilon^{\nu\lambda\rho\sigma}
    u_\lambda \dow_\rho u_\sigma 
    - \frac{\tilde\beta_1}{\varpi} \Delta^{\mu\nu} \dow_{\nu} T 
    - \frac{\tilde\beta_2}{\varpi} \Delta^{\mu\nu} \dow_\nu \frac{\varpi}{T}
    \nn\\
  &\qquad
    + \Delta^{\mu}{}_\nu \epsilon^{\nu\lambda\rho\sigma}
    \lb \frac{T\tilde\beta_3}{\varpi^2}
    u_\lambda \dow_\rho \frac{\varpi h_\sigma}{T}
    + \frac1T\nabla_\rho \lb \frac{T\tilde\beta_3}{\varpi}
    u_\lambda h_\sigma \rb \rb
    + \mathcal{O}(\dow^2)~~, \nn\\
  E^\mu
  &= - \half \epsilon_{\mu\nu\rho\sigma} u^\nu J^{\rho\sigma} \nn\\
  &= - q_\times \zeta^\mu
  - \lb q'_\times \zeta^\mu \zeta^\nu + \bar q P^{\mu\nu} \rb \bar \zeta_\nu
    - R^\mu
    + h^\mu \delta s
    + \epsilon^{\mu\nu} m_\nu + \mathcal{O}(\dow^2)~~.
\end{align}
Unlike the MHD mapping in \cref{the-B-map}, the magnetic fields do get a
non-hydrostatic contribution in a magnetically dominated plasma. Note that for
$E^\mu$ to be $\mathcal{O}(\dow)$, we need to set $q_\times = 0$. Using the map
between free energy currents in the two formulations given in
\cref{FreeEMap_1SFBCP}, we can find a mapping for hydrostatic free-energy
densities according to
\begin{equation}
  \mathcal{N}_{\text{BCP}}
  = \mathcal{N}_{\text{1SF}}
  + B^\mu \zeta_\mu
  + E_\mu \bar\zeta^\mu~~.
\end{equation}
Plugging in the expressions for $B^\mu$ and $E^\mu$ from above, this trivially
leads to the hydrostatic sector mapping given in \cref{mBCP_hs_map}.

To map the respective non-hydrostatic sector transport coefficients, we need to
explicitly compare the constitutive relations for the energy-momentum tensor in
the two formulations, along with the map $M^{\mu\nu} = -
\half \epsilon^{\mu\nu\rho\sigma} \xi_{\rho\sigma}$. After an involved algebra,
we find
\begin{align}
  T^{\mu\nu}_{\text{BCP}}
  &= T^{\mu\nu}_{\text{1SF}}
    \nn\\
  &\qquad
    - \lb \delta f - \frac{\rho\delta\rho}{\dow\rho/\dow\varpi}
    - \delta\mathcal{F} \rb \Delta^{\mu\nu}
    - \lb \delta\tau + \varpi\delta\rho - \delta\mathcal{T} \rb h^{\mu}h^{\nu} \nn\\
  &\qquad
    - 2(\ell^{(\mu} - \varpi n^{(\mu} - \mathcal{L}^{(\mu}) h^{\nu)}
    - \lb t^{\mu\nu} - \mathcal{T}^{\mu\nu} \rb
    + \mathcal{O}(\dow^2) ~~, \nn\\
  M^{\mu\nu}
  &= - \half\epsilon^{\mu\nu\rho\sigma} \xi_{\rho\sigma} \nn\\
  &\qquad 
    + 2 \lb \delta\mathcal{R} - (\alpha'_{EB}B^2 + \alpha_{EE}) \delta s \rb u^{[\mu} h^{\nu]} 
    + \lb \delta\mathcal{S} + \frac{\delta \rho}{\dow\rho/\dow\varpi} \rb
    \epsilon^{\mu\nu} \nn\\
  &\qquad
    - 2 h^{[\mu} \lb \mathcal{M}^{\nu]}
    + \frac{\varpi}{\rho} \epsilon^{\nu]\sigma} n_{\sigma} \rb
    - 2 u^{[\mu} \lb \mathcal{N}^{\nu]} + \alpha_{EE} \epsilon^{\nu]\lambda} m_\lambda \rb
    + \mathcal{O}(\dow^2)~~.
\end{align}
Various non-hydrostatic corrections appearing here are defined in
\cref{eSF_nhs_corrections,mBCP_nhs_corrections}. Since for the map to work the
last two lines in both the expressions above must vanish, this trivially leads
to the mapping in the non-hydrostatic sector given in \cref{mBCP_nhs_map}.

\section{Comparison with the effective action approach}
\label{app:action}

In this appendix we perform a comparison between the work of
\cite{Glorioso:2018kcp} and the equilibrium partition function construction that
we provided in \cite{Armas:2018atq}. Additionally, we use the construction of
\cite{Glorioso:2018kcp} in order to generalise their results so as to obtain an
ideal order effective action for the one-form hydrodynamic theories of
sec.~\ref{sec:ordinary} (unbroken phase) and sec.~\ref{sec:super} (fully broken
phase).

Following \cite{Glorioso:2018wxw, Glorioso:2018kcp}, we introduce a ``fluid spacetime'' with coordinates $\sigma^a$. A
point on this spacetime represents a ``fluid element'' parametrised by
$\sigma^{i=1,2,3}$ at some choice of internal time $\sigma^0$. On this fluid
spacetime, we define the coordinate fields $x^\mu(\sigma)$ which represent the
physical spacetime coordinates of the fluid element. Under a spacetime
diffeomorphism $\chi^\mu(x)$, these fields transform as
\begin{equation}
  x^\mu(\sigma) \to x^\mu(\sigma) + \chi^\mu(x(\sigma))~~.
\end{equation}
When the fluid is charged under a U(1) zero-form symmetry, we also associate
with every fluid element a phase field $\phi(\sigma)$. In the case of a one-form
symmetry, we instead introduce a one-form phase $\varphi_a(\sigma)$ as in \cite{Glorioso:2018kcp}. These
phases do not transform under spacetime diffeomorphisms,\footnote{We can pushforward
  these phases onto the physical spacetime as $\phi(x) = \phi(\sigma(x))$ and
  $\varphi_\mu(x) = \frac{\dow \sigma^a(x)}{\dow x^\mu} \varphi_a(\sigma(x))$,
  which have the expected transformation properties
  $\delta_\scX \phi(x) = \lie_\chi \phi(x) - \Lambda^\chi(x)$ and
  $\delta_\scX \varphi_\mu(x) = \lie_\chi \varphi_\mu(x) -
  \Lambda^\chi_\mu(x)$. In this case, the field $\varphi_\mu$, already introduced in \cite{Armas:2018atq}, coincides with that defined in \eqref{eq:introvarphimu}.}  but shift under the respective gauge transformations
$\Lambda^\chi(x)$ and $\Lambda^\chi_\mu(x)$
\begin{equation}
  \phi(\sigma) \to \phi(\sigma) - \Lambda^\chi(x(\sigma)), \qquad
  \varphi_a(\sigma)  \to \varphi_a(\sigma)
  - \frac{\dow x^\mu(\sigma)}{\dow \sigma^a} \Lambda^\chi_\mu(x(\sigma))~~.
\end{equation}
The fields $x^\mu(\sigma)$ together with $\phi(\sigma)$, or $\varphi_a(\sigma)$ for
the one-form case, form the effective dynamical fields of hydrodynamics.  Given
the background fields $g_{\mu\nu}(x)$, $A_\mu(x)$, and $b_{\mu\nu}(x)$ on the
physical spacetime, we can define their pullbacks onto the fluid spacetime as
\begin{align}\label{effAction_invariant_fields}
  h_{ab}(\sigma)
  &= \frac{\dow x^\mu(\sigma)}{\dow \sigma^a}
  \frac{\dow x^\nu(\sigma)}{\dow \sigma^b} g_{\mu\nu}(x(\sigma))~~, \nn\\
  B_a(\sigma)
  &= \frac{\dow x^\mu(\sigma)}{\dow \sigma^a} A_\mu(x(\sigma))
    + \frac{\dow \phi(\sigma)}{\dow \sigma^a}~~, \nn\\
  B_{ab}(\sigma)
  &= \frac{\dow x^\mu(\sigma)}{\dow \sigma^a}
    \frac{\dow x^\nu(\sigma)}{\dow \sigma^b} b_{\mu\nu}(x(\sigma))
    + \frac{\dow\varphi_b(\sigma)}{\dow \sigma^a}
    - \frac{\dow\varphi_a(\sigma)}{\dow \sigma^b}~~.
\end{align}
These fields have been defined such that they are invariant under the symmetry transformations of the
physical spacetime. In fact, they constitute the most general invariants made
out of dynamical and background fields.

Given these elements, we wish to construct a Wilsonian effective action for hydrodynamics involving
the fields in \cref{effAction_invariant_fields}, with certain symmetries imposed
on the fluid spacetime, so that we can recover the hydrodynamic dynamical
equations via a variational principle~\cite{Glorioso:2018wxw}. The physical
picture to keep in mind is that every distinct fluid element, parametrised by
$\sigma^i$, is evolving along the internal time $\sigma^0$. We expect the
hydrodynamic description to be invariant under an arbitrary relabelling of the
fluid elements and the choice of internal time for each fluid element, leading
to the symmetries
\begin{equation}\label{fluidSTredef}
  \sigma^a \to \sigma^a + f^a(\vec\sigma)~~.
\end{equation}
Note that we are not allowing for a time-dependent redefinition of $\sigma^a$,
since we require each fluid element and its choice of
internal time to stay the same as it moves through time. The transformations
\cref{fluidSTredef} are the most general fluid spacetime diffeomorphisms which
leave the internal time vector $\dow/\dow\sigma^0$ invariant.

In addition, we allow each fluid element to independently choose the associated
U(1) phase, leading to the shift symmetry
\begin{equation}\label{phase_redef}
  \phi(\sigma) \to \phi(\sigma) + \lambda(\vec\sigma)~~, ~~
  \varphi_a(\sigma) \to \varphi_a(\sigma) + \lambda_a(\vec\sigma)~~.
\end{equation}
Note that we are also requiring the choice of phase to remain the same as the
fluid element moves through time. We expect the symmetries \bref{phase_redef} to
hold when the underlying U(1) symmetry is not spontaneously broken. To motivate
this, let us consider the zero-form case first. At each point $p = (\sigma^a_p)$
in the fluid spacetime, we can define a charged operator
\begin{equation}\label{eq:0form-vertex-operator}
  V_p = \exp(i\phi(\sigma_p))~~.
\end{equation}
Under the shift \bref{phase_redef}, these operators admit a phase rotation
\begin{equation}
  V_p \to \exp(i\lambda(\vec\sigma_p))\, V_p~~ ,
\end{equation}
which is independent for every fluid element, but remains fixed as the charged
operator moves through time. When the symmetry is spontaneously broken, the
system picks a random preferred phase in the ground state and the respective
shift symmetry in \cref{phase_redef} should be dropped. In this case, the phase
pushforward onto the physical spacetime $\phi(x) = \phi(\sigma(x))$ acts as the
Goldstone mode of the broken symmetry, and we are led to the physics of
zero-form superfluid dynamics.

In the one-form case, on the other hand, the charged operators are defined over
non-local ``strings'' of fluid elements. Let us consider a space-like curve $C$
in the fluid spacetime defined in terms of an internal length parameter $\ell$
as $\sigma^a = \sigma^a_C(\ell)$. We can then define the operator
\begin{equation}\label{eq:1form-vertex-operator}
  V_C = \exp\lb i\int_C \varphi_a(\sigma) \,\df \sigma^a \rb
  = \exp\lb i\int \varphi_a(\sigma_C(\ell))
  \frac{\df\sigma_C^a(\ell)}{\df\ell} \,\df \ell \rb~~.
\end{equation}
Under the shift \bref{phase_redef}, this charged operator acquires a phase rotation given by
operator
\begin{equation}
  V_C \to \exp\lb i\int \lambda_a(\vec\sigma_C(\ell))
  \frac{\df\sigma_C^a(\ell)}{\df\ell} \,\df \ell \rb \, V_C~~,
\end{equation}
which is independent for every string of fluid elements, but remains fixed if a
string moves uniformly in time: $\sigma^0_C(\ell) \to \sigma^0_C(\ell) + \tau$,
where $\tau$ is independent of $\ell$. Using the analogy with the zero-form
case, we understand that when the shift symmetry \bref{phase_redef} is dropped,
the system picks up a preferred one-form phase in its ground state spontaneously
breaking the symmetry. The pushforward of the one-form phase
$\varphi_\mu(x) = \frac{\dow \sigma^a(x)}{\dow x^\mu} \varphi_a(\sigma(x))$ can
be identified with the Goldstone mode of this broken symmetry. Interestingly, in
this case there is another choice available to us. We can require the choice of
phase to be fixed under a non-uniform movement of the string in time:
$\sigma^0_C(\ell) \to \sigma^0_C(\ell) + \tau(\ell)$, which implies dropping the
time component of the one-form shift in \cref{phase_redef} setting
$\lambda_0(\sigma) = 0$. Since the $\varphi_0(\sigma)$ component of the phase does
not admit any redefinition in this case, we can interpret its pushforward onto the
physical spacetime $\varphi(x) = \varphi_0(\sigma(x))$ as a scalar
Goldstone. This is the partial symmetry breaking of one-form hydrodynamics
eluded to in \cref{sec:stringfluids}.\footnote{The effective action framework
  of~\cite{Glorioso:2018kcp} for MHD/string fluids deals with this partially
  broken picture of one-form hydrodynamics where $\lambda_0(\sigma) = 0$. The
  authors rightly note that the pullback of the full one-form phase
  $\varphi_\mu(x)$ is not a Goldstone in this picture, as we see that the
  one-form symmetry is only partially broken. However, the authors do not
  identify the pullback of the time component $\varphi(x)$ as a Goldstone mode
  either. Note that v1 of~\cite{Glorioso:2018kcp} on arXiv has a typo in
  equation (2.18) as we confirmed with the authors: the shift symmetry is only
  imposed in the spatial directions.}

Having identified the dynamical degrees of freedom and symmetries, one can construct
the most generic hydrodynamic effective action arranged in
a derivative expansion leading to a particular subsector of non-dissipative
constitutive relations. We do not repeat this exercise here and we encourage interested readers
to consult the relevant papers
such as~\cite{Glorioso:2018wxw,Glorioso:2018kcp}. However, to make contact with the
hydrodynamic formulation used in the bulk of this paper, it is instructive to
map the dynamical degrees of freedom in the two pictures. Starting with the
symmetry unbroken phase, we can identify the hydrodynamic fields
$\scB = (\beta^\mu, \Lambda^\beta)$ or $\scB = (\beta^\mu, \Lambda^\beta_\mu)$
introduced around \cref{eq:hydro0fields} and \cref{eq:mumu} respectively as
\begin{equation}
  \beta^\mu(x) = \frac{\dow x^\mu(\sigma)}{\dow\sigma^0}\bigg|_{\sigma=\sigma(x)}, \qquad
  \Lambda^\beta(x) =
  \frac{\dow\phi(\sigma)}{\dow\sigma^0}\bigg|_{\sigma=\sigma(x)},
  \qquad
  \Lambda^\beta_\mu(x) = \frac{\dow\sigma^a(x)}{\dow x^\mu}
  \frac{\dow\varphi_a(\sigma)}{\dow\sigma^0}\bigg|_{\sigma=\sigma(x)}.
\end{equation}
These are invariant under the fluid spacetime symmetries in
\cref{fluidSTredef,phase_redef}. In terms of the conventional fields, we
equivalently have
\begin{gather}
  u^\mu(x) = \frac{1}{\sqrt{-h_{00}(\sigma(x))}} \frac{\dow
    x^\mu(\sigma)}{\dow\sigma^0}\bigg|_{\sigma=\sigma(x)}, \qquad
  T(x) = \frac{1}{\sqrt{-h_{00}(\sigma(x))}}~~, \nn\\
  \mu(x) = \frac{B_0(\sigma(x))}{\sqrt{-h_{00}(\sigma(x))}}~~, \qquad
  \mu_\mu(x) = 
  \frac{\frac{\dow\sigma^a(x)}{\dow x^\mu} B_{0a}(\sigma(x))
  + \frac{\dow}{\dow x^\mu} \varphi_0(\sigma(x))}{\sqrt{-h_{00}(\sigma(x))}}~~.
\end{gather}
As noted in \cref{sec:setup}, the one-form chemical potential $\mu_\mu(x)$ is
not gauge-invariant.

When the symmetry is spontaneously broken and \cref{phase_redef} is relaxed, we
can identify the respective Goldstone modes and superfluid velocity as
additional fluid spacetime invariants
\begin{gather}
  \phi(x) = \phi(\sigma(x))~~,~~
  \xi_{\mu}(x) = \frac{\dow\sigma^a(x)}{\dow x^\mu} B_a(\sigma(x))~~, \nn\\
  \varphi_\mu(x) = \frac{\dow \sigma^a(x)}{\dow x^\mu} \varphi_a(\sigma(x))~~,~~
  \xi_{\mu\nu}(x) = \frac{\dow\sigma^a(x)}{\dow x^\mu}
  \frac{\dow\sigma^b(x)}{\dow x^\nu}B_{ab}(\sigma(x))~~.
\end{gather}
Interestingly, the respective Josephson equations $u^\mu \xi_\mu = \mu$ and
$u^\nu\xi_{\nu\mu} = \mu_\mu - T \dow_\mu(\beta^\nu\varphi_\nu)$ given in
\cref{sec:Josephson} are automatically satisfied. Finally, in the case when the
one-form symmetry is only partially broken, the respective scalar Goldstone and
string fluid variables can be read out as
\begin{equation}
  \varphi(x) = \varphi_0(\sigma(x))~~,~~
  \varpi h_\mu = \frac{\dow\sigma^a(x)}{\dow x^\mu}
  \frac{ B_{0a}(\sigma(x))}{\sqrt{-h_{00}(\sigma(x))}}~~.
\end{equation}

\subsubsection*{Order parameter}
The question of whether a global symmetry is spontaneously broken or unbroken can
be articulated in terms of an order parameter charged under the symmetry. In the
zero-form case, such an order parameter is provided by the expectation value of
the vertex operator \cref{eq:0form-vertex-operator}, i.e.
\begin{equation}
  \left\langle \exp(i\phi(\sigma_p)) \right\rangle~~.
\end{equation}
If this happens to be non-zero when computed within the effective action
framework of hydrodynamics, we understand that the symmetry is spontaneously
broken and we are in the superfluid phase, otherwise the symmetry is unbroken and we
are in the ordinary fluid phase. A similar construction can be extended to
one-form symmetries using \cref{eq:1form-vertex-operator} to obtain an order
parameter
\begin{equation}
  \left\langle \exp\lb i\int_C \varphi_a(\sigma) \,\df \sigma^a \rb \right\rangle~~.
\end{equation}
If, for large spacelike loops, the expectation value scales as the perimeter of
the loop we are in the symmetry broken phase, otherwise we are in the symmetry
unbroken or partially broken phase. This order parameter will not distinguish
between the partially broken and unbroken phases of one-form symmetry. If we were
at equilibrium, we could obtain a plausible operator that will make such distinction by
integrating over the Euclidean time circle
\begin{equation}
  \left\langle \exp\lb -\int_{S^1} \varphi_a(\vec\sigma) \,\df \sigma_E^a \rb
  \right\rangle
  = \left\langle \exp\lb \frac{i}{T_0} \varphi_0(\vec\sigma) \rb \right\rangle~~.
\end{equation}
Generically, there is no notion of preferred time outside thermal equilibrium to
define such an order parameter but within the regime of hydrodynamics, we can
use the fluid velocity to define this operator\footnote{The effective action
  construction of~\cite{Glorioso:2018kcp} contains a gauge symmetry
  $\varphi_a(\sigma) \to \varphi_a(\sigma) + \dow_a \Lambda(\sigma)$, which
  doesn't leave the out-of-equilibrium order parameter in
  \cref{eq:noneqb_order_param} invariant. Arriving at a correct gauge-invariant
  order parameter might need some more work which we leave for future
  considerations.  We thank P. Glorioso and D. T. Son for pointing this out to
  us.}
\begin{equation}\label{eq:noneqb_order_param}
  \left\langle \exp\lb \frac{i}{T} u^\mu \varphi_\mu(\sigma) \rb \right\rangle
  = \left\langle \exp\lb i \varphi(\sigma) \rb \right\rangle~~.
\end{equation}
Whether or not this operator is the required order parameter can be settled by
computing it within the effective field theory outlined in this appendix. We
leave it here as a speculative note and plan to come back to this question in
the future.

\section{Discrete symmetries}
\label{app:CPT}

If the physical system in question is invariant under certain discrete
symmetries, like chirality (parity) or CPT, on top of the continuous Poincar\'e
and zero/one-form symmetries, they can be used to further constraint the number
of allowed transport coefficients. The formulations of one-form hydrodynamics
and hot electromagnetism involves distinct sets of conserved quantities and are
mapped to each other via a Hodge duality operation, therefore discrete
symmetries in the respective pictures do not map to each other
trivially. Already in \cref{sec:stringfluidcorrections}, we discussed the action
of CP symmetry in string fluids, later noting in \cref{sec:MHD-String} that
CP-preserving string fluids map to parity or $\rmP_{\text{EM}}$-preserving
sector of MHD. Therefore, we devote this appendix to a more careful treatment of
discrete symmetries in hot electromagnetism and one-form hydrodynamics.

\begin{table}[t]
  \begin{subtable}[t]{0.31\textwidth}
    \flushright
    \begin{tabular}[t]{ccccc}
      \toprule
      & C & P & T & CPT \\
      \midrule
      $T^{tt}$, $g_{tt}$ & $+$ & $+$ & $+$ & $+$ \\
      $T^{ti}$, $g_{ti}$ & $+$ & $-$ & $-$ & $+$ \\
      $T^{ij}$, $g_{ij}$ & $+$ & $+$ & $+$ & $+$ \\
      \midrule
      $J^t$, $A_t$      & $-$ & $+$ & $+$ & $-$ \\
      $J^i$, $A_i$      & $-$ & $-$ & $-$ & $-$ \\
      \midrule
      $J^{ti}$, $b_{ti}$ & $-$ & $-$ & $+$ & $+$ \\
      $J^{ij}$, $b_{ij}$ & $-$ & $+$ & $-$ & $+$ \\
      \bottomrule
    \end{tabular}
  \end{subtable}
  \begin{subtable}[t]{0.3\textwidth}
    \centering 
    \begin{tabular}[t]{ccccc}
      \toprule
      & C & P & T & CPT \\
      \midrule
      $u^t$ & $+$ & $+$ & $+$ & $+$ \\
      $u^i$ & $+$ & $-$ & $-$ & $+$ \\
      $T$   & $+$ & $+$ & $+$ & $+$ \\
      \midrule
      $\mu$ & $-$ & $+$ & $+$ & $-$ \\
      $\phi$  & $-$ & $+$ & $-$ & $+$ \\ 
      \midrule
      $\mu_t$ & $-$ & $+$ & $-$ & $+$ \\
      $\mu_i$   & $-$ & $-$ & $+$ & $+$ \\
      $\phi_t$, $\varphi$ & $-$ & $+$ & $+$ & $-$ \\
      $\phi_i$ & $-$ & $-$ & $-$ & $-$ \\
      \bottomrule
    \end{tabular}
  \end{subtable}
  \begin{subtable}[t]{0.36\textwidth}
    \flushleft
    \begin{tabular}[t]{ccccc}
      \toprule
      & C & P & T & CPT \\
      \midrule
      $\xi_t$ & $-$ & $+$ & $+$ & $-$ \\
      $\xi_i$ & $-$ & $-$ & $-$ & $-$ \\
      $F_{ti}$, $E_i$, $B^t$  & $-$ & $-$ & $+$ & $+$ \\
      $F_{ij}$, $B^i$, $E^t$ & $-$ & $+$ & $-$ & $+$ \\
      \midrule
      $\varpi$ & $+$ & $+$ & $+$ & $+$ \\
      $h_{i}$, $\xi_{ti}$, $\zeta_i$, $\bar\zeta^t$  & $-$ & $-$ & $+$ & $+$\\
      $\xi_{ij}$, $\bar\zeta^i$, $h_t$, $\zeta_t$ & $-$ & $+$ & $-$ & $+$\\
      $H_{tij}$ & $-$ & $+$ & $+$ & $-$ \\
      $H_{ijk}$ & $-$ & $-$ & $-$ & $-$ \\
      \bottomrule
    \end{tabular}
  \end{subtable}
  \caption{Transformation properties of various quantities under the discrete
    symmetries C, P, and T. The first table summarises properties of conserved
    currents and the associated sources, the second table of dynamical fields,
    while the third table of various derived quantities.}
  \label{tab:CPT}
\end{table}

In order to do so, the first step is to define the action of discrete symmetries on the field
content. We introduce three operations: charge conjugation C, parity P, and
time-reversal T. Their action on the conserved currents and field content is
summarised in \cref{tab:CPT}. Note that $h^\mu$ transforms as a vector under
parity, which under the duality operation gets mapped to an axial-vector
$B^\mu$. Therefore, on the electromagnetism side, the parity operation is
actually defined in terms of the one-form discrete symmetries as
$\text{P}_{\text{EM}} = \text{CP}$. The holds for the time-reversal
operator, and we find
\begin{equation}
  \text{C}_{\text{EM}} = \text{C}~~, \qquad
  \text{P}_{\text{EM}} = \text{CP}~~, \qquad
  \text{T}_{\text{EM}} = \text{CT}~~.
\end{equation}
The charge conjugation operator, of course, is the same in both 
pictures. Interestingly, the full $(\text{CPT})_{\text{EM}}$ is mapped to
CPT in the one-form picture.

With \cref{tab:CPT} in place, we can easily work out the nature of various
transport coefficients under CP (i.e. $\text{P}_{\text{EM}}$) and CPT. Firstly,
all the transport coefficients in ordinary one-form hydrodynamics are
CP-even. For string fluids, we have already discussed the CP properties of
various transport coefficients in \cref{sec:stringfluidcorrections}. Lastly, for
generic one-form superfluids, since the transport coefficients can arbitrarily
depend on a zero-derivative CP-odd scalar ($\zeta\cdot\bar\zeta$), no terms in
the constitutive relations have a definite CP behaviour.

As for CPT, it is easy to check that all zero derivative tensor structures in
any phase of one-form hydrodynamics are CPT-even. Consequently, all
one-derivative transport coefficients are CPT-odd. Before we draw any
conclusions from this result, it is worth noting that the shear and bulk viscosity
terms in neutral relativistic hydrodynamics are CPT-odd as well (or equivalently
PT-odd due to neutrality). This is not surprising due to the dissipative nature
of these coefficients. However, this CPT is distinct from the ``microscopic''
CPT that is implemented, not at the level of the constitutive relations, but
more fundamentally at the level of an effective action, hydrostatic partition
function, or correlation functions (see for
instance~\cite{Glorioso:2018wxw}). In the hydrostatic sector, microscopic CPT-invariance
requires that all the CPT-violating terms in the hydrostatic partition function
vanish. Since all the hydrostatic one-derivative scalars are CPT-odd, all
the hydrostatic transport coefficients are turned off by requiring microscopic CPT-invariance in all
the phases of one-form hydrodynamics. Due to the map between and MHD and string fluids in eqs.~\eqref{eq:maphydro} and \ref{eq:mapnonhydro},
microscopic CPT-invariance implies, in particular, that the chemical potential $\mu_0$ in MHD must vanish.
In the non-hydrostatic sector, on the other
hand, microscopic CPT can be implemented using Onsager's relations. For instance, for
string fluids, these constraints have been worked out in
\cref{sec:Kubostring}. We leave a more detailed analysis of microscopic CPT in
generic one-form superfluids to a future work.

\makereferences
\providecommand{\href}[2]{#2}\begingroup\raggedright\endgroup


\begin{thebibliography}{10}

\bibitem{davidson2001introduction}
P.~Davidson, \emph{An Introduction to Magnetohydrodynamics}.
\newblock Cambridge Texts in Applied Mathematics. Cambridge University Press,
  2001.

\bibitem{goedbloed2004principles}
J.~Goedbloed, J.~Goedbloed and S.~Poedts, \emph{Principles of
  Magnetohydrodynamics: With Applications to Laboratory and Astrophysical
  Plasmas}.
\newblock Cambridge University Press, 2004.

\bibitem{Hernandez:2017mch}
J.~Hernandez and P.~Kovtun, \emph{{Relativistic magnetohydrodynamics}},
  \href{http://dx.doi.org/10.1007/JHEP05(2017)001}{\emph{JHEP} {\bfseries 05}
  (2017) 001}, [\href{https://arxiv.org/abs/1703.08757}{{\ttfamily
  1703.08757}}].

\bibitem{Bhattacharya:2011tra}
J.~Bhattacharya, S.~Bhattacharyya, S.~Minwalla and A.~Yarom, \emph{{A Theory of
  first order dissipative superfluid dynamics}},
  \href{http://dx.doi.org/10.1007/JHEP05(2014)147}{\emph{JHEP} {\bfseries 05}
  (2014) 147}, [\href{https://arxiv.org/abs/1105.3733}{{\ttfamily 1105.3733}}].

\bibitem{Bhattacharyya:2012nq}
S.~Bhattacharyya, \emph{{Constraints on the second order transport coefficients
  of an uncharged fluid}},
  \href{http://dx.doi.org/10.1007/JHEP07(2012)104}{\emph{JHEP} {\bfseries 07}
  (2012) 104}, [\href{https://arxiv.org/abs/1201.4654}{{\ttfamily 1201.4654}}].

\bibitem{Kovtun:2012rj}
P.~Kovtun, \emph{{Lectures on hydrodynamic fluctuations in relativistic
  theories}}, \href{http://dx.doi.org/10.1088/1751-8113/45/47/473001}{\emph{J.
  Phys.} {\bfseries A45} (2012) 473001},
  [\href{https://arxiv.org/abs/1205.5040}{{\ttfamily 1205.5040}}].

\bibitem{Jensen:2012jh}
K.~Jensen, M.~Kaminski, P.~Kovtun, R.~Meyer, A.~Ritz and A.~Yarom,
  \emph{{Towards hydrodynamics without an entropy current}},
  \href{http://dx.doi.org/10.1103/PhysRevLett.109.101601}{\emph{Phys. Rev.
  Lett.} {\bfseries 109} (2012) 101601},
  [\href{https://arxiv.org/abs/1203.3556}{{\ttfamily 1203.3556}}].

\bibitem{Banerjee:2012iz}
N.~Banerjee, J.~Bhattacharya, S.~Bhattacharyya, S.~Jain, S.~Minwalla and
  T.~Sharma, \emph{{Constraints on Fluid Dynamics from Equilibrium Partition
  Functions}}, \href{http://dx.doi.org/10.1007/JHEP09(2012)046}{\emph{JHEP}
  {\bfseries 09} (2012) 046},
  [\href{https://arxiv.org/abs/1203.3544}{{\ttfamily 1203.3544}}].

\bibitem{Haehl:2014zda}
F.~M. Haehl, R.~Loganayagam and M.~Rangamani, \emph{{The eightfold way to
  dissipation}},
  \href{http://dx.doi.org/10.1103/PhysRevLett.114.201601}{\emph{Phys. Rev.
  Lett.} {\bfseries 114} (2015) 201601},
  [\href{https://arxiv.org/abs/1412.1090}{{\ttfamily 1412.1090}}].

\bibitem{Haehl:2015pja}
F.~M. Haehl, R.~Loganayagam and M.~Rangamani, \emph{{Adiabatic hydrodynamics:
  The eightfold way to dissipation}},
  \href{http://dx.doi.org/10.1007/JHEP05(2015)060}{\emph{JHEP} {\bfseries 05}
  (2015) 060}, [\href{https://arxiv.org/abs/1502.00636}{{\ttfamily
  1502.00636}}].

\bibitem{Bhattacharyya:2008jc}
S.~Bhattacharyya, V.~E. Hubeny, S.~Minwalla and M.~Rangamani, \emph{{Nonlinear
  Fluid Dynamics from Gravity}},
  \href{http://dx.doi.org/10.1088/1126-6708/2008/02/045}{\emph{JHEP} {\bfseries
  02} (2008) 045}, [\href{https://arxiv.org/abs/0712.2456}{{\ttfamily
  0712.2456}}].

\bibitem{Rangamani:2009xk}
M.~Rangamani, \emph{{Gravity and Hydrodynamics: Lectures on the fluid-gravity
  correspondence}},
  \href{http://dx.doi.org/10.1088/0264-9381/26/22/224003}{\emph{Class. Quant.
  Grav.} {\bfseries 26} (2009) 224003},
  [\href{https://arxiv.org/abs/0905.4352}{{\ttfamily 0905.4352}}].

\bibitem{Glorioso:2017lcn}
P.~Glorioso, H.~Liu and S.~Rajagopal, \emph{{Global Anomalies, Discrete
  Symmetries, and Hydrodynamic Effective Actions}},
  \href{http://dx.doi.org/10.1007/JHEP01(2019)043}{\emph{JHEP} {\bfseries 01}
  (2019) 043}, [\href{https://arxiv.org/abs/1710.03768}{{\ttfamily
  1710.03768}}].

\bibitem{Haehl:2018lcu}
F.~M. Haehl, R.~Loganayagam and M.~Rangamani, \emph{{Effective Action for
  Relativistic Hydrodynamics: Fluctuations, Dissipation, and Entropy Inflow}},
  \href{http://dx.doi.org/10.1007/JHEP10(2018)194}{\emph{JHEP} {\bfseries 10}
  (2018) 194}, [\href{https://arxiv.org/abs/1803.11155}{{\ttfamily
  1803.11155}}].

\bibitem{Jensen:2018hse}
K.~Jensen, R.~Marjieh, N.~Pinzani-Fokeeva and A.~Yarom, \emph{{A panoply of
  Schwinger-Keldysh transport}},
  \href{http://dx.doi.org/10.21468/SciPostPhys.5.5.053}{\emph{SciPost Phys.}
  {\bfseries 5} (2018) 053},
  [\href{https://arxiv.org/abs/1804.04654}{{\ttfamily 1804.04654}}].

\bibitem{Armas:2015ssd}
J.~Armas, J.~Bhattacharya and N.~Kundu, \emph{{Surface transport in
  plasma-balls}}, \href{http://dx.doi.org/10.1007/JHEP06(2016)015}{\emph{JHEP}
  {\bfseries 06} (2016) 015},
  [\href{https://arxiv.org/abs/1512.08514}{{\ttfamily 1512.08514}}].

\bibitem{Armas:2016xxg}
J.~Armas, J.~Bhattacharya, A.~Jain and N.~Kundu, \emph{{On the surface of
  superfluids}}, \href{http://dx.doi.org/10.1007/JHEP06(2017)090}{\emph{JHEP}
  {\bfseries 06} (2017) 090},
  [\href{https://arxiv.org/abs/1612.08088}{{\ttfamily 1612.08088}}].

\bibitem{Jensen:2014ama}
K.~Jensen, \emph{{Aspects of hot Galilean field theory}},
  \href{http://dx.doi.org/10.1007/JHEP04(2015)123}{\emph{JHEP} {\bfseries 04}
  (2015) 123}, [\href{https://arxiv.org/abs/1411.7024}{{\ttfamily 1411.7024}}].

\bibitem{Banerjee:2015hra}
N.~Banerjee, S.~Dutta and A.~Jain, \emph{{Null Fluids - A New Viewpoint of
  Galilean Fluids}},
  \href{http://dx.doi.org/10.1103/PhysRevD.93.105020}{\emph{Phys. Rev.}
  {\bfseries D93} (2016) 105020},
  [\href{https://arxiv.org/abs/1509.04718}{{\ttfamily 1509.04718}}].

\bibitem{Schubring:2014iwa}
D.~Schubring, \emph{{Dissipative String Fluids}},
  \href{http://dx.doi.org/10.1103/PhysRevD.91.043518}{\emph{Phys. Rev.}
  {\bfseries D91} (2015) 043518},
  [\href{https://arxiv.org/abs/1412.3135}{{\ttfamily 1412.3135}}].

\bibitem{Grozdanov:2016tdf}
S.~Grozdanov, D.~M. Hofman and N.~Iqbal, \emph{{Generalized global symmetries
  and dissipative magnetohydrodynamics}},
  \href{http://dx.doi.org/10.1103/PhysRevD.95.096003}{\emph{Phys. Rev.}
  {\bfseries D95} (2017) 096003},
  [\href{https://arxiv.org/abs/1610.07392}{{\ttfamily 1610.07392}}].

\bibitem{Armas:2018ibg}
J.~Armas, J.~Gath, A.~Jain and A.~V. Pedersen, \emph{{Dissipative hydrodynamics
  with higher-form symmetry}},
  \href{http://dx.doi.org/10.1007/JHEP05(2018)192}{\emph{JHEP} {\bfseries 05}
  (2018) 192}, [\href{https://arxiv.org/abs/1803.00991}{{\ttfamily
  1803.00991}}].

\bibitem{Grozdanov:2018ewh}
S.~Grozdanov and N.~Poovuttikul, \emph{{Generalized global symmetries in states
  with dynamical defects: The case of the transverse sound in field theory and
  holography}}, \href{http://dx.doi.org/10.1103/PhysRevD.97.106005}{\emph{Phys.
  Rev.} {\bfseries D97} (2018) 106005},
  [\href{https://arxiv.org/abs/1801.03199}{{\ttfamily 1801.03199}}].

\bibitem{Armas:2018atq}
J.~Armas and A.~Jain, \emph{{Magnetohydrodynamics as superfluidity}},
  \href{http://dx.doi.org/10.1103/PhysRevLett.122.141603}{\emph{Phys. Rev.
  Lett.} {\bfseries 122} (2019) 141603},
  [\href{https://arxiv.org/abs/1808.01939}{{\ttfamily 1808.01939}}].

\bibitem{Grozdanov:2018fic}
S.~Grozdanov, A.~Lucas and N.~Poovuttikul, \emph{{Holography and hydrodynamics
  with weakly broken symmetries}},
  \href{http://dx.doi.org/10.1103/PhysRevD.99.086012}{\emph{Phys. Rev.}
  {\bfseries D99} (2019) 086012},
  [\href{https://arxiv.org/abs/1810.10016}{{\ttfamily 1810.10016}}].

\bibitem{Gaiotto:2014kfa}
D.~Gaiotto, A.~Kapustin, N.~Seiberg and B.~Willett, \emph{{Generalized Global
  Symmetries}}, \href{http://dx.doi.org/10.1007/JHEP02(2015)172}{\emph{JHEP}
  {\bfseries 02} (2015) 172},
  [\href{https://arxiv.org/abs/1412.5148}{{\ttfamily 1412.5148}}].

\bibitem{Gammie:2003rj}
C.~F. Gammie, J.~C. McKinney and G.~Toth, \emph{{HARM: A Numerical scheme for
  general relativistic magnetohydrodynamics}},
  \href{http://dx.doi.org/10.1086/374594}{\emph{Astrophys. J.} {\bfseries 589}
  (2003) 444--457}, [\href{https://arxiv.org/abs/astro-ph/0301509}{{\ttfamily
  astro-ph/0301509}}].

\bibitem{Lake:2018dqm}
E.~Lake, \emph{{Higher-form symmetries and spontaneous symmetry breaking}},
  \href{https://arxiv.org/abs/1802.07747}{{\ttfamily 1802.07747}}.

\bibitem{Hofman:2018lfz}
D.~M. Hofman and N.~Iqbal, \emph{{Goldstone modes and photonization for higher
  form symmetries}},
  \href{http://dx.doi.org/10.21468/SciPostPhys.6.1.006}{\emph{SciPost Phys.}
  {\bfseries 6} (2019) 006},
  [\href{https://arxiv.org/abs/1802.09512}{{\ttfamily 1802.09512}}].

\bibitem{Glorioso:2018kcp}
P.~Glorioso and D.~T. Son, \emph{{Effective field theory of
  magnetohydrodynamics from generalized global symmetries}},
  \href{https://arxiv.org/abs/1811.04879}{{\ttfamily 1811.04879}}.

\bibitem{landau2013fluid}
L.~Landau and E.~Lifshitz, \emph{Fluid Mechanics}.
\newblock No.~v. 6. Elsevier Science, 2013.

\bibitem{Loganayagam:2011mu}
R.~Loganayagam, \emph{{Anomaly Induced Transport in Arbitrary Dimensions}},
  \href{https://arxiv.org/abs/1106.0277}{{\ttfamily 1106.0277}}.

\bibitem{Jain:2016rlz}
A.~Jain, \emph{{Theory of non-Abelian superfluid dynamics}},
  \href{http://dx.doi.org/10.1103/PhysRevD.95.121701}{\emph{Phys. Rev.}
  {\bfseries D95} (2017) 121701},
  [\href{https://arxiv.org/abs/1610.05797}{{\ttfamily 1610.05797}}].

\bibitem{Jain:2018jxj}
A.~Jain, \emph{{A universal framework for hydrodynamics}}.
\newblock PhD thesis, Durham U., CPT, 2018-06-06.

\bibitem{Caldarelli:2010xz}
M.~M. Caldarelli, R.~Emparan and B.~Van~Pol, \emph{{Higher-dimensional Rotating
  Charged Black Holes}},
  \href{http://dx.doi.org/10.1007/JHEP04(2011)013}{\emph{JHEP} {\bfseries 04}
  (2011) 013}, [\href{https://arxiv.org/abs/1012.4517}{{\ttfamily 1012.4517}}].

\bibitem{Son:2007vk}
D.~T. Son and A.~O. Starinets, \emph{{Viscosity, Black Holes, and Quantum Field
  Theory}},
  \href{http://dx.doi.org/10.1146/annurev.nucl.57.090506.123120}{\emph{Ann.
  Rev. Nucl. Part. Sci.} {\bfseries 57} (2007) 95--118},
  [\href{https://arxiv.org/abs/0704.0240}{{\ttfamily 0704.0240}}].

\bibitem{Grozdanov:2017kyl}
S.~Grozdanov and N.~Poovuttikul, \emph{{Generalised global symmetries and
  magnetohydrodynamic waves in a strongly interacting holographic plasma}},
  \href{http://dx.doi.org/10.1007/JHEP04(2019)141}{\emph{JHEP} {\bfseries 04}
  (2019) 141}, [\href{https://arxiv.org/abs/1707.04182}{{\ttfamily
  1707.04182}}].

\bibitem{Kovtun:2016lfw}
P.~Kovtun, \emph{{Thermodynamics of polarized relativistic matter}},
  \href{http://dx.doi.org/10.1007/JHEP07(2016)028}{\emph{JHEP} {\bfseries 07}
  (2016) 028}, [\href{https://arxiv.org/abs/1606.01226}{{\ttfamily
  1606.01226}}].

\bibitem{Armas:2012ac}
J.~Armas, J.~Gath and N.~A. Obers, \emph{{Black Branes as Piezoelectrics}},
  \href{http://dx.doi.org/10.1103/PhysRevLett.109.241101}{\emph{Phys. Rev.
  Lett.} {\bfseries 109} (2012) 241101},
  [\href{https://arxiv.org/abs/1209.2127}{{\ttfamily 1209.2127}}].

\bibitem{Armas:2013aka}
J.~Armas, J.~Gath and N.~A. Obers, \emph{{Electroelasticity of Charged Black
  Branes}}, \href{http://dx.doi.org/10.1007/JHEP10(2013)035}{\emph{JHEP}
  {\bfseries 10} (2013) 035},
  [\href{https://arxiv.org/abs/1307.0504}{{\ttfamily 1307.0504}}].

\bibitem{Bhattacharya:2011eea}
J.~Bhattacharya, S.~Bhattacharyya and S.~Minwalla, \emph{{Dissipative
  Superfluid dynamics from gravity}},
  \href{http://dx.doi.org/10.1007/JHEP04(2011)125}{\emph{JHEP} {\bfseries 04}
  (2011) 125}, [\href{https://arxiv.org/abs/1101.3332}{{\ttfamily 1101.3332}}].

\bibitem{Armas:2016mes}
J.~Armas, J.~Gath, V.~Niarchos, N.~A. Obers and A.~V. Pedersen, \emph{{Forced
  Fluid Dynamics from Blackfolds in General Supergravity Backgrounds}},
  \href{http://dx.doi.org/10.1007/JHEP10(2016)154}{\emph{JHEP} {\bfseries 10}
  (2016) 154}, [\href{https://arxiv.org/abs/1606.09644}{{\ttfamily
  1606.09644}}].

\bibitem{Glorioso:2018wxw}
H.~Liu and P.~Glorioso, \emph{{Lectures on non-equilibrium effective field
  theories and fluctuating hydrodynamics}},
  \href{http://dx.doi.org/10.22323/1.305.0008}{\emph{PoS} {\bfseries TASI2017}
  (2018) 008}, [\href{https://arxiv.org/abs/1805.09331}{{\ttfamily
  1805.09331}}].

\end{thebibliography}
\end{document}